\documentclass[letter]{aa}
\usepackage[varg]{txfonts}
\usepackage{graphicx}
\usepackage{physics}
\usepackage{amsmath,amssymb}
\usepackage{bm}
\usepackage{array}
\newcommand{\dm}{\partial_\mu}
\usepackage{amsfonts}
\usepackage{amssymb}
\DeclareMathOperator{\Var}{var}
\DeclareMathOperator{\Cov}{cov}
\newcommand{\eq}[1]{Eq.~\eqref{#1}}
\usepackage{tikz} 
\usepackage{tikz-3dplot} 

\pdfpageattr {/Group << /S /Transparency /I true /CS /DeviceRGB>>}

\definecolor{Green}{RGB}{0,130,56}
\definecolor{emerald}{rgb}{0.3,0.85,0.2}
\definecolor{royalblue}{RGB}{65,105,225}
\definecolor{dodgerblue}{RGB}{30,144,255}

\usepackage{changes}
\definechangesauthor[name={Mohammad}, color=red]{MF}
\definechangesauthor[name={Pierre}, color=teal]{PA}
\definechangesauthor[name={Gwenael}, color=orange]{GB}
\definechangesauthor[name={Jacques}, color=Green]{JL}

\makeatletter
\renewcommand*\aa@pageof{, page \thepage{} of \pageref*{LastPage}}
\makeatother

\newcommand\MF[1]{{#1}}
\newcommand\MFR[2]{{#1}}
\newcommand\PA[1]{{#1}}
\newcommand\PAR[2]{{#1}}
\newcommand\PAN[2]{}
\newcommand\GB[1]{{#1}}
\newcommand\GBN[2]{}
\newcommand\GBR[2]{{#1}}
\newcommand\JL[1]{{#1}}

\bibpunct{(}{)}{;}{a}{}{,} % to follow the A&A style
\begin{document}
\title{The Resonant Tidal Evolution of the Earth-Moon Distance}
\author{Mohammad Farhat\inst{1}
\and Pierre Auclair-Desrotour\inst{1}
\and Gwena\"{e}l Bou\'{e}\inst{1}
\and Jacques Laskar\inst{1}}

\institute{IMCCE, CNRS, Observatoire de Paris, PSL University, Sorbonne Université, 77 Avenue Denfert-Rochereau, 75014, Paris, France}

\date{}
\abstract{Due to tidal interactions in the Earth-Moon system, the spin of the Earth slows down and the Moon drifts away. This recession of the Moon is now measured with great precision, but it has been realised, more than fifty years ago, that simple tidal models extrapolated back in time lead to an age of the Moon that is by far incompatible with the geochronological and geochemical evidence. In order to evade this \MFR{problem}{paradox}, more elaborate models have been proposed, taking into account the oceanic tidal dissipation. \MFR{However, these models did not}{but none so far has been able to} fit both the estimated lunar age and the present rate of lunar recession \MFR{simultaneously}{}. Here we present a physical model that reconciles these two constraints and yields a unique solution of the tidal history. This solution fits \MFR{}{remarkably well} \JL{well} the available geological proxies for the history of the Earth-Moon system  and consolidates the cyclostratigraphic method. The resulting evolution involves multiple crossings of resonances in the oceanic dissipation that are associated with significant and rapid variations in the lunar orbital distance, the Earth’s length of the day, and the Earth’s obliquity. }
\maketitle
\section{Introduction}
Due to the tidal interplay in the Earth-Moon system, the spin of the Earth brakes with time and the Earth-Moon distance increases  \citep{Darwin1879} 
at a present rate of $3.830\pm0.008$ cm/year that is measured using Lunar Laser Ranging  (LLR)   \citep{williams2016secular}. \MFR{There exists a rich narrative exploring the long term evolution of the system \citep{goldreich1966history, mignard1979evolution, touma1994evolution, neron1997long} and the dynamical constraints on the origin of the Moon \citep{touma1998resonances,cuk2019early}. Among all, it has been established that simple }{Simple} 
 tidal models starting with  the present recession rate and integrated backward in time predict a close encounter in the Earth-Moon system  within less than 1.6 billion years (Ga)  \citep{gerstenkorn1967controversy,macdonald1967evidence}. This is clearly not compatible with the estimated age of the Moon of $4.425\pm0.025$ Ga   \citep{maurice2020long},
which suggests that the present rate of rotational energy dissipation is much larger than it has typically been over the Earth's history.
To bypass this difficulty, empirical models have been fitted to  the available geological evidences
of the past rotational state of the Earth  \citep{walker1986,waltham_milankovitch_2015}, acquired  through the analysis of
paleontological  data (e.g.  \citep{williams2000geological}), sedimentary records of tidal rhythmites  \citep{williams1997precambrian,sonett1998neoproterozoic,williams2000geological,eriksson2000quantifying,de2017lunar}, or Milankovitch cyclostratigraphic sequences  \citep{meyers2018proterozoic,huang2020astronomical,sorensen2020astronomically,Lantink2021}. However, such models bring very little physical insight, and  the remaining uncertainty of the  geological data itself does not prevent circular arguments.

Major progress  was achieved with  the elaboration of oceanic tidal models. These models present a tidal frequency-dependent dissipation behavior  \citep{longuet1968eigenfunctions,platzman1984normal,muller2008large}, which allows for the encounter of high dissipation  resonant states during the Earth's history  \citep{webb1980tides,auclair2018oceanic,tyler2021tidal}.  However, there exists no effective model stemming from controlled analytical formulations that fits both the presently measured rate of lunar recession, and the estimated age of the Moon.  

Besides the dependence on the Earth's rotation rate, the varying continental configuration has also played a role in enhancing the oceanic resonances, or in exciting additional ones  \citep{platzman1983world,ooe1989effects,tyler2021tidal}. Paleo-dissipation might have also varied significantly during ice ages, as areas of continental shelves  vary with sea level  \citep{griffiths2009modeling,arbic2010coupled}. However, both ice ages and basin geometry cycles  \citep{boulila2018long} have much smaller periodicities compared to the Earth's age. Moreover, accurately accounting for such level of realism is hindered by the accumulating uncertainty in deep-time modelling. One has thus to compromise between the practicality of effective models with simplified geometries  \citep{webb1980tides,hansen1982secular,tyler2021tidal}, and the realism of costly numerical models that depend on paleogeographic reconstructions  \citep{green2017explicitly,daher2021long}. 

\MFR{Here we undertake a systematic exploration  of the time-varying tidal dissipation in the oceans and propose }{Here we present} a physical model that overlaps the two mentioned limits \MFR{(Section \ref{sec:oceanic_model})}{}. With a minimum number of free parameters, we constrain our model to only fit the two most certain points of the lunar evolution history: the present rate of lunar recession  and the lunar age \MFR{(Section \ref{sec:Constraining_parameters})}{}. This provides a unique solution to the Earth-Moon separation history \MFR{(Section \ref{sec:history_of_res})}{}. In this study, we \MFR{focus on the computation of the tidal response of the Earth, considering}{ consider} a reconstruction of  the continental drift up to one billion years ago,  followed by a  smooth transition towards a global ocean planet. \MFR{We accompany this computation with a reduced dynamical model of the system that captures the skeletal structure of the long term evolution based on robust features of the tidal response. However, we}{We}  anticipate this model to be the backbone of a fully spatial dynamical evolution in the \MFR{}{Earth-Moon} system \MFR{(more on that in Appendix \ref{app:orbital_dynamics}). The orbital solution that we produce demonstrates the robustness of the cyclostratigraphic machinery and further suggests interesting intervals for future investigations (Section \ref{sec:future_geo}) .    }{and to provide a reference for any geoscientific study that requires a history of the Earth-Moon distance.}

\section{Oceanic Model} \label{sec:oceanic_model}
\begin{figure}
    \includegraphics[width=\linewidth]{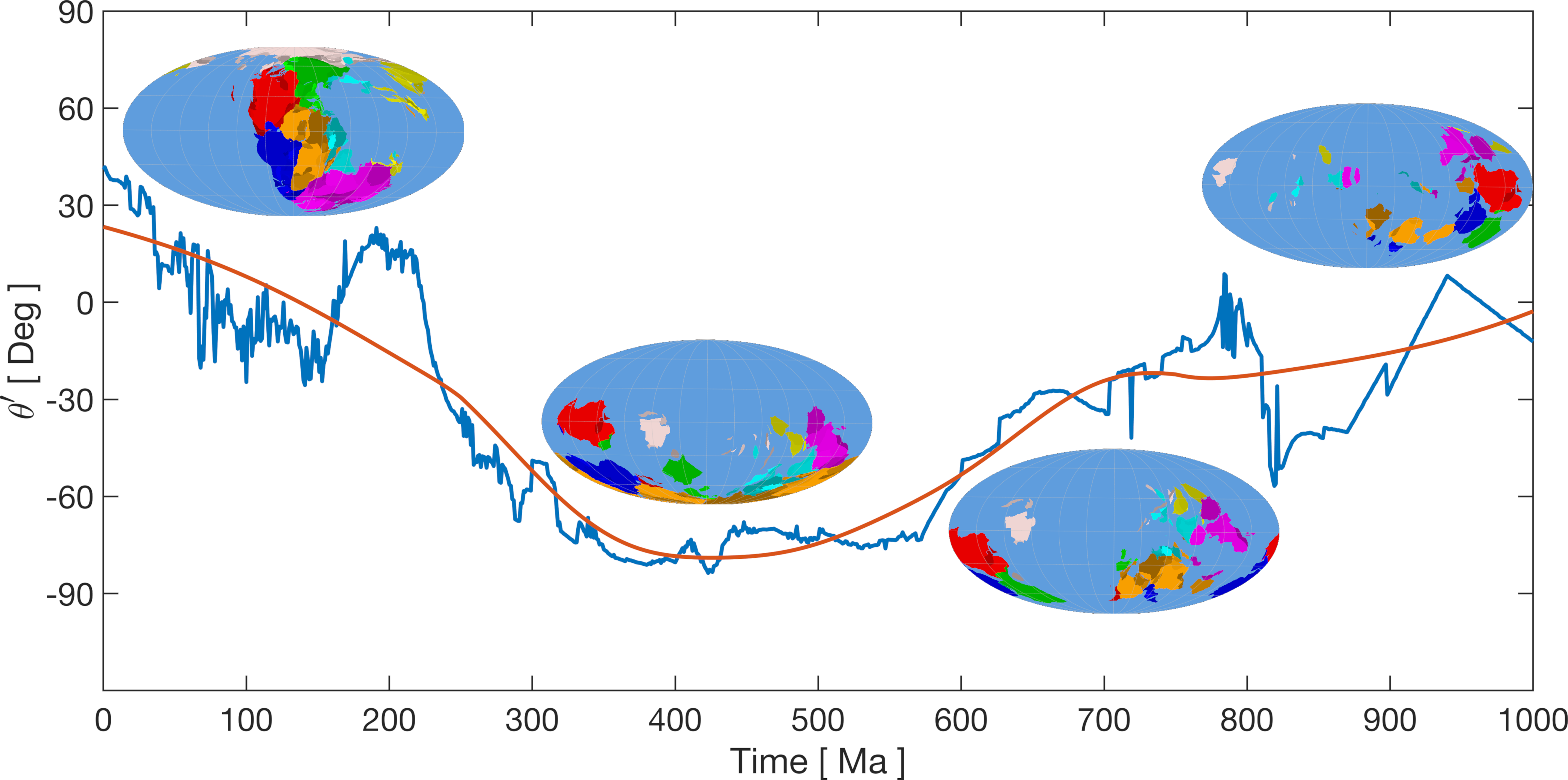}
    \caption{The temporal evolution of the latitude of the surface ``paleo-barycenter" over the last one billion years. The plate tectonics reconstruction is adopted from  \cite{merdith2021extending}, which establishes the first kinematically continuous tectonic motion model across multiple super-continental cycles. The evolution is smoothed in red using a moving polynomial regression filter with a window of 200 Myr. In our effective model, this curve maps the evolution of the center of the hemispheric continental cap that transitions from being symmetric about the equator during the Mesozoic, to being almost polar during the Paleozoic. }
    \label{geographic_center_evolution}
\end{figure}

%340
We compute the tidal response of the oceans and the solid-Earth to luni-solar semi-diurnal forcing, both combined with mimetic continental drift driven by plate tectonics. We focus on the dependence of dissipation on the Earth's spin rate. We combine two analytical approaches that describe long wavelength barotropic tidal flows over shallow spherical and hemispherical shells. The spherical shell describes a global ocean that we assume had existed in the earliest eons of the lifetime of the Earth  \citep{motoyama2020tidal}. Evidence on the existence of an early ocean is provided by the analysis of detrital zircon around $4.4$ Ga  \citep{wilde2001evidence}, the evidence on the interaction between the ocean and continental crust 4 billion years ago  \citep{mojzsis1996evidence}, and by the records of oxygen isotope composition of seawater  \citep{peck2001oxygen,johnson2020limited}. The  ``globality" of this ocean is justified by the analysis of continental crust growth curves based on geochemical evidence in zircon crystallization ages  \citep{dhuime2012change,hawkesworth2020evolution}. In compliance with these  curves, we consider that a hemispherical oceanic shell takes over in the most recent times. In our model, the center of this hemispheric continental cap  follows the evolution of the paleogeographic center. In doing so, we emphasize on the role of ``continentality'' in the tidal response, while avoiding the under-sampling of geometric scenarios due to theoretical limitations  \citep{hansen1982secular,tyler2021tidal} or due to uncertainties in plate tectonic models  \citep{matthews2016global,daher2021long}.  To compute this evolution, we adopt the recently developed paleogeographic reconstructions that cover the recent billion years  \citep{merdith2021extending}. Post-processing these reconstructions allows us to produce the latitudinal evolution of the center of the continental cap captured in Figure  \ref{geographic_center_evolution}. The tidal frequencies at which oceanic resonances are excited and the amplitudes of these resonances vary with the surface position of the hemispherical ocean (Figure \ref{TPW}). Super-continental formations and breakups  have thus their imprints on the predicted lunar recession rate. 
%307

\begin{figure*}[t]
    \centering
    \includegraphics[width=\textwidth]{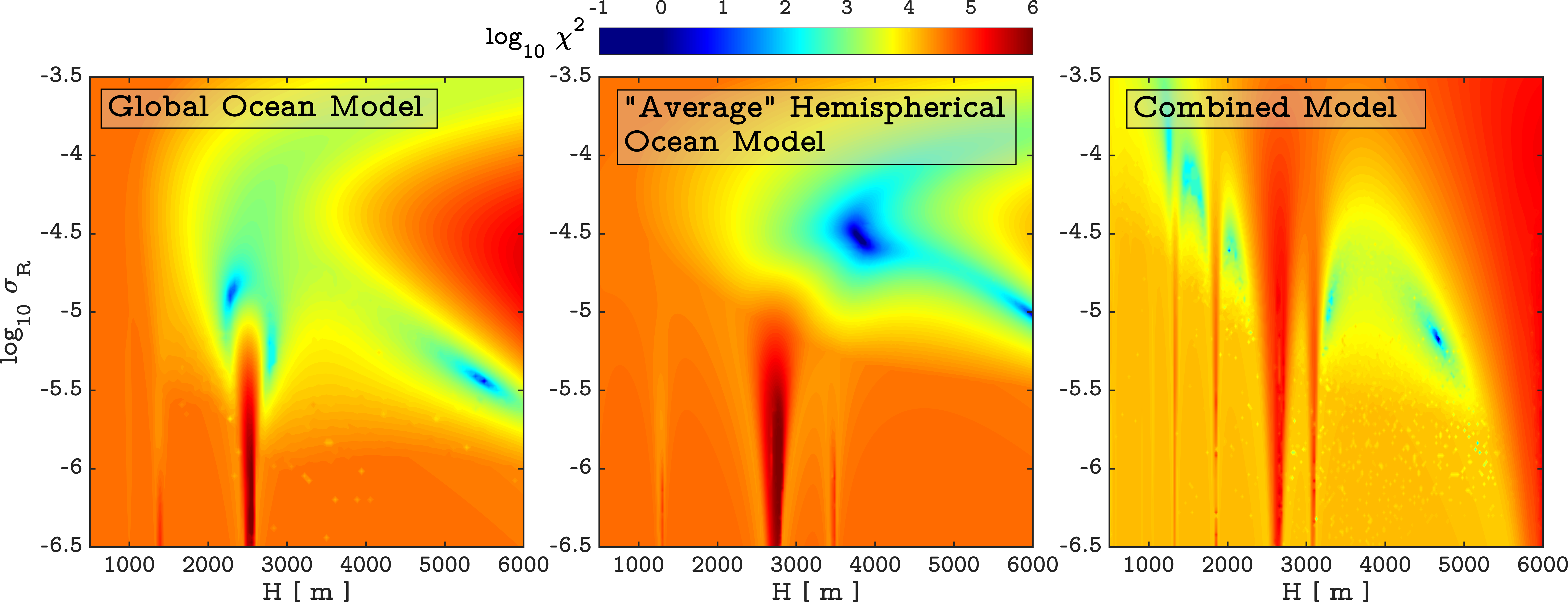}
    \caption{Misfit surfaces of $\chi^2$ for the three studied geometric models. The past dynamical evolution of the Earth-Moon system is reconstructed for the shown ranges of our two free model parameters $H$ and $\sigma_{\rm R}$. The misfit is established using the currently measured lunar recession rate via LLR, and the lunar age    (Appendix \ref{app:parameters_fit}). The three models differ in the imposed geometry of the oceanic shell over the geological history, with the combined model featuring more physical realism that the other two. The numerical results of this analysis are summarized in (Table \ref{table_misfit}). The dynamical evolution associated with each of the misfit minima is plotted in terms of the lunar semi-major axis in Figure  \ref{a_EM}, length of the day in Figure  \ref{LOD}, and obliquity and precession frequency in Figure  \ref{pre}. }
    \label{misfit_surfaces}
\end{figure*}

The dynamical evolution of the Earth-Moon system is  coupled to the tidal flows computation, which is also dependent on the chosen oceanic geometry. For the global ocean, the tidal torque is computed by solving the modified Laplace tidal equation using the Hough functions as eigenfuctions  (see Appendix \ref{global_response} and \cite{auclair2019final}). The oceanic dissipation is parametrized by an effective frequency $\sigma_{\rm R}$ that globally models the bottom friction and the conversion of barotropic flows into internal gravity waves, both mechanisms amounting to $\sim 91\%$ of the total dissipation  \citep{carter2008energetics}. This frequency $\sigma_{\rm R}$ can also be interpreted as the inverse of a dissipation timescale $\tau$ that quantifies the time needed to deplete the kinetic energy budget of tidal oscillations after switching  the forcing off. Although $\sigma_{\rm R}$ is probably a function of local topography, its spatial variation can be averaged out longitudinally over the Earth's rotation and latitudinally over precession and plate tectonics. The second free parameter in our model is the uniform effective oceanic thickness $H$. The imprints of these  two parameters on the tidal response spectrum are distinguishable: variations in $H$ smoothly shifts the positions of the resonant peaks while slightly varying their amplitudes. In contrast, variations in $\sigma_{\rm R}$ can completely reshape the tidal spectrum,  amplifying the resonant peaks by orders of magnitude when $\sigma_{\rm R}$ decreases, or completely absorbing the resonant peaks into the background spectrum otherwise (Figure \ref{compare_global_hemi_torque}). For the hemispherical geometry, we adopt the analytical approach of   \cite{webb1980tides} (see Appendix \ref{hemi_response}), in which the tidal solution is expanded in spherical harmonics (Figure \ref{fig:eigenfunctions}). In both geometries, we take into account the effect of the deformation of the solid part of the Earth adopting an Andrade rheology  \citep{Castillo-Rogez,renaud2018increased} (Appendix \ref{section_coupling}).

\section{Constraining effective parameters} \label{sec:Constraining_parameters}
 Assuming a reduced planar orbital model   (Appendix \ref{app:orbital_dynamics}), we compute the evolution of the Earth-Moon system that results from the luni-solar semi-diurnal tidal torque for ranges of values of our effective parameters $(H,\sigma_{\rm R})$. We do so for three models that ascend in realism: a global ocean model across the full geological history (similar to   \cite{tyler2021tidal}); an ``average'' hemispherical ocean model across the full geological history (similar to   \cite{webb1982tides}), for which the response at any tidal frequency is averaged over all possible oceanic positions on the sphere; and  our combined model that starts at the present with the hemispherical ocean evolving with the mimetic continental drift, then switches to the global ocean. For every constructed history of the Earth-Moon separation, we compute the  chi-squared $\chi^2$, taking only two data points into account: the well constrained lunar age of $4.425\pm0.025$ Ga   \citep{maurice2020long}, and the currently measured rate of lunar recession of $3.830\pm0.008$ cm/year   \citep{williams2016secular}. Misfit surfaces of $\chi^2$  for the three models are shown in Figure  \ref{misfit_surfaces}.  Two $\chi^2$ local minima exist for the global oceanic response. However, one of them corresponds to an unreasonably large average oceanic depth $H\approx 5500$ m, leaving us with a global minimum of $(H, \log_{10}\sigma_{\rm R})$= (2273 m, -4.89), where $\sigma_{\rm R}$ is in s$^{-1}$. The global minimum in the ``average'' hemispherical ocean model corresponds to   $(H, \log_{10}\sigma_{\rm R})$= (3816 m, -4.54), which is close to the average depth of the pacific ocean  \citep{amante2009etopo1}. For the combined model, the global minimum corresponds to $(H, \log_{10}\sigma_{\rm R})$=(4674 m, -5.19), where $H$ here is the thickness for the hemispherical phase of the model and twice that of the global ocean phase during earlier eons   (Appendix \ref{app:continental_drift}). The switch between the two geometries occurs at $t_{\rm switch}$, which is implicitly determined by the dynamical integrator   (Appendix \ref{app:continental_drift}). For the best fit solution, $t_{\rm switch}=3.25$ Ga, in agreement with suggestions  \citep{dhuime2012change,hawkesworth2020evolution} of the existence of a global ocean until $\sim2.5$ Ga. If one assumes that the oceanic volume is conserved in time, the best fit $H$ of the combined model corresponds to a volume of $1.19\times10^{18}$ m$^3$, only $10\%$ off the presently estimated value of $1.33\times10^{18}$ m$^3$ by global relief models  \citep{amante2009etopo1}.  The fitted dissipation frequency $\sigma_{\rm R}$ corresponds to a decay time $\tau=43.1$ hr, consistent with real oceanic studies \citep{GARRETT1971493,webb1973age} that suggest a range between 24 and 60 hr (or $\log_{10}\sigma_{\rm R}\in[-4.93,-5.33]$). The best fit values for the combined model correspond to a lunar trajectory characterized by a present rate of recession $\Dot{a}_0= 3.829$ cm/yr, and an impact time at $4.431$ Ga (Table \ref{table_misfit}).

\section{The Earth-Moon Separation: A History of Surfing Resonances}
\label{sec:history_of_res}
\begin{figure}
    \centering
    \includegraphics[width=\linewidth]{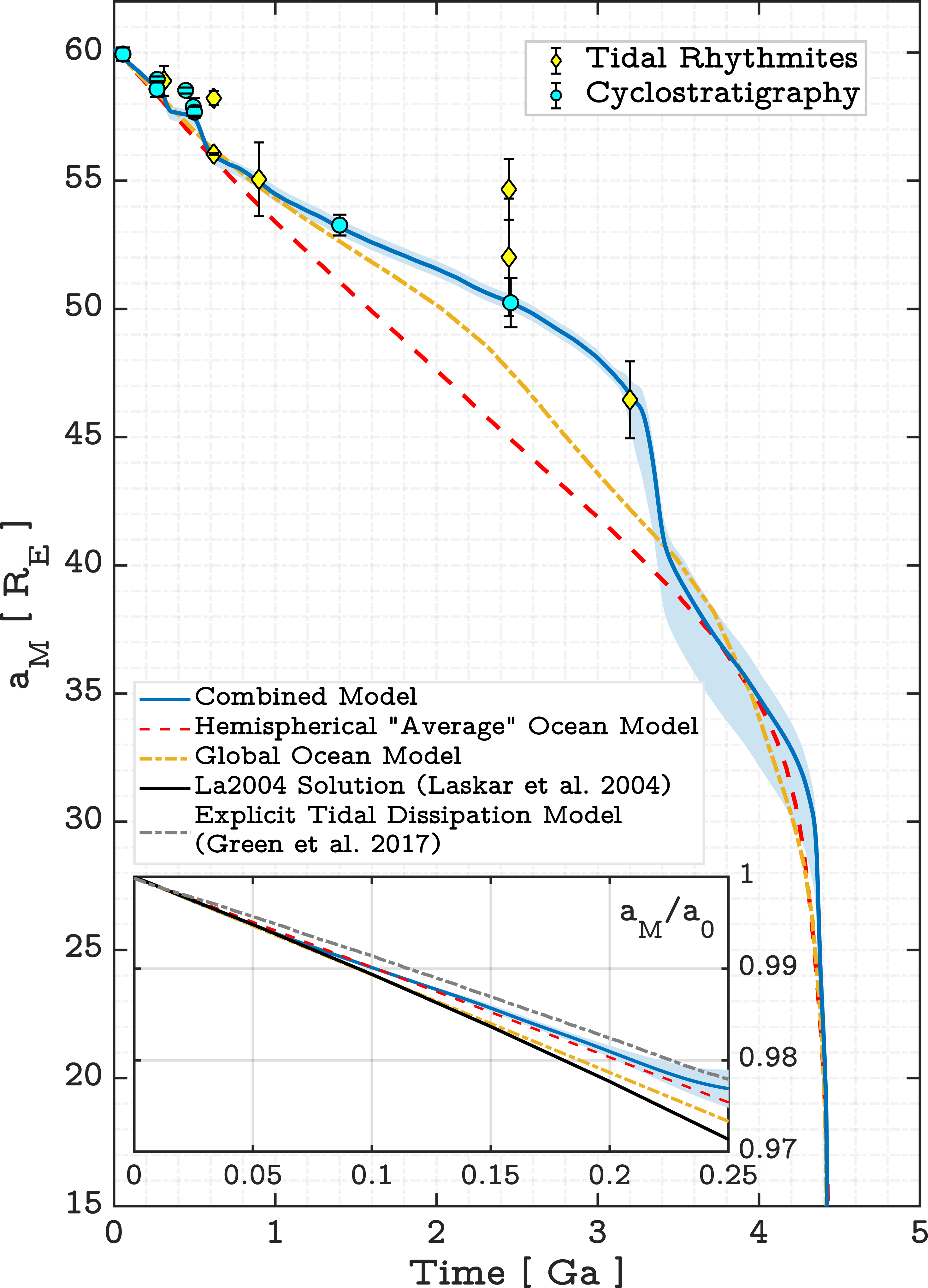}
    \caption{The evolution of the lunar semi-major axis with time. The Earth-Moon separation $a_{\rm M}$ is plotted for the three studied models taking the best fit values of the free parameters ($H,\sigma_{\rm R})$ as described in Figure  \ref{misfit_surfaces} and in the main text. Plotted on top of the evolution curves are geological inferences of $a_{\rm M}$ from cyclostratigraphy and tidal laminae data (Tables \ref{table:cyclo} and \ref{table:ry}). The shaded envelope  corresponds to $2\sigma$-uncertainty in the fitted parameters of the combined model  (Appendix \ref{app:parameters_fit}). In the narrow window we zoom over the  most recent 250 Myrs of the evolution, and we compare to the evolution corresponding to explicit numerical tidal modelling using paleogeographic reconstructions \citep{green2017explicitly}, and the prediction of the  numerical solution La2004 \citep{laskar2004long}.  }
    \label{a_EM}
\end{figure}

\begin{figure}
    \centering
    \includegraphics[width=\linewidth]{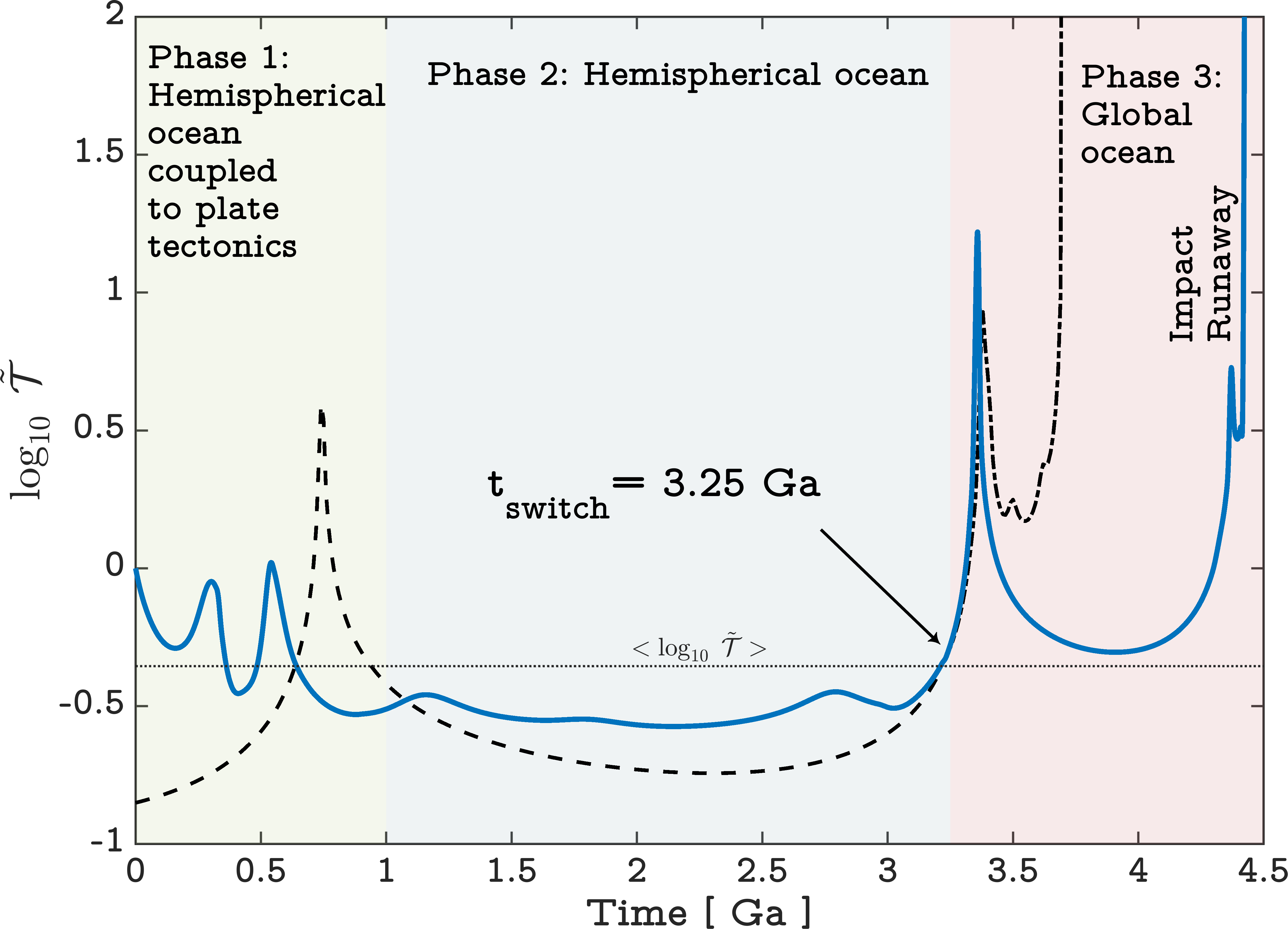}
    \caption{A history of the tidal torque.  The logarithm of the  semi-diurnal tidal torque of the Earth  (normalized by its present value: $\Tilde{\mathcal{T}}= \mathcal{T}/\mathcal{T}(t=0)$) is plotted as a function of time. The solid curve corresponds to the torque of the combined model that involves three phases: in the first phase, a hemispherical ocean migrates on the surface of the Earth following the evolution of the continental barycenter of Figure  \ref{geographic_center_evolution}. Lacking a continuous plate tectonics model beyond 1 Ga, in Phase 2 we fix the hemispherical ocean to its configuration at 1 Ga to avoid modelling discontinuities. It is noteworthy that the attenuated tidal torque over this phase is not due to the fixed  oceanic position but due to the tidal response occupying the non-resonant background of the spectrum for the tidal frequencies associated with this interval. Beyond $t_{\rm switch}$, we enter Phase 3 of the model with  the global ocean configuration. Dashed and dashed-dotted curves correspond respectively to the global and hemispherical oceanic torques that are ignored over the specified intervals by the selective combined model. }
    \label{torque_M_combined}
\end{figure}

For each of the global minima of the misfit parametric studies, we plot the evolution of the Earth-Moon distance in Figure  \ref{a_EM}. On top of the evolution, we spread a compilation of geological proxies from tidal rhythmites \citep{walker1986,sonett1998neoproterozoic,williams2000geological,eriksson2000quantifying,de2017lunar} and cyclostratigraphy \citep{meyers2018proterozoic,huang2020astronomical,sorensen2020astronomically,Lantink2021} (Tables \ref{table:cyclo} and \ref{table:ry}). The associated evolution of the Earth's length of the day (LOD), precession frequency, and obliquity are plotted in Figures  \ref{LOD}, \ref{pre}. The three models are constrained at the end points,  thus differences arise mostly in between. To better elaborate on the models' discrepancies, we plot in Figure  \ref{torque_M_combined} the temporal evolution of the tidal torque (normalized by its present value) associated with the combined model. Being directly proportional to tidal dissipation, the long term evolution of the torque is characterized by a non-monotonic increase, characteristic of the shrinking Earth-Moon separation, and interrupted by multiple crossings of resonances. The distribution of resonances in the hemispherical configuration ($t<t_{\rm switch}$) is less regular than that in the global configuration ($t>t_{\rm switch}$) (see also Figs. \ref{compare_global_hemi_torque} and \ref{hemispheric_pure_vs_solid} for a global description of the tidal response spectrum). Each resonance crossing in the torque generates an inflection point in the evolution of $a_{\rm M}$, which depends on the  width and to a lesser degree on the amplitude of the resonance peak \citep{auclair2014impact}. Figure \ref{torque_M_combined} depicts a critical feature of the combined model: starting with the hemispherical geometry at the present locates the torque around a resonance peak, which provides a  higher dissipation rate than for the global ocean configuration. This models the anomalous present rate of dissipation attributed to the blocking of westward tidal propagation by the current continental distribution and the effect of enhanced dissipation by continental shelves \citep{arbic2009tidal}. The first phase  of the model involves two major resonances between the present and $700$ Ma, resulting in cascade falls of $a_{\rm M}$ of $2.8R_{\rm E}$ within 330 Myr. These resonances  are associated with rapid variations of the Earth's obliquity (Figure  \ref{pre}) that could have triggered major climatic events. We observe that the first resonance overlaps  with the Palaeozoic oxygenation event ($\sim350$ Ma), while the second overlaps with the Neoproterozoic major oxygenation event ($\sim600$ Ma) and the Cambrian Explosion \citep{wood2019integrated}. Possible correlation between the Earth's LOD and the benthic ecosystem should thus be considered  \citep{klatt2021possible}. The second resonance peak is almost half an order of magnitude lower than that in the global configuration. This is an essential feature of the combined model for preserving the lunar angular momentum budget at this stage to better match the cyclostratigraphic proxy estimates at 1.4 and 2.5 Ga, which clearly cannot be explained by the other more dissipative models considered in Figure  \ref{misfit_surfaces}.

Following these resonances, the torque enters a long non-resonant interval associated with the intrinsic tidal response occupying the background of the spectrum (Figure  \ref{compare_global_hemi_torque}). This ``dormant'' torque phase covers the interval of the so-called ``boring billion years'' associated with stabilized rates of atmospheric oxygenation \citep{alcott2019stepwise}. Entering the oceanic global geometry phase of the combined model occurs at 3.25 Ga, namely after covering all significant super-continental cycles, although $t_{\rm switch}$ is implicitly determined by the dynamical integrator  (Appendix \ref{app:continental_drift}). Samples of continental growth curves predict a fast decay in continental crust volume beyond $t_{\rm switch}$ \citep{sun2019crustal,hawkesworth2020evolution}. After switching to the global ocean response spectrum, the torque passes through a major resonance around 3.35 Ga, resulting in a significant and abrupt drop in $a_{\rm M}$ of $6.5R_{\rm E}$ within $250$ Myr. Beyond this age, the evolution follows again the tidal dissipation background spectrum before terminating with the impact.

\section{A new target for geological studies}
\label{sec:future_geo}
\begin{figure}
    \centering
    \includegraphics[width=\linewidth]{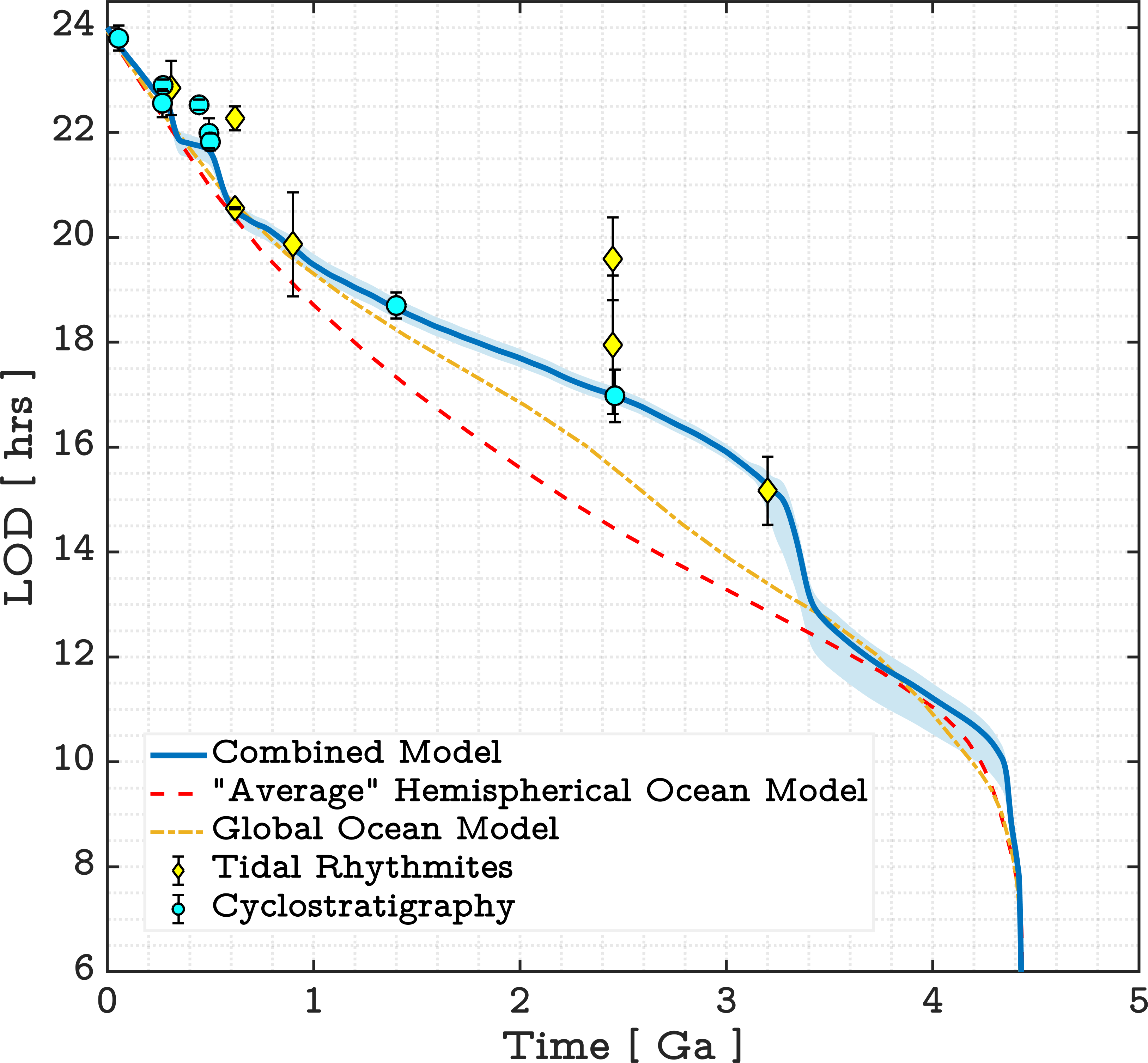}
    \caption{Evolution of the Earth's length of the day with time. Similar to Figure  \ref{a_EM}, but now for the LOD evolution associated with the three studied oceanic models. Geological data on the LOD are summarized in Tables \ref{table:cyclo} and \ref{table:ry}.   }
    \label{LOD}
\end{figure}

\begin{figure}
    \centering
    \includegraphics[width=\linewidth]{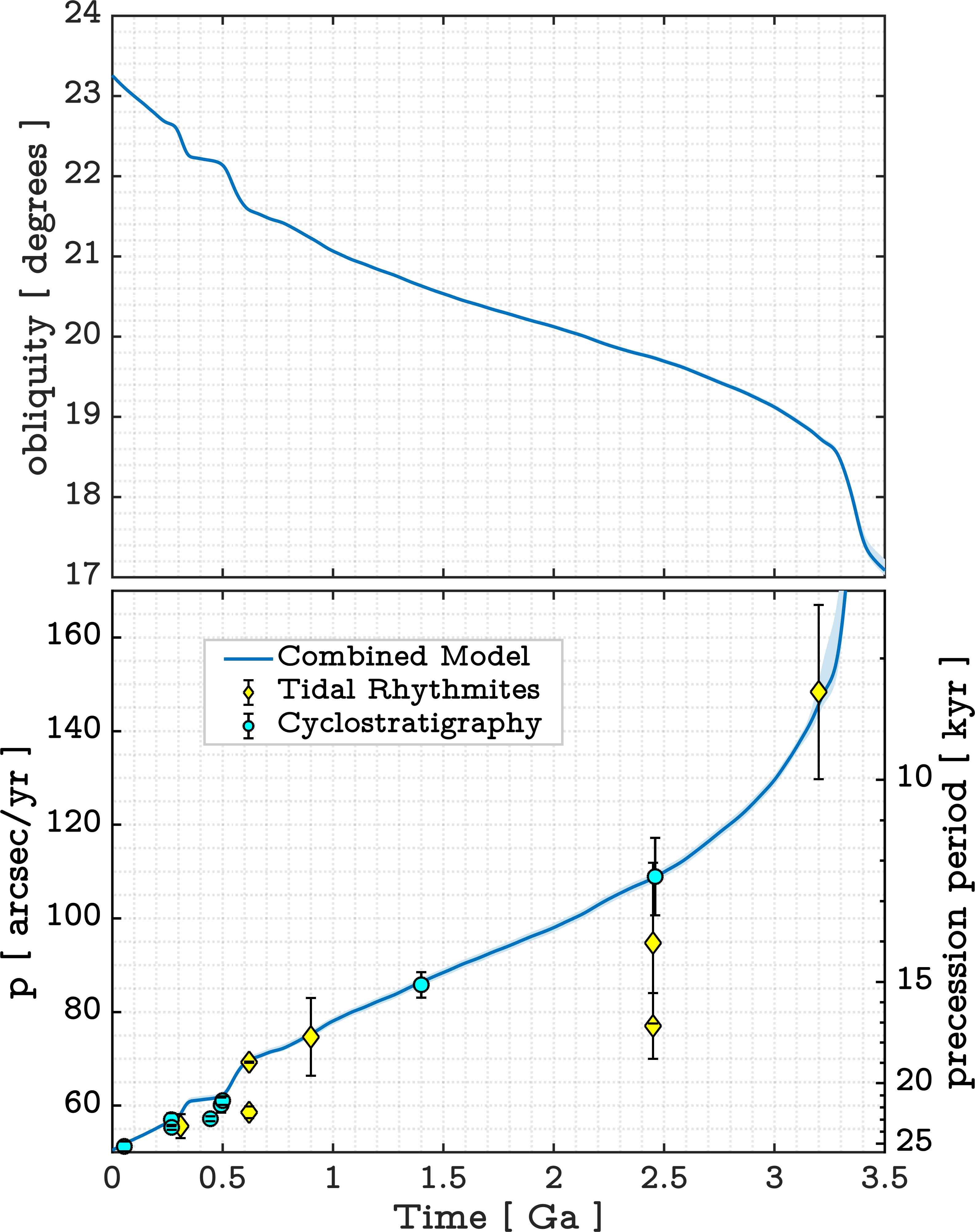}
    \caption{Evolution of the Earth's obliquity, precession frequency, and precession period with time. The evolution of $a_{\rm M}$ (Figure \ref{a_EM}) and LOD (Figure  \ref{LOD}) are used to compute the evolution of obliquity and precession by  \eqref{eq:eps} and \eqref{precessioneq}. The geological data of the precession frequency from tidal rhythmites and cyclostratigraphy are also plotted on top of the curve (Tables \ref{table:cyclo} and \ref{table:ry}). We note that the precession frequency is the directly measured observable in cyclostratigraphy. }
    \label{pre}
\end{figure}

In this article we built the first semi-analytical physical model that fits the most accurate constraints in the Earth-Moon evolution: the present tidal dissipation rate and the age of the Moon. We have deliberately avoided to fit our model to any of the available geological data. In an amazing way, the unique solution of our combined model is a nearly  perfect match to a large set of those geological data (Figure  \ref{a_EM} and  Figs. \ref{LOD}, \ref{pre}). This solution will provide a new target for geological studies. It clearly validates the cyclostratigraphic approach,  which estimates the Earth's precession frequency from stratigraphic sequences  \citep{meyers2018proterozoic,huang2020astronomical,sorensen2020astronomically,Lantink2021} (Table \ref{table:cyclo}). In particular, the cyclostratigraphic evaluation of the Earth-Moon distance at $2 459\pm1.3$ Ma in the Joffre banded iron formations (BIF) \citep{Lantink2021} is in remarkable agreement with our model, compared to the equivalent estimates deciphering tidal rhythmites in the ($\sim 2450$ Ma) Weeli Wooli BIF  in Australia \citep{walker1986,williams2000geological}. Our target curve can probably now be used to elaborate robust procedures for the analysis of these tidal rhythmites  that led sometimes to divergent interpretations \citep{walker1986,sonett1998neoproterozoic,williams2000geological} (Table \ref{table:ry}). We obtain a striking fit with the estimate of $a_{\rm M}$ at 3.2 Ga obtained through the analysis of the Moodies group rhythmites \citep{eriksson2000quantifying,de2017lunar}, but we do not deny that this agreement could be coincidental, and a new analysis of these sections, associated with  cyclostratigraphic estimates, is certainly welcome.
We expect that large  progress will be made in the near future with the analysis of many 
cyclostratigraphic records, which  could then be used to constrain even more our physical model. Of particular interest are the sequences that occur during the resonant states (or in their vicinity), corresponding to the steep slopes in Figure  \ref{a_EM}.
Finally, as this model provides a coherent history of the Earth-Moon distance, it  can  also be used to constrain the time scale of  lunar formation scenario \citep{cuk2016tidal}. \JL{This coherence between the geological data and the present scenario for the Earth-Moon evolution will also promote the use of these geological data, and in particular of the cyclostratigraphic geological data as a standard observational window for recovering the past history of the solar system.}

\begin{acknowledgements} 
We thank Ma\"{e}lis Arnould for her help with the plate tectonics model and the \texttt{GPlates} software and Matthias Sinnesael for  discussions on the geological data. We are grateful to Margriet Lantink and coworkers 
for the  communication of  their results on the Joffre sequence before publication and allowing us to include their data point  
in the present work. This project has been supported by the 
French Agence Nationale de la Recherche (AstroMeso ANR-19-CE31-0002-01)
and by the European Research Council (ERC) 
under the European Union’s Horizon 2020 research and innovation program (Advanced Grant AstroGeo-885250). 
This work was granted access to the HPC resources of MesoPSL financed by the Region 
Île-de-France and the project Equip@Meso (reference ANR-10-EQPX-29-01) of the 
programme Investissements d’Avenir supervised by the Agence Nationale pour la Recherche. 
\end{acknowledgements}
\begin{contributions}
MF conducted the simulations and drafted the paper. PAD brought expertise in oceanic tides and GB in solid tides. JL initiated the study and supervised it. All contributed to the study at all stages. All contributed to the writing of the paper.
\end{contributions}

\bibliographystyle{aa} % style aa.bst
%\bibliography{bibliography} % your references Yourfile.bib

\begin{appendix}

\section{Orbital dynamics}\label{app:orbital_dynamics}
For the reconstruction of the Earth-Moon distance, we use a reduced secular dynamical model describing the exchange of angular momentum between the Earth's rotation and the lunar orbital motion, ignoring the Earth's obliquity, lunar eccentricity, and lunar inclination. \MF{This simplification allows for a systematic understanding of the hierarchically complex contributions of multiple intervening players.} 
\JL{This is justified by considering that large values of  the inclination or eccentricity of the Moon could only be reached when the Earth-Moon distance is smaller than $30 R_{\rm E}$ } \citep{goldreich1966history,touma1994evolution,rubincam2016tidal}, \MF{which corresponds to the very early stage of the system (the Moon spends $97\%$ of its lifetime beyond this limit; see Figure \ref{a_EM}).}  Thus, this reduced model provides the skeleton of the secular evolution in the system around which full spatial dynamics can flesh; \JL{this latter would also require to extend the oceanic tidal model to the obliquity component, which could be the 
task of a next stage of this work. }{} Other effects such as climate friction  \citep{levrard2003climate} and core-mantle coupling \citep{neron1997long,touma2001nonlinear} are ignored, as well as early history resonances \citep{touma1998resonances} and halts of tidal interaction due to Laplace plane transitions  \citep{cuk2016tidal}.  {} Under these assumptions, the governing dynamical system of equations reads
\begin{align}
\frac{dL_\Omega}{dt}&= -\left(\mathcal{T}_{\rm M} + \mathcal{T}_{\rm S}\right) \ , \\
    \frac{dL_{\rm M}}{dt}&= \mathcal{T}_{\rm M} \ ,
\end{align}
where $\mathcal{T}_{\rm M}$ is the lunar semi-diurnal tidal torque coupling between the oceanic and the solid response of the Earth, and $\mathcal{T}_{\rm S}$ is its solar counterpart. The orbital angular momentum of the Moon  $L_M=\beta\sqrt{G(M_{\rm E}+M_{\rm M}) a_{\rm M}}$, where $\beta=M_{\rm E}M_{\rm M}/(M_{\rm E}+M_{\rm M})$ is the Earth-Moon system's reduced mass. The rotational angular momentum of the Earth is defined as $L_\Omega= C(\Omega)\Omega$, with the time varying principal moment of inertia given by \citep{goldreich1966history}
\begin{equation}    \label{eq:C}
    C(\Omega)= C(\Omega_0) + \frac{2 k_2^f R_{\rm E}^5 }{9G}(\Omega^2-\Omega_0^2) .
\end{equation}
% \begin{equation}
%     C(\Omega)= C_0 + \frac{2k_2^f R_{\rm E}^5\Omega^2}{9G}.
% \end{equation}
Here $k_2^f$ is the second degree fluid Love number of centrifugal/tidal deformation, and $G$ is the gravitational constant. 
The differential equation is integrated backwards in time using  Runge-Kutta 9(8) method. The tidal torque computation is coupled to the orbital integrator and is computed simultaneously at each step. It takes the model parameters $(H,\sigma_{\rm R})$ as input, and the system's variables $a_{\rm M}$ and $\Omega$ to compute the tidal frequency, and consequently the coupled tidal response.  

Once the lunar semi-major axis ($a_{\rm M}$) and the rotation speed of the Earth ($\Omega$) are determined, we compute the obliquity of the Earth ($\epsilon$) and the precession frequency \citep{laskar2004long} $p$ as derived quantities.
Starting with equations (40) and (46) from \cite{correia_tidal_2010} in the case of  zero eccentricity, we obtain 
\begin{equation}
\frac{d\epsilon}{dt} = \frac{K n}{C(\Omega)\Omega} \sin \epsilon \left(  \frac{\Omega}{2n_{\rm M}} \cos \epsilon-1 \right)  ,
\end{equation}
and
\begin{equation}
\frac{da_{\rm M}}{dt} = \frac{2K }{\beta a_{\rm M}} \left(  \frac{\Omega}{n_{\rm M}}\cos \epsilon -1 \right) ,
\end{equation}
that is  
\begin{equation}
\frac{d \epsilon}{d a_{\rm M}} = \frac{\beta n_{\rm M} a_{\rm M}}{4C(\Omega)\Omega}   \sin \epsilon \frac{ \Omega\cos \epsilon  -2n_{\rm M} }{ \Omega\cos \epsilon  -n_{\rm M}} .
\label{eq:eps}
\end{equation}
We note that the tidal response parameter,  $K$, disappears from the equations. This would also be the case if $K$ depended on $\Omega$.  
The obliquity evolution equation  (\ref{eq:eps}) is integrated using the values of $a_{\rm M}$ and $\Omega$ that result from the tidal flows and orbital dynamics coupled system. The precession frequency $p$ is then derived using equations (6) and (8) from  \cite{laskar2004long} with zero eccentricity and inclination, that is 
\begin{equation}\label{precessioneq}
    p=\frac{3}{2}\left(\frac{GM_{\rm S}}{a_{\rm E}^3} + \frac{GM_{\rm M}}{a_{\rm M}^3}\right)E_d(\Omega_0)\frac{\Omega}{\Omega_0^2}\cos \epsilon .
\end{equation}
In \eqref{eq:eps} and \eqref{precessioneq}, the used constant values for the Earth's radius $R_{\rm E},$ the gravitational constant of the Moon $GM_{\rm M}$, and the Sun $GM_{\rm S}$, the mass ratio $M_{\rm E}/M_{\rm M}$, the rotational velocity $\Omega_0$, the Earth's semi-major axis $a_{\rm E}$, and the inertia parameter $C(\Omega_0)/M_{\rm E} R_{\rm E}^2 $ are adopted from INPOP21  \citep{INPOP21a}. The dynamical ellipticity at the origin of date, $E_d(\Omega_0)=0.003243$, is determined from the initial conditions for the obliquity ($\epsilon_0)$ and precession $(p_0)$ adopted from the La2004 solution \citep{laskar2004long}. All values of used parameters are summarized in  Table \ref{table_values}.

This derivation of the obliquity and precession frequency evolutions is only valid in the limit of a distant Moon, namely when the Moon is beyond its Laplace radius  \citep{boue2006precession,farhat2021laplace} and its Laplace plane is the ecliptic rather than the Earth's equatorial plane  \citep{tremaine2009satellite}. In our $a_{\rm M}$ evolution of Figure \ref{a_EM}, the Laplace regime transition occurs very early in the evolution ($t>4 $Ga), thus in  Figure  \ref{pre}, we plot the evolution of the precession frequency and obliquity between the present and $3.5$ Ga. We also scatter on the curve the geological inferences of the precession frequency, which in the case of cyclostratigraphy is the direct observable (Table \ref{table:cyclo}).

\section{Continental drift and oceanic geometry shifting}\label{app:continental_drift}
Our dynamical integrator allows for variations in the oceanic geometry, be it a variation in the position of the oceanic hemisphere, or a shift between the hemispheric and global oceanic configurations. In the combined model, the first phase (Figure  \ref{torque_M_combined}) starts with the center of the continental cap following the evolution of the geographic center over the recent billion years. For this, we adopt a recent model, which reconstructs a kinetically continuous history of plate tectonics \citep{merdith2021extending}. The geographic center is traced by computing the surface projection of the ``barycenter'' of the continental distribution. This allows for a higher level of realism in oceanic  modelling. Beyond 1 Ga, and due to the lack of plate tectonic data, the model continues with the position of the ocean at 1 Ga. To post process the continental drift evolution, and to produce the time-sliced sketches of Figure \ref{geographic_center_evolution}, we used the  \texttt{GPlates} open-source reconstruction software  \citep{boyden2011next,gurnis2012plate}.  

At $1.5$ Ga, the integrator starts  computing simultaneously the tidal response of a global oceanic geometry, with uniform thickness $H_{\rm global}=H/2$, in order to guarantee oceanic volume conservation when we switch between the geometries. However, the hemispherical response remains the one accounted for in the dynamical evolution. While simultaneously computing both, the code detects when they equate, and switches to the global configuration identifying this time as $t_{\rm switch}$. The physical outcome of this process is guaranteeing a better compliance with continental crust growth curves \citep{dhuime2012change,hawkesworth2020evolution}, and thus avoiding effects arising from blocking of westward tidal propagation or enhanced continental dissipation at continental shelves \citep{arbic2009tidal}. The mathematical outcome of this process is evident in Figure \ref{torque_M_combined} in guaranteeing a smooth dynamical evolution of $a_{\rm M}$ (Figure \ref{a_EM}) without any discontinuities and modelling artifacts. For the misfit minimum of our combined model $t_{\rm switch}=3.25$ Ga. 
\begin{figure}
    \centering
    \includegraphics[width=.9\linewidth]{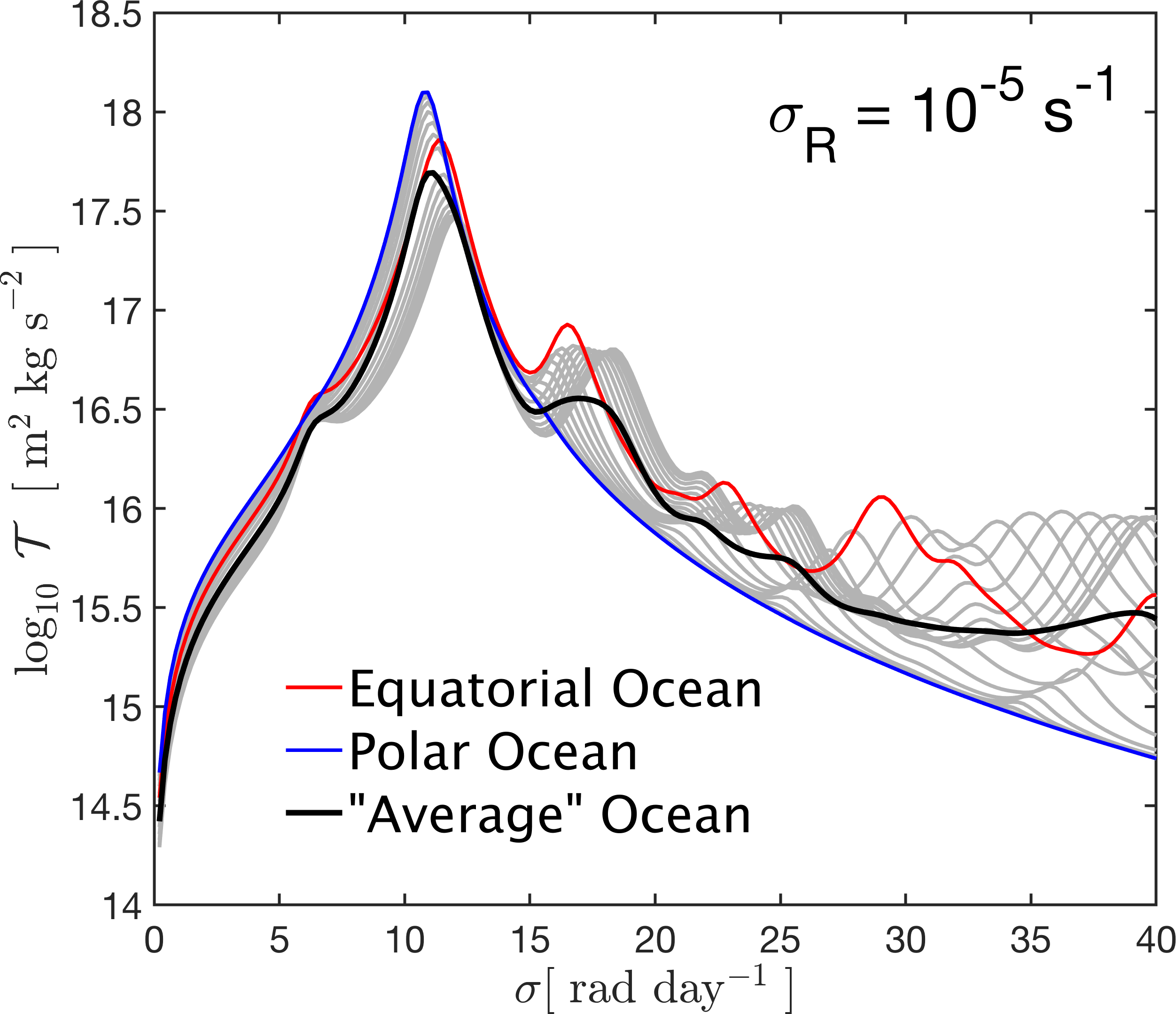}
    \caption{The drifting effect of the continental cap on the oceanic response: The tidal torque of a hemispheric ocean is plotted as a function of the forcing semi-diurnal frequency for different positions of the center of the ocean. With longitudinal symmetry, the latter is defined by the latitude of the oceanic center which evolves according to Figure \ref{geographic_center_evolution}. The drifting effect on the resonances ranges from position shifting and attenuation for small forcing frequencies, to major distortion in the spectrum at larger frequencies. Extreme distortion occurs in the polar oceanic scenario: the major resonance around $11$ rad/day reaches a maximum relative to other configurations, and the rest of the resonances are absorbed into the background leaving a unimodal spectrum. This behavior makes it important to take  into account the  position of the hemispherical cap into the model (Figure \ref{geographic_center_evolution}). }
    \label{TPW}
\end{figure}

\section{Parameters fit}
\label{app:parameters_fit}
\begin{table}
    \centering
    \begin{tabular}{p{1.7cm}p{1.85cm} p{1.9cm} p{1.85cm}}
        \hline\hline
    \textbf{Model} & \textbf{Global Ocean} & \textbf{Hemispherical Ocean} & \textbf{Combined Model} \\
    \hline
        $\sigma_{\rm R}$ [ s$^{-1} ]$ & $1.2770\times10^{-5}$   &$2.8860\times10^{-5}$ & $6.4417\times10^{-6}$\\
        $H$  [ m ]&2273 & 3816& 4674\\
        $\Dot{a}_0$[cm yr$^{-1}]$ &3.833 &3.828 &3.829\\
        $t_{\rm f}$  [ Ga ] &4.422 &4.432 &4.431\\
        $\chi^2$ &0.0775 &0.0705 &0.0345\\
      \hline
    \end{tabular}
    \caption{Summary of the misfit analysis showing the best fit values of the two free parameters $\sigma_{\rm R}$ and $H$ for each of the three studied models, along with the corresponding value of $\chi^2$, and the resulting lunar recession rate at the present $\Dot{a}_0$ and the impact time $t_{\rm f}$.}
    \label{table_misfit}
\end{table}
To construct the misfit surfaces of Figure  \ref{misfit_surfaces}, we compute the evolution for each pair of $(H,\sigma_{\rm R})$ on the two-dimensional grid. The present rate of lunar recession $\Dot{a}_0$ and the impact time $t_{\rm f}$ are then extracted for each evolution sample, and the mean square weighted deviation $\chi^2$ (Table 1) is then computed as 
\begin{equation}
    \chi^2 = \frac{1}{2} \left[\left( \frac{\Dot{a}_0-\Dot{a}_0^{\rm LLR}}{\sigma^{\rm LLR}} \right)^2 +  \left( \frac{t_{\rm f}-t_{\rm f}^{\rm geo}}{\sigma^{\rm geo}} \right)^2\right]\ ,
\end{equation}
where we use Lunar Laser Ranging (LLR) estimates of lunar orbital recession \citep{williams2016secular}: $\Dot{a}_0^{\rm LLR}\pm\sigma^{\rm LLR}= 38.30 \pm 0.08$ mm/year; and geochemical estimates of lunar formation time \citep{maurice2020long}: $t_{\rm f}^{\rm geo}\pm\sigma^{\rm geo}= 4.425 \pm 0.025$ Ga. The maximum likelihood detection problem is further optimized by fitting the surface around the minimum by a parabola to avoid the limitation of the grid resolution. 

We evaluate the uncertainties on the fitted parameters from those on the observables following the standard propagation of uncertainty method. Because of the absence of correlation between the two data $\dot a_0$ and $t_\mathrm{f}$, the entries of the variance matrix,
\begin{equation}
\Sigma = 
\begin{bmatrix}
\Var(H) & \Cov(H,\sigma_{\rm R}) \\[0.5em]
\Cov(H,\sigma_{\rm R}) & \Var(\sigma_{\rm R})
\end{bmatrix}\,,
\end{equation}
are given by
\begin{subequations}
\begin{eqnarray}
\Var(H) &=& \left(\frac{\partial H}{\partial \dot a_0}\right)^2\left(\sigma^{\rm LLR}\right)^2 + \left(\frac{\partial H}{\partial t_\mathrm{f}}\right)^2\left(\sigma^{\rm geo}\right)^2\,, \\[0.5em]
\Var(\sigma_{\rm R}) &=& \left(\frac{\partial \sigma_{\rm R}}{\partial \dot a_0}\right)^2\left(\sigma^{\rm LLR}\right)^2 + \left(\frac{\partial \sigma_{\rm R}}{\partial t_\mathrm{f}}\right)^2\left(\sigma^{\rm geo}\right)^2\,, \\[0.5em]
\Cov(H,\sigma_{\rm R}) &=& \frac{\partial H}{\partial \dot a_0}\frac{\partial \sigma_{\rm R}}{\partial \dot a_0} \left(\sigma^{\rm LLR}\right)^2 + \frac{\partial H}{\partial t_\mathrm{f}}\frac{\partial \sigma_{\rm R}}{\partial t_\mathrm{f}}\left(\sigma^{\rm geo}\right)^2\,.
\end{eqnarray}
\end{subequations}

The partial derivatives entering in these formulae are computed numerically from the fit of $(\dot a_0 \pm \sigma^{\rm LLR}, t_\mathrm{f})$ and $(\dot a_0, t_\mathrm{f} \pm \sigma^{\rm geo})$.
The marginal uncertainties on the parameters $H$ and $\sigma_{\rm R}$ are $\sigma_H = \Var(H)^{1/2}= 32.75 $ m and $\sigma_{\sigma_{\rm R}} = \Var(\sigma_{\rm R})^{1/2}= 0.2631  \times 10^{-6}$ s$^{-1}$, respectively. We use the variance matrix $\Sigma$ to evaluate the $2\sigma$-confidence ellipsoid around the best fit parameters $(H,\sigma_{\rm R})$. The Earth-Moon distance $a_M$ and the length of the day $\mathrm{LOD}$ have been integrated for 25 pairs of parameters $(H,\sigma_{\rm R})$ chosen at the boundary of this $2\sigma$-confidence region. Their envelop represents the $2\sigma$-uncertainty area plotted in shaded blue in (Figure \ref{a_EM}) and (Figs. \ref{LOD} and \ref{pre}).

\section{Geological data}
\MFR{In tables \ref{table:cyclo} and \ref{table:ry}, we compile geological data-sets that provide historical snapshots of the past rotational state of the Earth and the lunar orbital distance. }{}
\begin{table*}
\centering
\newcolumntype{R}{ >{${}}r<{{}$} }
\newcolumntype{C}{ >{${}}c<{{}$}}
\begin{tabular}{l l R R R R R R}
\hline\hline                        
\rule{0pt}{1em}
           dataset  & 
           \multicolumn{1}{c}{ Reference }  & 
           \multicolumn{1}{c}{ T [ Ga ]  }  & 
           \multicolumn{1}{c}{ $p$ [ $\text{arcsec} / \text{yr}$ ] }          &
           \multicolumn{1}{c}{$a_{\rm M}$ [ $R_{\rm E}$ ]}             & 
            \multicolumn{1}{c}{$a_{\rm M}$ [ km ]}             & 
           \multicolumn{1}{c}{LOD [ hr ]}        \\     
\hline
\rule{0pt}{1em}
 IC &\cite{laskar2004long}   &  0.000     &   \bf 50.467718 \phantom{.0}   &  60.142611\phantom{.0}  &383598  \phantom{\pm 10000}             &     24.00  \phantom{\pm 0.100}     \\
 Walvis Ridge &\cite{meyers2018proterozoic}  &  
 0.055     &   \bf     51.28\pm   1.02       &      59.94 \pm  0.26       &         382284   \pm  1650   &     23.80     \pm  0.24 \\
Lucaogou(a) &\cite{huang2020astronomical} &
 0.268     &   \bf     57.01\pm   1.37       &      58.57 \pm   0.31       &        373538  \pm   2000   &     22.56    \pm   0.27 \\
 Lucaogou(b) &\cite{huang2020astronomical}   &  
 0.270     &   \bf     55.36\pm   0.51       &      58.95 \pm   0.12       &        375967  \pm    750\phantom{0}   &     22.90    \pm   0.11 \\
 Yangtze Block &\cite{zhong_late_2020}   &  0.445     &   \bf     57.19\pm   0.53       &      58.52 \pm   0.12       &        373277  \pm    750\phantom{0}   &     22.53    \pm   0.10 \\
 Alum shale &\cite{sorensen2020astronomically}  &
 0.493     &   \bf     60.11\pm   1.59       &      57.88 \pm   0.34       &        369153  \pm   2200   &     21.99    \pm   0.28 \\
 Luoyixi &\cite{fang_cyclostratigraphy_2020}  &
 0.500     &   \bf     61.06\pm   0.94       &      57.67 \pm   0.20       &        367854  \pm   1300   &     21.82    \pm   0.16 \\
 Xiamaling &\cite{meyers2018proterozoic}  &
 1.400     &   \bf     85.79\pm   2.72       &      53.27 \pm   0.41       &        339777  \pm   2600   &     18.70    \pm   0.25 \\
 Joffre &\cite{Lantink2021}  &
 2.460     &   \bf    108.91\pm   8.28       &      50.24 \pm   0.96       &        320452  \pm   6100   &     16.98    \pm   0.50 \\
\hline

\end{tabular}
\caption{Cyclostratigraphic data. In boldface are the direct observables, i.e. here the precession frequency $p$ in arcsec/yr. The time of observation $T$ is in the second column. The semi-major axis of the Moon ($a_{\rm M}$) (in Earth radius ($R_{\rm E}$) or in km), and the length of the solar day (LOD), in hours, are derived from the observed quantities using the model that is presented in the text. These values may thus differ from the corresponding values published in the referenced publications. IC denotes the initial conditions \citep{laskar2004long}. The two values (a,b) for the Lucaogou dataset correspond to different analyses (a): TimeOptMCMC \citep{meyers2018proterozoic}; (b) obliquity and precession cycle counting \citep{huang2020astronomical}. Whenever it is specified in the original publication, the uncertainty in $p$ is set to $2\sigma$. The uncertainty of the other variables is propagated through the nominal solution of the present study.}   
 \label{table:cyclo} 
\end{table*}

\begin{table*}
\centering
\newcolumntype{R}{ >{${}}r<{{}$} }
\newcolumntype{C}{ >{${}}c<{{}$}}
\begin{tabular}{l l R R R R R R}
\hline\hline                        
\rule{0pt}{1em}
           dataset  & 
            \multicolumn{1}{c}{ Reference }  & 
            \multicolumn{1}{c}{ T [ Ga ]  }  &
           \multicolumn{1}{c}{ smo/yr}          &
           \multicolumn{1}{c}{ $p$ [ $\text{arcsec} / \text{yr}$ ] }          &
           \multicolumn{1}{c}{$a_{\rm M}$ [ $R_{\rm E}$ ]}             & 
           \multicolumn{1}{c}{LOD [ hr ]}        \\     
\hline
\rule{0pt}{1em}
IC &\cite{laskar2004long}           &  0.000  &    13.4289  \phantom{000.}      &50.467718 \phantom{00.}    & 60.142611\phantom{0.} &  24.00 \phantom{00000.} \\
Mansfield &\cite{sonett1998neoproterozoic}    &  0.310   &  \bf  13.86 \pm  0.21&   55.60 \pm 2.55 \phantom{0}  &     58.89\pm  0.59   &   22.85 \pm 0.52  \\
Elatina &\cite{sonett1998neoproterozoic}    &  0.620   &  \bf  14.93 \pm  0.01&   69.24 \pm 0.13 \phantom{0}  &     56.04\pm  0.03   &   20.56 \pm 0.02  \\
Elatina &\cite{williams1997precambrian,williams2000geological}     &  0.620   &  \bf  14.10 \pm  0.10&   58.55 \pm 1.24\phantom{0}   &     58.22\pm  0.28   &   22.27 \pm 0.23  \\
Cottonwood &\cite{sonett1998neoproterozoic}     &  0.900   &  \bf  15.33 \pm  0.60&   74.68 \pm 8.32  \phantom{0} &     55.06\pm  1.44   &   19.87 \pm 0.99  \\
Weeli Wolli &\cite{williams1990tidal,williams2000geological}     &  2.450   &  \bf  16.70 \pm  1.10&   94.71 \pm 17.19  &     52.01\pm  2.29   &   17.95 \pm 1.32  \\
Weeli Wolli &\cite{walker1986}     &  2.450   &  \bf  15.50 \pm  0.50&   77.04 \pm 7.03 \phantom{0}  &     54.66\pm  1.18   &   19.59 \pm 0.79  \\
Moodies Group &\cite{eriksson2000quantifying}   &  3.200   &                      &  148.36 \pm 18.61  & \bf 46.45\pm  1.50   &   15.17 \pm 0.65  \\
&\cite{de2017lunar}&&&&&\\
\hline

\end{tabular}
\caption{Tidal rhythmites data. In boldface are the observables. In general, the observable is the number of synodic lunar months per year or in an equivalent way, as quoted here, the number of sidereal lunar months per year (col. 3). The values are issued from the referenced publications (col. 1). For the Moodies Group, we could not infer this quantity from the original publication, and the corresponding estimate of the lunar semi-major axis was taken from  \cite{de2017lunar}. The semi-major axis $a_{\rm M}$ is obtained through Kepler's law ($n_{\rm M}^2a_{\rm M}^3=G(M_{\rm M}+M_{\rm E})$). As for the cyclostratigraphic data (Table.\ref{table:cyclo}), all other quantities ($p$, LOD)  are derived from the observed quantities using the model that is presented in the text. These values may thus differ from the corresponding values published in the referenced publications. The uncertainty of the observables are propagated to the derived variables through the nominal solution of the present study. The values at the origin ($T=0$) are from \cite{laskar2004long}. It should be noted that the present value of sidereal lunar months per year and lunar semi-major axis provided here for $T=0$ differs from some published value because we consider here averaged values, which should be the case for such long term studies (see Figure 18 from  \cite{laskar2004long}).}    
 \label{table:ry} 
\end{table*}

\begin{table*}[]
    \centering
    \begin{tabular}{p{9cm}p{8cm}  }
        \hline\hline
    \textbf{Parameter}& \textbf{Value} \\
    \hline
      Surface gravitational acceleration $(g)$  & 9.81 m s$^{-2}$ \\
      Earth radius  $(R)$  & 6378.1366 km \\
       Solar gravitational constant $(GM_{\rm S})$  &2.959122082853813556248$\times 10^{-4}$ AU$^3$ day$^{-2}$  \\
        Earth-Moon gravitational constant $(G[M_{\rm E}+M_{\rm M}])$  & 8.997011395221144381906$\times 10^{-10}$ AU$^3$ day$^{-2}$ \\
        Earth to Moon mass ratio $(M_{\rm E}/M_{\rm M})$ &81.30056789872074318737\\ 
        Uniform oceanic density $(\rho_{\rm oc})$ & 1022 kg m$^3$ \\
        Andrade characteristic time ($\tau_{\rm {\rm A}}$) &  $2.19\times 10^{4}$ yr \\
        Andrade rheological exponent ($\alpha_{\rm A}$) & 0.25\\
        Average rigidity of the deformable mantle ($\mu_{\rm E}$) & $17.3\times 10^{10}$ Pa  \\
        Average viscosity of the deformable mantle ($\eta_{\rm E}$) & $3.73\times 10^{21}$ Pa s\\
         Present day mean lunar semi-major axis ($a_0$) & 60.142611 $R_{\rm E}$\\
         Present day mean sidereal length of the day ($LOD_{\rm s}$) & 23.934468 hr\\
         Present day mean obliquity ($\epsilon_0$) & $23.2545^\circ$ \\
         Present day mean precession frequency ($p_0$) & $50.467718$ arcsec yr$^{-1}$\\
         Earth's semi-major axis $(a_{\rm E})$ & $1.495978707\times10^8$ km\\
         Earth's inertia parameter  ($C_0/(M_{\rm E}R^2)$) & 0.3306947357075918999972 \\
         Earth's fluid Love number  ($k_2^f)$ & 0.93 \\
      \hline
    \end{tabular}
    \caption{Values of constant parameters used in the numerical implementation of the theory. Astronomical values are adopted from INPOP21 \citep{INPOP21a}. Oceanic and rheological parameters are adopted from \citet{auclair2019final} and \cite{gerkema2008introduction}. The average rigidity is computed from the PREM model \citep{dziewonski1981preliminary}, while the average viscosity is computed from mantle viscosity inversions in \cite{lau2016inferences}. The initial conditions of the orbital integration are the mean elements from the La2004 astronomical solution \citep{laskar2004long}.} 
    \label{table_values}
\end{table*}

\section{The tidal response of a hemispherical ocean}\label{hemi_response}
This appendix builds towards computing the tidal response of a hemispherical ocean on the surface of the Earth. The formalism is heavily based on earlier works \citep{longuet1970free,webb1980tides} describing the free oscillations and the tidal response of a hemispherical ocean symmetric about the equator, and we expand here on it by adopting the true polar wander scenario \citep{webb1982tides} to solve for a general oceanic position. We note that  the mathematical formulation of the referenced works \citep{longuet1970free,webb1980tides,webb1982tides} contains several misprints  that we correct here. 

In the co-planar problem under study (ignoring the Earth's obliquity and the lunar orbital inclination), we define a frame of reference co-rotating with the Earth with spin vector $\vec\Omega=\Omega \hat{z}$, $\Omega$ being the Earth's spin rate and $\hat{z}$ the unit vector along its figure axis. In this frame, we use the spherical coordinates $(r,\theta,\lambda)$ denoting the radius, the co-latitude, and the longitude respectively, and their corresponding unit vectors $(\hat{r}, \hat{\theta}, \hat{\lambda})$\PAN{Just a comment: for the longitude, we use $\varphi$ in previous works instead of $\lambda$. For unit vectors we use $\textbf{\rm e}_\varphi$.}{ok}. We start with \PAR{the linearised system of equations that describe}{the governing system of equations describing} the conservation of momentum and mass in a tidally forced shallow oceanic layer \citep{matsuyama2014tidal}
\begin{subequations}
\label{momentum_continuity}
\begin{align}\label{momentum1} 
    &\partial_t\Vec{u}+\sigma_{\rm R}\Vec{u}+\Vec{f}\cross\Vec{u}+g\grad\zeta = g\grad\zeta_\mathrm{eq},\\
    &\partial_t\zeta +\grad\cdot \PAR{\left( H\Vec{u} \right)}{H\Vec{u}}=0, \label{continuity1}
\end{align}
\end{subequations}
where $\Vec{u}= u_\theta\hat{\theta}+u_\lambda \hat{\lambda}$ is the horizontal velocity field, $g$ is the gravitational acceleration at the surface, $\zeta$ is the oceanic depth variation, $\zeta_\mathrm{eq}$ is the equilibrium depth variation, $H$ is the uniform oceanic thickness (the first of only two free parameters in our model), and $\sigma_{\rm R}$ is the Rayleigh (or linear) drag frequency \citep{matsuyama2014tidal,auclair2018oceanic}, an effective dissipation parameter characterizing the damping of the oceanic tidal response by dissipative mechanisms (the second free parameter in our model). On Earth, $\sigma_{\rm R}$ mainly accounts for the conversion of barotropic tidal flows into internal gravity waves, which represents nearly $85\%$ of the total dissipation for the actual lunar semi-diurnal oceanic tide (see for e.g. \cite{carter2008energetics}). For this mechanism, the Rayleigh drag frequency can actually be related to physical parameters such as the Brunt-Väisälä frequency, which quantifies the stability of the ocean's stratification against convection (see for e.g. \cite{gerkema2008introduction}), or \PAR{}{to }the length-scale of topographical patterns at the oceanic floor \citep{bell1975topographically, palmer1986alleviation}. In \eq{momentum_continuity}, the Coriolis parameter $\Vec{f}$ is given by \PAN{Mind that $\Vec f$ is not the Coriolis force, but twice the radial component of the spin vector. Also, be aware that you are implicitly assuming the so-called traditional approximation when you ignore the vertical direction (both the vertical components of the Coriolis forces and the components induced by vertical motions are ignored).}{ok}
\begin{equation} \label{Corilios_def}
        \Vec{f}=2\Omega\cos\theta\hat{r},
\end{equation}
the horizontal gradient operator $\grad$ is defined as \PAN{In order to avoid confusion (partial derivatives are not applied to unit vectors), I would permute the factors and write the equation as $\grad = R^{-1} \left[ \hat{\theta} \partial_{\theta} + \hat{\lambda} \left(\sin \theta \right)^{-1} \partial_{\lambda} \right]$. Also, use different types of brackets in factorised expressions: for instance, $ f = a \left\{ b + c \left[ d + e \left( g + h  \right) \right]  \right\}$.}{ok}
\begin{equation}
   \grad = R^{-1} \left[ \hat{\theta} \partial_{\theta} + \hat{\lambda} \left(\sin \theta \right)^{-1} \partial_{\lambda} \right],
\end{equation}
and the horizontal divergence of the velocity field $\grad \cdot \vec u$ as
\begin{equation}
    \grad \cdot \Vec{u} = \left( R \sin \theta \right)^{-1} \left[ \partial_\theta\left( \sin \theta u_\theta\right) + \partial_\lambda u_\lambda \right] ,
\end{equation}
with $R$ being the Earth's radius.\PA{ Finally, we remark that the interaction of tidal flows with the mean flows of the oceanic circulation are ignored in the momentum equation (\eq{momentum1}).} 

For $\Vec{u}=\partial_t\Vec{x}$, where $\vec x$ is the horizontal tidal displacement field, we have \PAN{$\Vec{x}$ was not introduced before. Specify that this is the horizontal displacement field. In the equation below, use two types of brackets: $ \left[\partial_t^2 +(\sigma_{\rm R}+\Vec{f}\cross)\partial_t \right]\Vec{x}+g\left(\grad\zeta-\grad\Bar{\zeta}\right) =0$.}{ok}
\begin{equation}
 \left[\partial_t^2 +(\sigma_{\rm R}+\Vec{f}\cross)\partial_t \right]\Vec{x}+g\left(\grad\zeta-\grad{\zeta_\mathrm{eq}}\right) =0.
\end{equation}
%\PAR{Following \cite{proudman1920dynamical}, we use the Helmholtz theorem (see \cite{arfken1999mathematical}, Chapter 1) to decompose the horizontal displacement vector field  into a sum of a curl-free vector field $\grad\Phi$ and a divergence-free vector field $\grad\Psi\cross\hat{r}$, namely $\grad \cross \left( \grad \Phi \right) = \vec 0$ and $\grad \cdot \left( \grad\Psi\cross\hat{r} \right) = 0 $,}{}
\PA{Following \cite{proudman1920dynamical}, we use Helmholtz's theorem (e.g. \cite{arfken1999mathematical}, Chapter 1) to decompose the horizontal displacement vector field into 
\begin{equation} \label{horz_tidal_x}
     \Vec{x}=\grad\Phi +\grad\Psi\cross\hat{r},
\end{equation}
where $\grad\Phi$ is a curl-free vector field ($\grad \cross \left( \grad \Phi \right) = \vec 0$), and $\grad\Psi \cross \hat{r}$ a divergence-free vector field ($\grad \cdot \left( \grad\Psi\cross\hat{r} \right) = 0 $). In the above equation, we have introduced the divergent displacement potential~$\Phi$ and the rotational displacement streamfunction~$\Psi$ (e.g. \cite{GMW1983,webb1980tides,tyler2011tidal}), the latter accounting for the vortical component of the tidal displacement field (e.g. \cite{vallis2017atmospheric}). As discussed by \cite{FoxKemper2003}, while the Helmholtz decomposition is unique for infinite domains, this is not true for bounded domains such as hemispherical oceanic shells due to lack of additional physical constraints on the boundary conditions for either of the components of the sum. There are boundary conditions only on the total flux at coastlines. Impermeability is a typical boundary condition: the net flux normal to the coast is zero, which reads $\Vec{x} \cdot \hat{n} = 0$, where $\hat{n}$ designates the outward pointing unit vector defining the normal to the coast. Following \cite{webb1980tides,webb1982tides}, we assume that both components of the flux satisfy this condition, namely   
\begin{align}
\label{boundary_cond}
& \hat{n} \cdot \grad \Phi = 0,  & \hat{n} \cdot \left( \grad \Psi \cross \hat{r} \right) = 0. 
\end{align}
We note that the second condition of the above equation can be rewritten as $\left( \hat{r} \cross \hat{n} \right) \cdot \grad \Psi = 0 $, which implies that $\Psi$ is constant along the coastline (in the following, we set $\Psi = 0$ at the oceanic boundary). This condition thus means that the coastline corresponds to a closed streamfunction contour, which depicts a distinct gyre of the tidal flow. 

Although arbitrary, the assumption that both components of the flux satisfy the impermeability condition has been profusely used to study the dynamics of ocean basins because of its convenience relative to other possible conditions (e.g. \cite{GMW1983,Watterson2001,HH2020helmholtz}). Particularly, this assumption provides a unique decomposition apart from an arbitrary additive constant to each function, $\Phi$ and $\Psi$. Moreover, the second condition given by \eq{boundary_cond} enforces the orthogonality of the curl-free and divergence-free components of the tidal flow. By combining together the identity $\grad \cdot \left( \Phi \grad \Psi \cross \hat{r} \right) = \left( \grad \Psi \cross \hat{r} \right) \cdot \grad \Phi$ and Gauss' theorem (e.g. \cite{arfken1999mathematical}), 
\begin{equation}
    \int_{\mathcal{O}} \grad \cdot \left( \Phi \grad \Psi \cross \hat{r} \right) d A = \oint_{\partial \mathcal{O}} \Phi \left( \grad \Psi \cross \hat{r} \right) \cdot \hat{n} \, d \ell,
\end{equation}
with $dA$ and $d\ell$ being infinitesimal area element of the hemispherical oceanic domain $\mathcal{O}$ and length element of the coastline $\partial \mathcal{O}$, respectively, we obtain
\begin{equation}
    \int_{\mathcal{O}} \left(  \grad \Phi \right) \cdot \left( \grad \Psi \cross \hat{r} \right)  d A = \oint_{\partial \mathcal{O}} \Phi \left( \grad \Psi \cross \hat{r} \right) \cdot \hat{n} \, d \ell.
\end{equation}
As the second condition of \eq{boundary_cond} enforces $\left( \grad \Psi \cross \hat{r} \right) \cdot \hat{n} = 0$ along the coastline, it follows that
\begin{equation}\label{helmholtz_ortho}
    \int_{\mathcal{O}} \left( \grad \Phi \right) \cdot \left( \grad \Psi \cross \hat{r} \right) d A = 0,
\end{equation}
meaning that the components $\grad \Phi$ and $\grad \Psi \cross \hat{r}$ each belong to one of the two orthogonal subspaces that form the space of horizontal displacements satisfying the assumed boundary conditions. We remark that the orthogonality of the Helmholtz decomposition is not necessarily verified in the general case since it is itself a consequence of the specific boundary condition chosen for the divergence-free component of the tidal flow. }

\PA{The functions $\Phi$ and $\Psi$ are expanded in terms of complete sets of eigenfunctions over the domain $\mathcal{O}$ such that
\begin{equation}\label{current_potential}
    \Phi(\theta,\lambda,t)= \sum_{r=1}^{\infty}p_r(t)\phi_r(\theta,\lambda),
\end{equation} 
\begin{equation} \label{stream_function}
      \Psi(\theta,\lambda,t)= \sum_{r=1}^{\infty}p_{-r}(t)\psi_r(\theta,\lambda).
\end{equation}
The eigenfunctions ($\phi_r,\psi_r)$ satisfy, over the oceanic domain ($\mathcal{O}$), the Helmholtz equations (e.g. \cite{riley1999mathematical}, Chapter 21)
\begin{align}\label{pot_stream_eigen}
    \grad^2\phi_r+\mu_r\phi_r&=0,\\
    \label{pot_stream_eigen2}
    \grad^2\psi_r+\nu_r\psi_r&=0,
\end{align}
and, along the coastline ($\partial \mathcal{O}$), the boundary conditions given by \eq{boundary_cond}, 
\begin{equation}\label{boundary_con}
    \hat{n} \cdot \grad\phi_r=0, \hspace{1cm} \psi_r=0,
\end{equation}
where we have introduced the horizontal Laplacian,
\begin{equation}
    \grad^2 =\left( R \sin \theta \right)^{-2} \left[ \sin \theta \,  \partial_\theta \left( \sin \theta \, \partial_\theta \right) + \partial_{\lambda \lambda} \right] ,
\end{equation}
and the real eigenvalues $\mu_{r}$ and $\nu_{r}$ associated with the eigenfunctions $\phi_{r}$ and $\psi_{r}$, respectively. We note that the eigenfunctions are normalized such that
\begin{equation}\label{phi_psi_ortho}
    \int_\mathcal{O}\phi_r\phi_sdA=\int_\mathcal{O}\psi_r\psi_sdA= \delta_{rs},
\end{equation}
the notation $\delta_{rs}$ referring to the Kronecker $\delta$-symbol $\delta_{rs}=1 $ for $r=s$ and $0$ otherwise.} Using these conditions, the eigenfunctions are defined as 
\PAN{Don't forget to specify the ranges where the indices $n$ and $m$ take their values ($n=m=0$ is excluded). Also, in addition with the expressions of $\phi_r$ and $\psi_r$ in terms of the spherical harmonics ($\phi_r = \Re \left( Y_{n,m} \right)$ and $\psi_r = \Im \left( Y_{n,m} \right)$, basically), give the reciprocal expression of the spherical harmonics in terms of the $\phi_r$ and $\psi_r$ (i.e. $Y_{n,m} = \phi_r + i \psi_r$) as you do for the second change of basis in Eqs. (S109-S110). It is important to emphasise that there are two changes of basis, and to specify the coefficients of the transformation matrices.}{ok}
\begin{align}\label{eigenfunctionsss}
    \phi_r&=\frac{\alpha_{n,m}}{R}P_n^m(\cos\theta)\cos m\lambda,\\
    \label{eigenfunctionsss2}
    \psi_r&=\frac{\alpha_{n,m}}{R}P_n^m(\cos\theta)\sin m\lambda,
\end{align}
with eigenvalues $\mu_r=\nu_r=n(n+1)/R^2$, and the normalization coefficient
\begin{equation}\label{normalizartion_alpha}
    \alpha_{n,m} = \sqrt{\frac{2n+1}{\pi}\frac{(n-m)!}{(n+m)!}\frac{1}{1+\delta_{m0}}}\PAR{.}{,}
\end{equation}
\PAR{In Eqs.~\eqref{eigenfunctionsss} and~\eqref{eigenfunctionsss2}, each}{where each} harmonic index $r$ of the eigenfunctions is associated with a degree $n$ and order $m$, and the expansion functions are the Associated Legendre Functions \citep{abramowitz1988handbook}. \GB{In the definition of $\phi_r$~\eqref{eigenfunctionsss}, $n \in \mathbb{N}$ and $m=0,1,...,n$ while in the expression of $\psi_r$~\eqref{eigenfunctionsss2}, $n\in\mathbb{N}^*$ and $m=1,2,...,n$. By convention,  we set $\psi_0=0$ hereafter}\PAN{I would rather call $n$ and $m$ the latitudinal and longitudinal degrees of spherical harmonics. But it is just my opinion. Besides, Abramowitz \& Stegun, which you refer to, call the $P_l^m$ the Associated Legendre Functions, and not the Associated Legendre Polynomials.}{ok}. \PA{Figure~\ref{fig:eigenfunctions} shows the eigenfunctions $\phi_r$ and $\psi_r$ for \PAR{$1 \leq m \leq n \leq 4$}{the first orders and degrees} and the streamlines of the associated tidal flows.}

\def\wbox{1cm}
\def\hraisebox{0.25\textwidth}
\def\verspacem{1.5cm}
\def\horspacen{1.55cm}
\begin{figure*}[htb]
\centering
\textsc{$\left\{ \phi_r \right\}$ \hspace{7cm} \textsc{$\left\{ \psi_r \right\}$}} \\
  \raisebox{\hraisebox}[1cm][0pt]{%
   \begin{minipage}{\wbox}%
   \textsc{$0$} \\[\verspacem] \textsc{$1$} \\[\verspacem] \textsc{$2$} \\[\verspacem] \textsc{$3$} \\[\verspacem] \textsc{$4$} 
\end{minipage}}
\includegraphics[width=.511\textwidth,trim = 0.0cm 0.0cm 0.0cm 0.cm,clip]{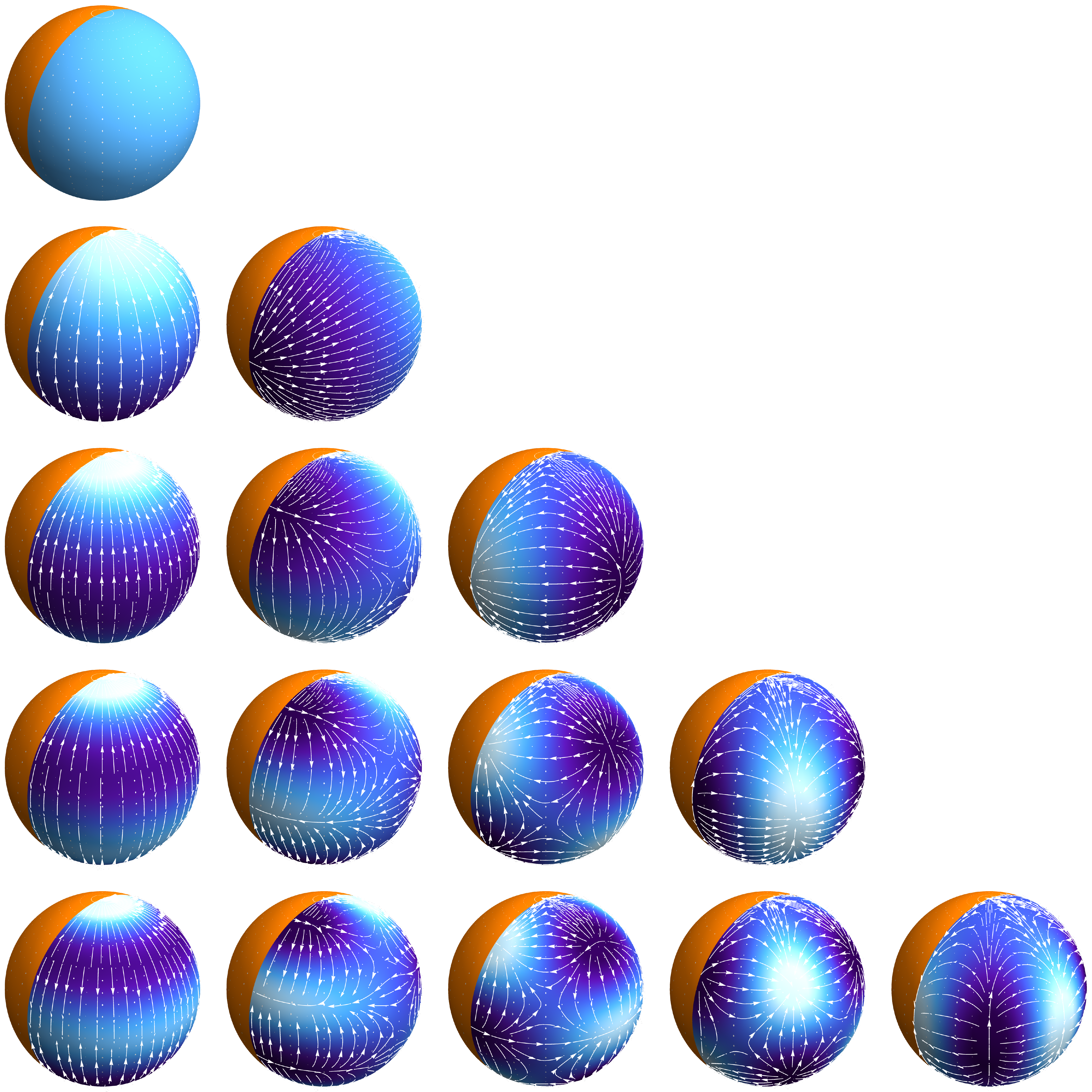}
\includegraphics[width=.409\textwidth,trim = 0.0cm 0.0cm 0.0cm 0.cm,clip]{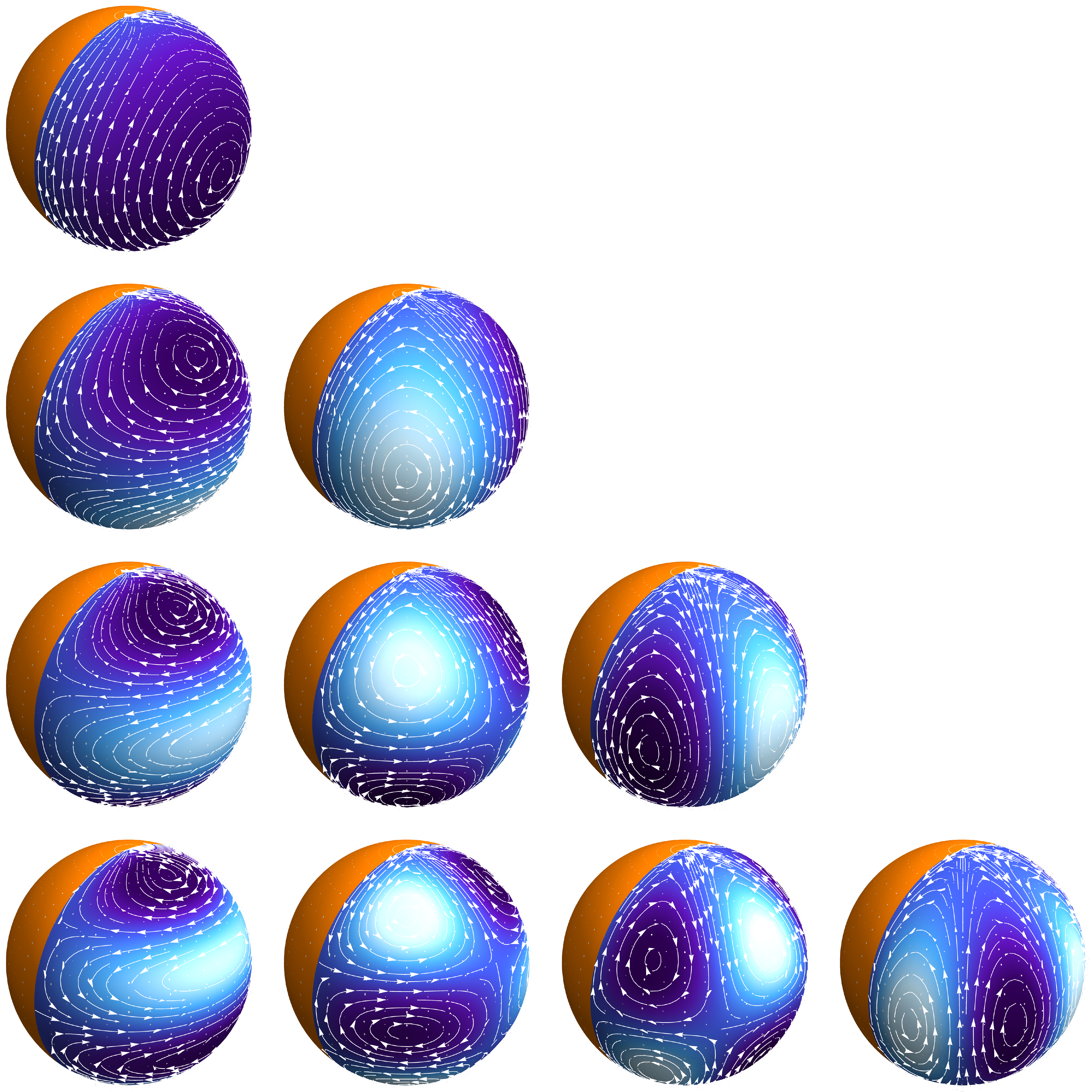}\\
\textsc{$n/ m$} \hspace{1.cm} $0$ \hspace{\horspacen} $1$ \hspace{\horspacen} $2$ \hspace{\horspacen} $3$ \hspace{\horspacen} $4$ \hspace{\horspacen} $1$ \hspace{\horspacen} $2$ \hspace{\horspacen} $3$ \hspace{\horspacen} $4$  \hspace{0.5cm}
\caption{\PA{Eigenfunctions $\phi_r$ (left) and $\psi_r$ (right) and associated tidal flows. The eigenfunctions defined by Eqs.~\eqref{eigenfunctionsss} and~\eqref{eigenfunctionsss2} are plotted over the hemispherical oceanic domain for $ 0 \leq n \leq 4 $ (from top to bottom) and $0 \leq m \leq n$ (from left to right). Bright or dark colors designate positive or negative values of the eigenfunctions, respectively. Streamlines indicate the tidal flows corresponding to $\grad \phi_r$ for the set $\left\{ \phi_r \right\}$ and to $\grad \psi_r \cross \hat{r}$ for the set $\left\{ \psi_r \right\}$.  } }
\label{fig:eigenfunctions}
\end{figure*}

The eigenfunctions $(\phi_r,\psi_r)$ can be split into two sets describing tidal solutions that are symmetric or anti-symmetric about the equator, and thus one can decide, based on the symmetry of the tidal forcing, on the associated set of eigenfunctions that need to be considered  using a classification scheme \citep{longuet1968eigenfunctions,longuet1970free} for the pairs $(n,m)$. However, in our model, where the ocean is no longer symmetric about the equator, both symmetric and anti-symmetric eigenfunctions are required.  Substituting the definitions of Eqs. \eqref{current_potential}, \eqref{stream_function}, and \eqref{pot_stream_eigen} into the continuity equation \eqref{continuity1}, one finds that \PAN{In multi-factor terms, put constant factors before functions. For instance: $\zeta=H\sum_{r=1}^\infty \mu_r p_r\phi_r$.}{ok} 
\begin{equation}\label{eq.zetasol}
  \zeta=H\sum_{r=1}^\infty \mu_r p_r\phi_r.
\end{equation}

What is left to complete the solution is to find the coefficients $p_r$ and $p_{-r}$, by substituting the series expansions in the momentum equation \eqref{momentum1} and multiplying by $\grad\phi_r$ and $\grad\psi_r\cross\hat{r}$, then integrating over the oceanic area. Starting with the former we get
\begin{align}\label{pr1}\nonumber
    \sum_{s=0}^\infty &\left(\partial_t^2p_s+\sigma_{\rm R}\partial_t p_s +gHp_s\mu_s -g{\zeta}_{\mathrm{eq},s}\right)\grad\phi_s\cdot\grad\phi_r \\\nonumber
    &+\left(\partial_t^2p_{-s}+\sigma_{\rm R}\partial_tp_{-s}\right)\left(\grad\psi_s\cross\hat{r}\right)\cdot\grad\phi_r 
    \\&+ \partial_t p_s\left(\Vec{f}\cross\grad\phi_s\right)\cdot\grad\phi_r+\partial_t p_{-s}\left[\Vec{f}\cross\left(\grad\psi_s\cross\hat{r}\right)\right]\cdot\grad\phi_r=0.
\end{align}
The product of the gradients of two eigenfunctions is computed using Green's first identity (e.g. \cite{strauss2007partial}, Chapter 7)
\begin{equation}
    \int_\mathcal{O}\grad\phi_s\cdot\grad\phi_r dA =\int_{\partial\mathcal{O}} \phi_s\left(\grad\phi_r\cdot\hat{n}\right) d\ell - \int_\mathcal{O}\phi_s\grad^2\phi_rdA.
\end{equation}
The first term on the right hand side vanishes as it includes the boundary condition at the coast (Eq. \ref{boundary_con}). The second term is computed using the eigenvalue equation (Eq. \ref{pot_stream_eigen}) and the normalization condition, thus
\begin{equation}
     \int_\mathcal{O}\grad\phi_s\cdot\grad\phi_r dA = \mu_r \delta_{rs}.
\end{equation}
Rearranging the other terms using vector identities we write Eq.\eqref{pr1} as
\begin{align}\label{pr2}\nonumber
    \sum_{s=0}^\infty &\left(\partial_t^2p_s+\sigma_{\rm R}\partial_t p_s +gHp_s\mu_s -g{\zeta}_{\mathrm{eq},s}\right)\mu_r \delta_{r,s} \\\nonumber
    &+\left(\partial_t^2p_{-s}+\sigma_{\rm R}\partial_tp_{-s}\right) \int_\mathcal{O}\left(\grad\psi_s\cross\hat{r}\right)\cdot\grad\phi_r \,dA \\\nonumber
    &+ \partial_t  p_s \int_\mathcal{O} \Vec{f}\cdot \left(\grad\phi_s\cross\grad\phi_r\right)dA\\
    &+\partial_t p_{-s}\int_\mathcal{O} \left(\Vec{f}\cdot\hat{r}\right) \left(\grad\phi_r\cdot\grad\psi_s\right)dA=0.
\end{align}
The third term vanishes due to orthogonality\PA{ (see \eq{helmholtz_ortho})}, and upon replacing the Coriolis term by its definition in \eq{Corilios_def} we are left with
\begin{align}\label{pr4}\nonumber
 &\left(\partial_t^2p_r+\sigma_{\rm R}\partial_t p_r +gHp_r\mu_r -g{\zeta}_{\mathrm{eq},r}\right)\mu_r \\\nonumber
 &- 2\Omega\sum_{s=1}^\infty \partial_t  p_s \int \cos\theta\hat{r}\cdot \left(\grad\phi_r\cross\grad\phi_s\right)dA\\
    &+2\Omega\sum_{s=1}^\infty\partial_t p_{-s}\int \cos\theta \left(\grad\phi_r\cdot\grad\psi_s\right)dA=0.
\end{align}
To close the system, we multiply the momentum equation with $\grad\psi_r\cross\hat{r}$ to get
 \begin{align}\label{pr5}\nonumber
 &\sum_{s=0}^\infty \left(\partial_t^2p_s+\sigma_{\rm R}\partial_t p_s +gHp_s\mu_s -g{\zeta}_{\mathrm{eq},s}\right)\grad\phi_s\cdot\left(\grad\psi_r\cross\hat{r}\right)\\\nonumber
  &+ \partial_t p_s\left(\Vec{f}\cross\grad\phi_s\right)\cdot\left(\grad\psi_r\cross\hat{r}\right)+\partial_t p_{-s}\left[\Vec{f}\cross\left(\grad\psi_s\cross\hat{r}\right)\right]\cdot\left(\grad\psi_r\cross\hat{r}\right)
  \\
    &+\left(\partial_t^2p_{-s}+\sigma_{\rm R}\partial_tp_{-s}\right)\left(\grad\psi_s\cross\hat{r}\right)\cdot\left(\grad\psi_r\cross\hat{r}\right)=0.
\end{align}   
Integrating \eq{pr5} over the area of the ocean and using basic vector product identities we get
\begin{align}
\nonumber\sum_{s=0}^\infty &\left(\partial_t^2p_s+\sigma_{\rm R}\partial_t p_s \right)\int\grad\psi_s\cdot\grad\psi_r dA \\\nonumber
&-\partial_t p_s\int\left(\Vec{f}\cdot\hat{r}\right)\left(\grad\phi_s\cdot\grad\psi_r\right)dA \\
  &- \partial_t p_{-s}\int\left(\Vec{f}\cdot\hat{r}\right)\hat{r}\cdot\left(\grad\psi_r\cross\grad\psi_s\right)dA      =0 ,  
\end{align}
where upon replacing the Coriolis term as before we have

\begin{align} \label{pr7} \nonumber
 &\partial_t^2p_{-r}+\sigma_{\rm R}\partial_t p_{-r}- \frac{2\Omega}{\nu_r}\sum_{s=1}^\infty \partial_t  p_s \int \cos\theta \left(\grad\phi_s\cdot\grad\psi_{r}\right)dA\\
&-\frac{2\Omega}{\nu_r}\sum_{s=0}^\infty\partial_t p_{-s}\int \cos\theta \hat{r}\cdot \left(\grad\psi_r\cross\grad\psi_s\right)dA=0. 
\end{align}
We identify in Eqs. \eqref{pr4} and \eqref{pr7} the so-called ``gyroscopic coefficients" (e.g. \cite{proudman1920dynamical1,proudman1920dynamical}) that are defined as
\begin{align} \nonumber \label{gyro1}
 \beta_{r,s}&=-\int_\mathcal{O} \cos\theta \, \hat{r} \cdot \left( \grad\phi_r\cross\grad\phi_s \right) dA,\\
    \nonumber
     \beta_{r,-s}&= \ \ \, \int_\mathcal{O} \cos\theta \, \grad\phi_r\cdot\grad\psi_s dA,   \\ \nonumber
         \beta_{-r,s}&=-\int_\mathcal{O} \cos\theta \, \grad\psi_r\cdot\grad\phi_s dA,\\
    \beta_{-r,-s}&=-\int_\mathcal{O} \cos\theta \, \hat{r} \cdot \left( \grad\psi_r\cross\grad\psi_s \right) dA,
\end{align}
with $\beta_{-s,r}= -\beta_{r,-s}$. These  coefficients carry the effect of rotational distortion to the tidal waves and the boundary conditions imposed by the coastlines. Using these definitions, Eqs. \eqref{pr4} and \eqref{pr7} form an infinite linear system in the coefficients $p_r(t)$ and $p_{-r}(t)$ that reads 
\begin{equation} 
\label{system_1_before}
    \partial_t^2p_r+\sigma_{\rm R}\partial_t p_r +gH\mu_rp_r -g{\zeta}_{\mathrm{eq},r} + \frac{2\Omega}{\mu_r}\sum_{s=-\infty}^{s=\infty}\beta_{r,s}\partial_tp_s =0,
\end{equation}
\begin{equation}
\label{system_2_before}
    \partial_t^2p_{-r} +\sigma_{\rm R}\partial_tp_{-r} +\frac{2\Omega}{\nu_r}\sum_{s=-\infty}^{s=\infty}\beta_{-r,s}\partial_t p_s =0.
\end{equation}
This system  will be transformed to the frequency domain (\ref{section_coupling}), and then truncated and solved spectrally as a function of the tidal forcing frequency. However, we pause here to extend the theory to the effects of self-attraction and loading between the ocean and the deforming mantle in order to have a complete self-consistent tidal response of the Earth. 

\section{Coupling the hemispheric oceanic response with solid deformation}
\label{section_coupling}
In the tidal theory under study, the solid component of the Earth is subject to viscoelastic deformation as a result of three contributions: the direct tidal effect of the tidal perturber, the loading effect of the perturbed oceanic shell, and the effect of gravitational self-attraction between the oceanic shell and the solid part \citep{farrell1972deformation,zahel1980mathematical}. If one were to take these into account when studying oceanic tides, $\zeta$ becomes a function of two moving surfaces: the free oceanic surface $\zeta_{\rm os}$, and the vertically deforming solid surface $\zeta_{\rm ss}.$ 

In the frame co-rotating with the Earth as defined in Appendix \ref{hemi_response}, the gravitational potential is expressed as \citep[\PA{e.g.}][]{auclair2019final} 
\begin{equation}\label{grav_pot_gen}
    {U}(\vec r,\vec r^{\,\prime})= \frac{GM}{|\vec r-\vec r^{\,\prime}|} - \frac{GM}{r^{\prime \, 2}}r\cos\theta\, ,
\end{equation}
where $G$ is the gravitational constant, $M$ is the mass of the tidal perturber (the Sun or the Moon), and $r^\prime$ is the distance between the Earth and the perturber. In the shallow ocean approximation, the tidal potential at the Earth's surface ($r=R$) is \PAN{Here the superscript T refers to ``tide''. Therefore, the tidal potential should be denoted by $\mathcal{U}^{\rm T}$ instead of $\mathcal{U}^T$.}{ok}
\begin{equation}
    {U}^{\rm T}(\theta,\lambda, \vec r^{\,\prime}) =   {U}(R,\theta,\lambda,\vec r^{\,\prime}) -\frac{GM}{r^\prime}\, ,
\end{equation}
where a constant offset was removed as it does not contribute to the tidal force. In the frequency domain, \GBR{the tidal potential $U^{\rm T}$}{the associated complex gravitational potential $U^{\rm T}$, such that $ \mathcal{U}^{\rm T}= \Re\{U^{\rm T}\}$,} is expanded spectrally in Fourier series and spatially in spherical harmonics\GB{, with complex coefficients $U_n^{m;s}$, as} \citep{kaula2013theory,auclair2019final} \PAN{So this means that we are already working in the Fourier domain here. One should introduce the relationship between Fourier's complex fields and the real fields.}{ok}
\begin{equation}\label{tidal_pot}
    U^{\rm T} = \sum_{n=2}^{\infty}\sum_{m=-n}^{n}\sum_{s=-\infty}^{\infty} U_n^{m;s} P_n^m(\cos\theta) \exp{i(\sigma_{m}^st+m\lambda)},
\end{equation}
where $s$ is an integer and the tidal forcing frequency $\sigma_{m}^s=m\Omega -s n_{\rm orb}$,  the frequency $n_{\rm orb}$ being the orbital mean motion of the tidal perturber.
In the absence of obliquity, the $n^{\rm th}$ harmonic of the tidal potential $U_n^{m;s}$ is given by  \citep{ogilvie2014tidal}
\begin{equation}\label{tid_pot_deco}
    U_n^{m;s} = \frac{GM}{a}\left(\frac{R}{a}\right)^n A_{n,m,s}(e),
\end{equation}
where $a$ is the semi-major axis of this perturber, and $A_{n,m,s}(e)$ are dimensionless functions of the orbital eccentricity of the perturber $e$ computed via the Hansen coefficients \citep{laskar2005note,correia2014deformation}\PAN{It seems that the notation used for eccentricity $e$ is not introduced before.}{ok}. In our study, we restrict the tidal potential to the dominant contribution of the semi-diurnal component identified by $n=m=s=2$, and corresponding to  the tidal frequency \PA{$\sigma_{2}^2=2(\Omega - n_{\rm orb})$}\PAN{The notation for the orbital frequency is $n_{\rm orb}$.}{ok}. For this component, and neglecting the small orbital eccentricity of the Sun and the Moon, $A_{2,2,2}(0)= \sqrt{3/5}$. \GB{Hereafter, we use $U^{\rm T}_n$ to represent a single harmonic $(n,m,s)$ of the tidal potential. This harmonic of degree $n$ is defined as
\begin{equation}
    U^{\rm T}_n = U_n^{m;s}P_n^m(\cos\theta)\exp\left\{i(\sigma^s_m t+m\lambda)\right\}\,.
\end{equation}
Moreover, in the following we write $\sigma$ instead of $\sigma_{m}^s$ to simplify the notation.} Subject to $U^{\rm T}_n$ \GB{only, the equilibrium oceanic depth would be $\bar\zeta = U_n^{\rm T}/g$. But the loading effect of the deforming oceanic shell adds to the tidal potential and they both affect the ocean surface $\bar\zeta_{\mathrm{os}}$ and the ocean floor corresponding to the solid surface $\bar\zeta_\mathrm{ss}$. The former takes the form} \citep{matsuyama2014tidal}
\begin{equation} \label{solid_surface}
    \bar\zeta_\mathrm{os}=\frac{ h^{\rm T}_nU^{\rm T}_n}{g}+ \sum_l  \frac{3\rho_{\rm oc}}{(2l+1)\rho_{\rm se}} h^{\rm L}_l\zeta_l,
\end{equation}
where $\rho_{\rm oc}$ and $\rho_{\rm se}$ stand for the uniform oceanic and solid Earth densities respectively. \GB{In this equation, the oceanic depth variation $\zeta$ is decomposed into spherical harmonics defined over the full sphere as
\begin{equation}\label{eq.zetaSH}
    \zeta_l(\theta,\lambda,t) = \sum_{m=-l}^l \zeta_l^m(t)P_l^m(\cos\theta)\exp(im\lambda)\,.
\end{equation}
Although $\zeta$ given in (\ref{eq.zetasol}) is only defined over the oceanic hemisphere, this decomposition over the whole sphere is required when applying the Love numbers. Using the orthogonality of spherical harmonics, and the fact that $\zeta(\theta,\lambda,t)=0$ over the continental hemisphere, Eq.~(\ref{eq.zetaSH}) also reads
\begin{align}\nonumber
    \zeta_l(\theta,\lambda,t) =& \frac{1}{2}\sum_{m=0}^l \alpha_{lm}^2 \int_{\mathcal{O}} \zeta(\theta',\lambda',t) P_l^m(\cos\theta)P_l^m(\cos\theta') \\&\times\cos(m(\lambda-\lambda'))\,d\Omega\,,
\end{align}
where the integral is computed over the solid angle $2\pi$ spanned by the ocean.
}
The second contribution to the equilibrium tide, which is due to the solid redistribution of mass, is
\begin{equation}\label{zeta_eq_corr}
   \Bar{\zeta}_\mathrm{ss} = (1+k_n^{\rm T})\frac{ U_n^{\rm T}}{g} + \sum_l  \frac{3\rho_{\rm oc}}{(2l+1)\rho_{\rm se}} (1+k_l^{\rm L})  \zeta_l.
\end{equation}
In equations~\eqref{solid_surface} and \eqref{zeta_eq_corr}, we used the tidal Love numbers $k_n^{\rm T}$ and $h_n^{\rm T}$, and the surface loading Love numbers $k_n^{\rm L}$ and $h_n^{\rm L}$, where the first of each set is the transfer function corresponding to the gravitational response, and the second codes for the vertical displacement.\PAR{ We emphasize that the Love numbers are defined here in the Fourier domain. Therefore, they correspond to the intrinsic mechanical impedances of the solid part that relate its visco-elastic tidal response to tidal forcings in the permanent regime, and they characterise both the elastic deformation of the body and its anelastic deformation resulting from energy dissipation due to viscous friction in the Earth's interior. In the general case, the four Love numbers ($k_n^{\rm T},h_n^{\rm T},k_n^{\rm L}$, and $h_n^{\rm L}$) can be computed from internal structure models (e.g.  \cite{Tobie2005icarus,Tobie2019aap,Bolmont2020}). In the present study, for the sake of simplicity, we use the closed-form solutions derived for a uniform solid interior \citep{munk1960rotation},}{}
\begin{equation} \label{love_def}
\left\{k_n^{\rm T},h_n^{\rm T},k_n^{\rm L},h_n^{\rm L}\right\}= \frac{1}{(1+\Tilde{\mu}_n)}\left\{\frac{3}{2(n-1)},\frac{2n+1}{2(n-1)},-1, -\frac{2n+1}{3} \right\},
\end{equation}
where $\Tilde{\mu}_n$ is a complex dimensionless effective shear modulus, with a form dependent on the chosen solid rheology  \citep{efroimsky2012tidal,renaud2018increased}. \PAR{To specify $\Tilde{\mu}_n$}{}, we consider an Andrade rheology \citep{andrade1910viscous}, which has the advantage over the commonly used Maxwell rheology in attenuating the rapid decay of the anelastic component of the deforming solid Earth for high tidal frequencies \PAR{\citep{Castillo-Rogez,auclair2019final}}{\citep{auclair2019final, Castillo-Rogez}}. This is particularly useful in avoiding an overestimation of the tidally dissipated energy of the solid part during early eons. For this rheology, $\Tilde{\mu}_n$ takes the form \PAR{\citep{findley,efroimsky2012tidal}}{\citep{efroimsky2012tidal,findley}}
\begin{equation}\label{effective_mu}
    \Tilde{\mu}_n = \frac{4(2n^2+4n+3)\pi R^4}{3nGM_{\rm E}^2} \frac{\mu_{\rm E}}{1+(i\sigma\tau_{\rm A})^{-\alpha_{\rm A}} \Gamma(1+\alpha_{\rm A})+(i\sigma\tau_{\rm M})^{-1}},
\end{equation}
where $M_{\rm E}$ is the mass of the Earth, $\mu_{\rm E}$  its average rigidity,  $\Gamma$ is the gamma function \citep{abramowitz1988handbook}; $\alpha_{\rm A}$ is a dimensionless rheological exponent determined experimentally \PAR{\citep{Castelnau,petit2010iers}}{\citep{petit2010iers,Castelnau}}; $\tau_{\rm A}$ is the anelastic Andrade timescale, and $\tau_{\rm M}$ the Maxwell relaxation time defined  as the ratio of viscosity to rigidity. For a volumetric average of the mantle's shear modulus $\mu_{\rm E}=17.3\times 10^{10}$ Pa, and volumetric average of viscosity deduced from inversions of  \cite{lau2016inferences}, we have $\tau_{\rm M}=685$ yrs. The values \PAR{$\alpha_{\rm A} = 0.25$ and  $\tau_{\rm A}= 2.19\times 10^{4}$ yrs}{} that we use in our model are adopted from  \cite{auclair2019final}. All used values of parameters are summarized in Table \ref{table_values}.

Taking the effect of solid Earth deformation into account, we replace the equilibrium tide $\zeta_\mathrm{eq}$ in the momentum equation \eqref{momentum1} by the difference $\Bar\zeta_\mathrm{os}-\Bar\zeta_\mathrm{ss}$ of Eqs.\eqref{solid_surface} and \eqref{zeta_eq_corr}, and we resolve it in the Fourier domain using the forcing tidal frequency $\sigma$. The modified momentum conservation equations now reads \PAN{This equation is a bit confusing because it is given in the temporal domain, while the tilt factors are defined in the Fourier domain.}{ok}
\begin{equation}
    \label{momentum_coupled}
   i\sigma\Vec{u}+\sigma_{\rm R}\Vec{u}+\Vec{f}\cross\Vec{u} = -g\grad\left(-\gamma_n^{\rm T}\Bar{\zeta} + \sum_l \gamma_l^{\rm L}\zeta_l\right),
\end{equation}
\GB{with $\Bar\zeta = U_n^{\rm T}/g$, and}
where the loading and tidal tilt factors are defined as  \citep{matsuyama2014tidal} 
\begin{equation}\label{tilt} 
     \gamma_n^{\rm T}= 1+k_n^{\rm T}-h^{\rm T}_n \hspace{.3cm};\hspace{.3cm} \gamma_l^{\rm L}=1-\frac{3\rho_{\rm oc}}{(2l+1)\rho_{\rm se}}(1+k_l^{\rm L} -h_l^{\rm L}).
\end{equation}
\PA{Just like the Love numbers, $\gamma_n^{\rm T}$ and $\gamma_l^{\rm L}$ are complex in the general case and tend to unity as the deformability of the solid and oceanic layers decreases.}
Now we get to the added contribution of the ocean-solid coupling to the linear system of $p_r$ and $p_{-r}$. Multiplying the added contribution of loading and self-attraction effects by $\grad\phi_r$ and $\grad\psi_r\cross\hat{r}$ then resolving the added terms in the frequency domain, after some manipulations we finally re-write the system of Eqs. \eqref{system_1_before} and \eqref{system_2_before} as
% \begin{align}\nonumber
%         &\partial_t^2p_r+\sigma_{\rm R}\partial_t p_r +gH\mu_r\left(1-\frac{\gamma_r^L}{2} \right)p_r -g\gamma^T_n\Bar{\zeta}_r + \frac{2\Omega}{\mu_r}\sum_{s=-\infty}^{s=\infty}\beta_{r,s}\partial_tp_s -\frac{1}{2}gH \sum_{\substack{s^\prime=1\\s^\prime\neq r}}^\infty \mu_{s^\prime}F_r^{s^\prime}p_{s^\prime}=0,\\
%         & \partial_t^2p_{-r} +\sigma_{\rm R}\partial_tp_{-r} +\frac{2\Omega}{\nu_r}\sum_{s=-\infty}^{s=\infty}\beta_{-r,s}\partial_t p_s =0,
% \end{align}
\begin{align}\nonumber
        -\sigma^2p_r-i\sigma\sigma_{\rm R} p_r +gH\mu_r\left(1-\frac{\gamma_r^{\rm L}}{2} \right)p_r -g\gamma^{\rm T}_n\Bar{\zeta}_r \\
        - \frac{2i\sigma\Omega}{\mu_r}\sum_{s=-\infty}^{s=\infty}\beta_{r,s}p_s -\frac{1}{2}gH \sum_{\substack{s^\prime=1\\s^\prime\neq r}}^\infty \mu_{s^\prime}F_r^{s^\prime}p_{s^\prime} & =0,\\
        -\sigma^2p_{-r} -i\sigma\sigma_{\rm R}p_{-r} -\frac{2i\sigma\Omega}{\nu_r}\sum_{s=-\infty}^{s=\infty}\beta_{-r,s}\partial_t p_s & =0,
\end{align}
where we have defined
\begin{equation} \label{F_matrix}
   F_r^{s^\prime} = 4 \alpha_{n,m}\alpha_{n^\prime,m^\prime}\sum_p\sum_q\gamma_p^{\rm L} q^2\alpha_{p,q}^2\frac{\mathcal{O}_{p,q}^{n,m}\mathcal{O}_{p,q}^{n^\prime,m^\prime}}{(q^2-m^2)(q^2-m^{\prime2})} \,,
\end{equation}
with $\mathcal{O}_{p,q}^{n,m}$ corresponding to the Gram matrix of the ALFs,
\begin{equation}\label{overlapdef}
    \mathcal{O}_{n,m}^{u,v} = \int_{-1}^1 P_n^m(\mu)P_u^v(\mu)d\mu,
\end{equation}
for which the method of computation is detailed in Appendix \ref{overlap_section}. Coupled to the orbital dynamical evolution, the tidal frequency $\sigma$ is determined at each point in time in the hemispherical phase of the model,  then the system is truncated at $r_{\rm max}$ and solved numerically (see Appendix \ref{section_tidal_response}). We re-write the linear system as \PAR{}{follows:}
\begin{align}\label{final_linear_system}
    &(a^{(1)} + a_r^{(2)})p_r + a_r^{(3)} \sum_{s=-\infty}^{s=\infty}\beta_{r,s}p_s+a^{(5)} \sum_{\substack{s^\prime=1\\s^\prime\neq r}}^\infty \mu_{s^\prime}F_r^{s^\prime}p_{s^\prime} = c_r ,\\
    &a^{(1)} p_r + a_r^{(4)} \sum_{s=-\infty}^{s=\infty}\beta_{r,s}p_s = 0,
\end{align}
where the first equation is for $r > 0$, and the second is for $r < 0$, and where \PA{we have introduced the coefficients} \PAN{You can put these coefficients on 2 or 3 lines instead of 6 (3 or 2 coefficients per line) to save space.}{ok}
\begin{align} \nonumber
    a^{(1)} &= -\sigma^2 -i\sigma\sigma_{\rm R}, \hspace{1cm} a^{(2)}_r = gH\mu_r(1-\gamma_r^{\rm L}/2), \\ \nonumber
      a^{(3)}_r &= -2i\sigma\Omega\mu_r^{-1}, \hspace{1.3cm} a^{(4)}_r = -2i\sigma\Omega\nu_{-r}^{-1},\\
       a^{(5)} &= -\frac{1}{2}gH,\hspace{2cm}  c_r = g\gamma_n^{\rm T}\Bar{\zeta_r}.
\end{align}

\section{The gyroscopic coefficients}\label{Gyro_ALPs}
The gyroscopic coefficients \PA{introduced in \eq{gyro1}} characterize the rotational distortion of tidal waves via the Coriolis force term and the effect of boundary conditions imposed by the oceanic geometry.  This coupling is dependent on the position of the ocean on the sphere and the relative position of the tidal perturber with respect to the tidally forced ocean. Since we are after a generic configuration describing the response of the oceanic hemisphere at any position, the expressions of \eq{gyro1} should be written for any frame rotating with the ocean. We start with the definition of the ALFs (Chapter 8 of  \cite{abramowitz1988handbook})
\begin{equation} \label{def_alp}
    P_n^m(\mu)= \frac{(-1)^m}{2^n  n!}(1-\mu^2)^{m/2}\dm^{n+m}(\mu^2-1)^n,
\end{equation}
which are solutions to the Legendre equation,
\begin{equation}\label{legendreeq}
   \dm\left[(1-\mu^2)\dm P_n^m\right]+\left[n(n+1)-\frac{m^2}{1-\mu^2}\right]P_n^m=0.
\end{equation}
Upon differentiation we obtain
\begin{equation}\label{derivative}
    \dm P_n^m= -\frac{m\mu }{1-\mu^2}P_n^m - \frac{P_n^{m+1}}{\sqrt{1-\mu^2}}.
\end{equation}
Substituting \eq{derivative} in \eq{legendreeq} we get the  recurrence relation
\begin{align}\nonumber
    &P_n^{m+2} - \frac{2m\mu^2(m+1)}{1-\mu^2}P_n^m - 2\mu(m+1)\dm P_n^m\\ &+ (n(n+1) -m(m+1))P_n^m =0,
\end{align}
which gives the  useful relation
\begin{equation}\label{mudmu}
    \mu\dm P_n^m = \frac{P_n^{m+2}}{2(m+1)} + \left[\frac{n(n+1)+m(m+1)}{2(m+1)}-\frac{m}{1-\mu^2}\right]P_n^m .
\end{equation}
From Eqs. (\ref{def_alp}-\ref{mudmu}), it is straightforward to obtain the \PAR{}{following }ALFs recurrence relations that are \PAR{necessary}{useful} to compute the integral equations of the gyroscopic coefficients,
\begin{align}\label{Rec1}
   \hspace{-.5cm}& \frac{\mu P_n^m}{\sqrt{1-\mu^2}}= -\frac{1}{2m} \left( P_n^{m+1} + (n-m+1)(n+m)P_{n}^{m-1}\right), \\\label{Rec2}
     &\frac{ P_n^m}{\sqrt{1-\mu^2}}= -\frac{1}{2m} \left( P_{n-1}^{m+1} + (n+m-1)(n+m)P_{n-1}^{m-1}\right),  \\\label{Rec4}
    & \sqrt{1-\mu^2}\dm P_n^m= -\frac{1}{2}P_{n}^{m+1}+\frac{1}{2}(n+m)(n-m+1)P_{n}^{m-1},\\\label{Rec5}
     &(1-\mu^2)\dm P_n^m = \frac{1}{2n+1}\left((n+1)(n+m)P_{n-1}^{m} \!- \!n(n\!-\!m+\!1)P_{n+1}^{m}\right)\!.
\end{align}
The theory of the hemispherical tidal response is based on an ocean bounded by two meridians\PAR{. Thus}{} for an oceanic center moving on the sphere, we rotate instead the spin axis relative to the center of the ocean, and accordingly the frame of the tidal perturber to maintain the coplanar configuration of the dynamical system. These rotations will enter the system through the Coriolis term, specifically through the gyroscopic coefficients, along with the tidal forcing term. We define an arbitrary rotation  $\{\theta,\lambda\}\rightarrow \{\theta^\prime,\lambda^\prime\}$ using an Eulerian rotation matrix of the form $\mathcal{R}_3(\alpha)\mathcal\,{R}_2(\beta)\mathcal\,{R}_3(\gamma)$, with ($0\leq\alpha\leq2\pi)$ and $(0\leq\beta\leq\pi)$, and we fix $\gamma=0$ (see  Fig. \ref{trans_fig}). For a vector $J$ defined as \PAN{The rotation angles $\alpha$, $\beta$, and $\gamma$ and there ranges need to be introduced here.}{ok}
\begin{align}\nonumber \label{vecJ}
    J &= \mathcal{R}_3(\alpha)\mathcal{R}_2(\beta)\begin{pmatrix}
0 & 0 & 1
\end{pmatrix}^T \\
&=\begin{pmatrix}
\sin\beta\cos\alpha & \sin\beta\sin\alpha & \cos\beta
\end{pmatrix}^T,
\end{align}
the transformed gyroscopic coefficients are
\begin{align}
    &\frac{R^2}{\alpha_r\alpha_s}\beta_{r,s}= J_z \,\beta_{r,s}^{(1)}+J_x\,\beta_{r,s}^{(2)}+J_y\,\beta_{r,s}^{(3)}, \\
    &\frac{R^2}{\alpha_r\alpha_s}\beta_{r,-s}=  J_z\,\beta_{r,-s}^{(1)}+J_x\,\beta_{r,-s}^{(2)}+J_y\,\beta_{r,-s}^{(3)},\\
    &\frac{R^2}{\alpha_r\alpha_s}\beta_{-r,-s}=  J_z\,\beta_{-r,-s}^{(1)}+J_x\,\beta_{-r,-s}^{(2)}+J_y\,\beta_{-r,-s}^{(3)},\\ \label{confused}
    &\beta_{-r,s}= -\beta_{s,-r},
\end{align}
where\PA{,} for $r$ associated with the harmonic pair of integers $(n,m)$ and $s$ associated with $(u,v)$, we  introduced the coefficients
\begin{align}
    \beta_{r,s}^{(1)}&= \frac{2}{m^2-v^2} \int\left[  v^2 P_u^v\dm P_n^m+ m^2P_n^m\dm P_u^v \right]\mu d\mu, \\
  \beta_{r,s}^{(2)}&= \frac{\pi}{4}   \int \left[ m K_{m,v}^{(1)} P_n^m \dm P_u^v \bar{\mu}^{1/2} -v K_{m,v}^{(2)} \dm P_n^m P_u^v \bar{\mu}^{1/2}\right]d\mu ,\\
   \beta_{r,s}^{(3)}&=  \int \left[ m K_{m,v}^{(3)} P_n^m \dm P_u^v \bar{\mu}^{1/2} -v K_{m,v}^{(4)} \dm P_n^m P_u^v \bar{\mu}^{1/2}\right]d\mu ,\\
    \beta_{r,-s}^{(1)}&= \frac{-2v}{m^2-v^2} \int\left[ \dm P_u^v\dm P_n^m \mu \bar{\mu}+ m^2 P_n^m P_u^v \frac{\mu}{\bar{\mu}} \right]d\mu,\\
  \beta_{r,-s}^{(2)}&= \frac{\pi}{4}   \int \left[  K_{m,v}^{(2)} \dm P_n^m \dm P_u^v \bar{\mu}^{3/2} -mv K_{m,v}^{(1)}  P_n^m P_u^v \bar{\mu}^{-1/2}\right]d\mu, \\
  \beta_{r,-s}^{(3)}&=  \int \left[  K_{m,v}^{(4)} \dm P_n^m \dm P_u^v \bar{\mu}^{3/2} -mv K_{m,v}^{(3)}  P_n^m P_u^v \bar{\mu}^{-1/2}\right]d\mu, \\
    \beta_{-r,-s}^{(1)}&= \frac{2mv}{m^2-v^2} \int\left[  P_u^v\dm P_n^m+ P_n^m\dm P_u^v \right]\mu d\mu, \\
  \beta_{-r,-s}^{(2)}&= \frac{\pi}{4}   \int \left[  v K_{m,v}^{(1)} \dm P_n^m P_u^v \bar{\mu}^{1/2}-m K_{m,v}^{(2)} P_n^m \dm P_u^v \bar{\mu}^{1/2}\right]d\mu, \\
   \beta_{-r,-s}^{(3)}&=  \int \left[ v K_{m,v}^{(3)} \dm P_n^m P_u^v \bar{\mu}^{1/2} -m K_{m,v}^{(4)} P_n^m \dm P_u^v \bar{\mu}^{1/2}\right]d\mu ,
\end{align}
with $\bar{\mu}=1-\mu^2$ and 
\begin{align} 
   K_{m,v}^{(1)}&= (1+\delta_{v,0})\delta_{m-v,1} -\delta_{m-v,-1}, \\ 
    K_{m,v}^{(2)}&=  K_{v,m}^{(1)}, \\
     K_{m,v}^{(3)}&= m \left(
     \frac{1}{m^2-(v^2+1)^2} + \frac{1}{m^2-(v^2-1)^2}\right),\\
     K_{m,v}^{(4)}&=  K_{v,m}^{(3)}.
\end{align}
Under this transformation, the latitude of the center of the ocean in the rotating frame is given by
\begin{equation} \label{colatitude}
    \cos\theta^\prime= \cos\theta\cos\beta + \sin\theta\sin\beta\cos(\lambda-\alpha).
\end{equation}
To compute the integrals involved in the gyroscopic coefficients, we make use of the essential condition\footnote{we note that this general condition is invalid in the case where $n=m=u=v=0$. However, this case is excluded here by the definition of the eigenfunctions in Eqs. \eqref{eigenfunctionsss} and \eqref{eigenfunctionsss2}.} (e.g.  \cite{longuet1970free}) \PAN{This is not true in the general case (e.g. $n=m=u=v=0$).}{ok}
\begin{equation}
  \left.  P_n^m P_u^v  \right|_{\mu=\pm1} =0,
\end{equation}
and we  use the overlap integral of two ALFs (Eq. \ref{overlapdef}), which we compute using the closed form relations provided in the following section. Now we have at hand all the elements to compute the gyroscopic coefficients harmonically. The final form of the three coefficients with superscript $^{(1)}$ are identical to those in  \cite{webb1980tides} and similar to those in  \cite{longuet1970free} up to certain misprints.  For the rest of the terms, the expressions given in  \cite{webb1982tides} involve numerous typographical errors and inconsistencies, so we provide here the full set of the gyroscopic coefficients. The coefficients $ \beta_{r,s}^{(1)}$ and  $ \beta_{r,s}^{(2)}$ read as
    \begin{align}\nonumber
         \beta_{r,s}^{(1)} = &\left[\frac{u(u+1)+v(v+1)}{v+1} -2v\frac{n(n+1)-u(u+1)+v}{m^2-v^2}   \right]\mathcal{O}_{n,m}^{u,v} \\
         &+ \frac{1}{v+1}\mathcal{O}_{n,m}^{u,v+2},
    \end{align}
\begin{subequations}
    \begin{align}
         \beta_{r,s}^{(2)}= &\frac{\pi}{4} \left[  m K_{m,v}^{(1)}\int   P_n^m \dm P_u^v \bar{\mu}^{1/2} d\mu -v K_{m,v}^{(2)}\int  \dm P_n^m P_u^v \bar{\mu}^{1/2}d\mu \right] \label{betna_rs_2_a} \\
        =    &\frac{\pi}{8}\left\{ m K_{m,v}^{(1)}\left[(u+v)(u-v+1)\mathcal{O}_{n,m}^{u,v-1} - \mathcal{O}_{n,m}^{u,v+1}\right]\right.\nonumber \label{beta_rs_2_b} \\
     &\qquad \left.{}  -vK_{m,v}^{(2)}\left[(n+m)(n-m+1)\mathcal{O}_{n,m-1}^{u,v} - \mathcal{O}_{n,m+1}^{u,v}\right] \right\},
    \end{align}
\end{subequations}
where we used  \eq{Rec4} for each integrand in \eq{betna_rs_2_a} to obtain \eq{beta_rs_2_b}. The coefficients $\beta_{r,s}^{(3)}$, $ \beta_{r,-s}^{(1)}$, and $ \beta_{r,-s}^{(2)}$ read as
\begin{align}
   \beta_{r,s}^{(3)}=   \frac{1}{2} &\left\{ m K_{m,v}^{(3)}\left[(u+v)(u-v+1)\mathcal{O}_{n,m}^{u,v-1} - \mathcal{O}_{n,m}^{u,v+1}\right]\right.\nonumber \\
 &\qquad \left.{}  -vK_{m,v}^{(4)}\left[(n+m)(n-m+1)\mathcal{O}_{n,m-1}^{u,v} - \mathcal{O}_{n,m+1}^{u,v}\right] \right\},
\end{align}
\begin{align}
   \beta_{r,-s}^{(1)}=    &\frac{-2v}{(m^2-v^2)(2n+1)} \left\{ (n+1)(n-1)(n+m) \mathcal{O}_{n-1,m}^{u,v} \right.\nonumber \\ &\qquad \left.{} + n(n+2)(n-m+1) \mathcal{O}_{n+1,m}^{u,v} \right\},
\end{align}
\begin{subequations}
    \begin{align}\label{b2r-s} 
        \beta_{r,-s}^{(2)}&= \frac{\pi}{4} \left[  K_{m,v}^{(2)}  \!\! \int  \!\! \dm P_n^m \dm P_u^v \bar{\mu}^{3/2}d\mu  -mv K_{m,v}^{(1)} \!\!\int \!\!  P_n^m P_u^v \bar{\mu}^{-1/2}d\mu\right] \\ \label{beta_r_-s_a}
       &= \frac{\pi}{4} \left[  K_{m,v}^{(2)} \!\!\int \!\! \dm P_n^m  \bar{\mu}  \dm P_u^v \bar{\mu}^{1/2}d\mu -mv K_{m,v}^{(1)} \!\! \int \!\!  P_n^m P_u^v \bar{\mu}^{-1/2}d\mu\right] \\ \nonumber
     &=  \frac{\pi K_{m,v}^{(2)} }{8(2n+1)} \Bigg\{ \\\nonumber
     &\,\,\,\,\,\,(n+1)(n+m) \left[ (u+v)(u-v+1)\mathcal{O}_{n-1,m}^{u,v-1} - \mathcal{O}_{n-1,m}^{u,v+1}\right]  \\\nonumber 
    &\,\,\,+n(n-m+1) \left[\mathcal{O}_{n+1,m}^{u,v+1} - (u+v)(u-v+1)\mathcal{O}_{n+1,m}^{u,v-1}   \right]\Bigg\} \\
    &\,\,\,+\frac{\pi v K_{m,v}^{(1)}}{8}\left\{ \mathcal{O}_{n-1,m+1}^{u,v} + (n+m-1)(n+m)\mathcal{O}_{n-1,m-1}^{u,v}\right\},
    \end{align}
\end{subequations}
where we used the recurrence relations of \eq{Rec4} and \eq{Rec5} to compute the first integral of \eq{beta_r_-s_a}, and the relation of \eq{Rec2} to compute the second integral. Finally, the remaining terms $ \beta_{r,-s}^{(3)}$, $  \beta_{-r,-s}^{(1)}$, $ \beta_{-r,-s}^{(2)}$, and $ \beta_{-r,-s}^{(3)}$ read as
\begin{align}\nonumber
    \beta_{r,-s}^{(3)}=& \frac{ K_{m,v}^{(4)} }{2(2n+1)} \Bigg\{ \\\nonumber
    &\,\,\,(n+1)(n+m) \left[ (u+v)(u-v+1)\mathcal{O}_{n-1,m}^{u,v-1} - \mathcal{O}_{n-1,m}^{u,v+1}\right] \\\nonumber  
    &+n(n-m+1) \left[\mathcal{O}_{n+1,m}^{u,v+1} - (u+v)(u-v+1)\mathcal{O}_{n+1,m}^{u,v-1}   \right]\Bigg\} \\
    &+\frac{ v K_{m,v}^{(3)}}{2}\left\{ \mathcal{O}_{n-1,m+1}^{u,v} + (n+m-1)(n+m)\mathcal{O}_{n-1,m-1}^{u,v}\right\},
\end{align}
%\item $ \beta_{-r,-s}^{(1)}$ as in  \cite{longuet1970free}:
\begin{equation}
    \beta_{-r,-s}^{(1)}= \frac{-2mv}{m^2-v^2} \mathcal{O}_{n,m}^{u,v},
\end{equation}
\begin{align}
   \beta_{-r,-s}^{(2)}=    &\frac{\pi}{8}\left\{ -m K_{m,v}^{(2)}\left[(u+v)(u-v+1)\mathcal{O}_{n,m}^{u,v-1} - \mathcal{O}_{n,m}^{u,v+1}\right]\right.\nonumber \\
 &\qquad \left.{}  +vK_{m,v}^{(1)}\left[(n+m)(n-m+1)\mathcal{O}_{n,m-1}^{u,v} - \mathcal{O}_{n,m+1}^{u,v}\right] \right\},
\end{align}
\begin{align}
   \beta_{-r,-s}^{(3)}=    \frac{1}{2}&\left\{ -m K_{m,v}^{(4)}\left[(u+v)(u-v+1)\mathcal{O}_{n,m}^{u,v-1} - \mathcal{O}_{n,m}^{u,v+1}\right]\right.\nonumber \\
 &\qquad \left.{}  +vK_{m,v}^{(3)}\left[(n+m)(n-m+1)\mathcal{O}_{n,m-1}^{u,v} - \mathcal{O}_{n,m+1}^{u,v}\right] \right\}.
\end{align}
%%%%%%%%%%%%%%%%%%%%%%%%%%%%%%%%%%%%%%%%%%%%%%%%

\begin{figure} 
\centering
\tdplotsetmaincoords{60}{130}
	\begin{tikzpicture}[scale=4]
	    \begin{scope}[opacity=0.1]
		\shade[ball color = white, opacity = 1] (0,0) circle (1cm);
		\end{scope}
		%\draw[black,fill=white,opacity=0.0] (0,0) circle (1cm);
	\begin{scope}[tdplot_main_coords]
		\tikzstyle{init} = [black]		
		\tikzstyle{prec} = [blue]		
		\tikzstyle{nuta} = [red]		
		\tikzstyle{base} = [thick,-stealth]
		\tikzstyle{angle} = [thick,-latex]	
		\tikzstyle{circle} = [thin,-stealth]	
		\coordinate (P) at ({0)},{-1},{0});
        \draw[thick, -stealth] (0,0,0) -- (P)node[anchor=south ]{$\boldsymbol{{O}}$};
		\def\epsi{10}
		\def\etheta{50}
		\def\rang{0.65}	
		
		\coordinate (O) at (0,0,0);
		\draw[thick,-stealth,init] (O) -- (1,0,0) node[anchor=north east]{$\boldsymbol{\hat{x}}$};
		\draw[thick,-stealth,init] (O) -- (0,1,0) node[anchor=north west]{$\boldsymbol{\hat{y}}$};
		\draw[thick,-stealth,init] (O) -- (0,0,1) node[anchor=south]{$\boldsymbol{\hat{s}}$};

		\tdplotsetrotatedcoords{\epsi}{0}{0}
		\draw[tdplot_rotated_coords,angle,prec] (O) --(1,0,0) node[anchor=north east]{$\boldsymbol{\hat{x}}^\prime$};
		\draw[tdplot_rotated_coords,angle,prec] (O) --(0,1,0) node[anchor=west]{$\boldsymbol{\hat{y}}^\prime$};
		\tdplotdrawarc[tdplot_rotated_coords,circle,prec]{(0,0,0)}{1}{0}{360}{}{}	
		\tdplotdrawarc[tdplot_rotated_coords,angle,prec]{(0,0,0)}{\rang}{90-\epsi}{90}{anchor=north east,prec}{$\boldsymbol{\alpha}$}	
		\tdplotdrawarc[tdplot_rotated_coords,angle,prec]{(0,0,0)}{\rang}{-\epsi}{0}{anchor=north east,prec}{$\boldsymbol{\alpha}$}	
		
% \tdplotdrawarc[->,color=black]{(0,0,0.8)}{0.1}{0}{340}{anchor=south west,color=black}{$\boldsymbol{\vec\Omega}$}
% \tdplotsetthetaplanecoords{0}
% \tdplotsetthetaplanecoords{50}
% \tdplotdrawarc[tdplot_rotated_coords,->,color=red]{(1,0,0.7)}{0.1}{120}{470}{anchor=south west,color=black}{}

		\tdplotsetrotatedcoords{\epsi}{\etheta}{0}
		\draw[tdplot_rotated_coords,base,nuta] (O) --(1,0,0) node[anchor=north east]{$\boldsymbol{\hat{x}}^{\prime\prime}$};
		\draw[tdplot_rotated_coords,base,nuta] (O) --(0,0,1) node[anchor=south east]{$\boldsymbol{\hat{s}}^{\prime\prime}$};
		\tdplotsetrotatedthetaplanecoords{0}
		\tdplotdrawarc[tdplot_rotated_coords,circle,nuta]{(0,0,0)}{1}{0}{360}{}{}		
		\tdplotdrawarc[tdplot_rotated_coords,angle,nuta]{(0,0,0)}{\rang}{90-\etheta}{90}{anchor=south west,nuta}{$\boldsymbol{\beta}$}		
		\tdplotdrawarc[tdplot_rotated_coords,angle,nuta]{(0,0,0)}{\rang}{-\etheta}{0}{anchor=south,nuta}{$\boldsymbol{\beta}$}
	\end{scope}
	\end{tikzpicture}
\caption{The adopted transformation scheme that allows recovering the tidal response of a hemispheric ocean with an arbitrary center on the sphere. We use an Eulerian transformation of the form $\mathcal{R}_3(\alpha)\mathcal{R}_2(\beta)\mathcal{R}_3(\gamma)$ with $\gamma=0$, allowing us to shift the latitude of the oceanic center $O$ by shifting the spin axis from $\hat{s}$ to $\hat{s}^{\prime\prime}$ in a true polar wander scenario \citep{webb1982tides}. } \label{trans_fig}
\end{figure}

\section{The overlap integral  $\mathcal{O}_{n,m}^{u,v}$}
\label{overlap_section}
We provide here a closed form solution for the computation of the overlap integral of \eq{overlapdef}. The procedure is assimilated from tools of angular momentum quantization \citep{Varshalovich}. Following  \cite{dong2002overlap}, and introducing the notation $q= v-m$, we have
\begin{equation}\label{overlap1}
    \mathcal{O}_{n,m}^{u,v} = C_{n,m}^{u,v}\sum_l (2l+1) \mathcal{D}(|q|,l) \cdot \begin{pmatrix} n & u&l \\ 0&0&0 \end{pmatrix}\begin{pmatrix} n&u& l \\ -m & v & m-v \end{pmatrix},
\end{equation}
where the factors $ C_{n,m}^{u,v}$ are given by
\begin{equation} \label{overlapC}
    C_{n,m}^{u,v} = (-1)^\kappa 2^{|q|-2}|q| \sqrt{\frac{(n+m)!(u+v)!}{(n-m)!(u-v)!}},
\end{equation}
and the coefficients $ \mathcal{D}(|q|,l)$ by
\begin{equation}\label{overlapD}
    \mathcal{D}(|q|,l) = \left[ 1+(-1)^{l+|q|}\right] \sqrt{\frac{(l-|q|)!}{(l+|q|)!} }\frac{\Gamma(l/2)\Gamma((l+|q|+1)/2)}{((l-|q|)/2)!\Gamma((l+3)/2)}.
\end{equation}
We note here that the phase $\kappa$ introduced in  \cite{dong2002overlap} as 
\begin{equation}
  \kappa = \begin{cases}
    m& \text{if } v \geq m,\\
    v             & \text{otherwise},
\end{cases}
\end{equation}
corrects the phase given in  \cite{mavromatis1999generalized} and  \cite{crease1966tables}, where the latter was used for the computation of the gyroscopic coefficients in  \cite{longuet1970free} and \cite{webb1980tides,webb1982tides}.

In \eq{overlap1}, the summation over $l$ runs for $|n-u|\leq l\leq (n+u) ; l\geq |q|;$ and $|l+n+u|$ is even.
Finally, the Wigner $3$-$\rm jm$ symbols are determined from   \cite{Varshalovich} by
\begin{align}\label{3jm}\nonumber
    \begin{pmatrix}
    a & b & c \\ d& e&f
    \end{pmatrix} 
    =& (-1)^{R_{21}+R_{31}+R_{32}}\left[\frac{R_{31}!R_{32}!R_{33}!R_{13}!R_{23}!}{(J+1)!R_{11}!R_{12}!R_{21}!R_{22}!}\right]^{1/2}\\
    &\times\sum_z(-1)^z \frac{(R_{21}+z)!(R_{11}+R_{31}-z)!}{z!(R_{31}-z)!(R_{23}-z)!(R_{13}-R_{31}+z)!},
\end{align}
where $J=a+b+c$, and $R_{ij}$ are the elements of the so-called Regge $\mathfrak{R}$-symbol \citep{regge1958symmetry} defined as
\begin{align}\label{overlapRegge}
    \nonumber 
    &R_{11}=-a+b+c, \hspace{.6cm }R_{12}= a-b+c,\hspace{.6cm }R_{13}= a+b-c, \\
     \nonumber 
    &R_{21}=  a+d , \hspace{1.6cm }R_{22}=  b+e,\hspace{1.3cm }R_{23}= c +f,\\     
    &R_{31}= a-d , \hspace{1.6cm }R_{32}= b-e,\hspace{1.3cm }R_{33}=c-f.
\end{align}
% \begin{equation}
%     \mathfrak{R}= \begin{Vmatrix}
%     -a+b+c & a-b+c & a+b-c \\
%     a+d & b+e & c +f \\
%     a-d & b-e & c-f
%     \end{Vmatrix}.
% \end{equation}
The summation in \eq{3jm} runs over all integer values of $z$ for which all the factorial arguments are non-negative.
Finally, we note that using \eq{overlap1}, $\mathcal{O}_{n,m}^{u,v}=0$ when $v=m$. In that case, we alternatively use
\begin{equation}
    \mathcal{O}_{n,m}^{u,m} = \frac{2}{2n+1}\frac{(n+m)!}{(n-m)!}\delta_{n,u}\,.
\end{equation}
This method for the computation of the overlap integral was verified numerically using MATLAB's ALFs package.
\section{The tidal forcing term $\Bar{\zeta_r}$}
As in  \cite{webb1980tides}, considering the equilibrium tide $\Bar{\zeta}$ to  have a unit root mean square amplitude, and to be driven by a spherical harmonics term 
\begin{equation}\label{SH_def}
    Y_p^q(\theta,\lambda) = \sqrt{\frac{2p+1}{4\pi}\frac{(p-q)!}{(p+q)!}}\,P_p^q(\cos\theta)\,\exp(iq\lambda)\,,
\end{equation}
with angular frequency $\sigma$, we have
\begin{equation}
    \Bar{\zeta}= \sqrt{2\pi}Y_{p}^{q}(\theta,\lambda)\exp(i\sigma t).
\end{equation}
Under the rotation of the coordinate system described by the Euler angles  ($\alpha,\beta,\gamma)$ (see Appendix \ref{Gyro_ALPs} and   Fig. \ref{trans_fig}), the spherical harmonics transform as \citep{Varshalovich} \PAN{The new colatitude and longitude, $\vartheta$ and $\varphi$, were not introduced.}{ok}
\begin{equation}
    Y_p^{s} (\theta^\prime,\lambda^\prime)= \sum_{q=-p}^p Y_p^q(\theta,\lambda)D_{s,q}^p(\alpha,\beta,\gamma),
\end{equation}
or
 \begin{equation} \label{Ypq_rotated}
     Y_p^{q} (\theta,\lambda)= \sum_{s=-p}^p Y_p^{s}(\theta^\prime,\lambda^\prime)D_{s,q}^{p*}(\alpha,\beta,\gamma).  
 \end{equation}
\PAR{where  $D_{q,s}^{p}$ designate the Wigner D-functions. These functions are themselves}{} the product of three functions \citep{Varshalovich}, each depending on one argument $\alpha, \beta,$ or $\gamma$,
\begin{equation}\label{wignerD}
    D_{s,q}^{p}(\alpha,\beta,\gamma)= e^{-iq\alpha}d_{sq}^{p}(\beta)e^{-is\gamma}.
\end{equation}
In this expression, the functions $d_{sq}^{p}(\beta)$ are given by
\begin{align}\nonumber
    d_{sq}^{p}(\beta) &= (-1)^{p-s}\left[ (p+q)!(p-q)!(p+s)!(p-s)!\right]^{1/2}\\
    &\times\sum_{j}(-1)^j \frac{(\cos\beta/2)^{q+s+2j}(\sin\beta/2)^{2p-q-s-2j}}{j!(p-q-j)!(p-s-j)!(q+s+j)!}\PA{,}
\end{align}
with $j$ running over all integer values for which the factorial arguments are positive. This sum involves $N+1$ terms, where $N$ is the minimum of $ (p+q),(p-q),(p+s),$ and $(p-s).$ Since we are studying the semi-diurnal tide\PAR{ ($p=q=2$),}{} we are left with one term only. Expanding the harmonic \PAR{factor $Y_p^{s}(\theta^\prime,\lambda^\prime)$ of \eq{Ypq_rotated}}{} in terms of the basis eigenfunctions we get \PA{the expression of the equilibrium oceanic depth variation in the rotated frame of reference,}
\begin{equation}
    \Bar{\zeta}=\sqrt{\frac{\pi R^2}{2}}\exp(i\sigma t) \,\sum_{s=-p}^p D_{s,q}^{p*}(\alpha,\beta,\gamma)\,(1+\delta_{s,0})^{1/2} \left[\phi_p^{s} + i\psi_p^{s}\right]\PA{.}
\end{equation}
\PAR{Then, invoking the definition of the component $\Bar{\zeta_r}$,}{} 
\begin{equation}
    \Bar{\zeta_r} = \int_\mathcal{O} \phi_r\Bar{\zeta} dA \PA{,}
\end{equation}
we get \PA{its expression in the rotated frame of reference,}
\begin{align}\label{tidalforcingterm}\nonumber
      \Bar{\zeta_r} =& \sqrt{\frac{\pi R^2}{2}} \exp(i\sigma t)\,\sum_{s=-p}^p D_{s,q}^{p*}(\alpha,\beta,\gamma)\,(1+\delta_{s,0})^{1/2}\\
      &\times\Bigg[\int_\mathcal{O}\phi_p^s\phi_n^mdA + i\int_\mathcal{O}\psi_p^{s}\phi_n^mdA\Bigg] \PA{,}
\end{align}
where \PA{the dot products of the eigenfunctions simplify to}
\begin{equation}
   \int_\mathcal{O}\phi_p^s\phi_n^mdA = \begin{cases}
   \delta_{n,p} ,& \text{if } s=m, \\
   (-1)^m\delta_{n,p} ,& \text{if } s=-m,\\
   0 ,&\text{otherwise},
   \end{cases}
\end{equation}
and
\begin{equation}
    \int_\mathcal{O}\psi_p^s\phi_n^mdA =\begin{cases}
    0 ,& \text{if } m+s ={\text{even}},\\
    \displaystyle\alpha_{p,s}\,\alpha_{n,m} \frac{2s}{s^2-m^2}\mathcal{O}_{p,s}^{n,m}              & \text{otherwise.}
\end{cases}
\end{equation}
We note that as the index $s$ takes negative values, we use
\begin{equation}
    P_p^{-s}= (-1)^s \frac{(p-s)!}{(p+s)!}P_p^s
\end{equation}
in the overlap integral of \eq{overlapdef}.

\section{\PAR{The tidal torque of a hemispherical ocean}{The Tidal Response}} \label{section_tidal_response}
Once the gyroscopic coefficients are computed, the linear system of the coefficients $(p_r,p_{-r})$ in \eq{final_linear_system} is truncated and solved numerically. What we are after is the tidal torque that enters in the dynamical equations of the Earth-Moon system. Two torques are involved as explained in the main text, and they depend on the rotational angular velocity of the Earth and the orbital frequency of the tidal perturber. Defining the tidal torque transferring power from the Earth's rotational momentum to the perturber's orbital angular momentum  by $\mathcal{T}$, the power lost by the Earth would be $\mathcal{T}\Omega$, and the power gained by the perturber is $\mathcal{T} n_{\rm orb}$. The difference between them is the dissipative work of the total tidal mass redistribution $\mathcal{W}_{\rm diss}$, thus
\begin{equation}\label{hemi_torque}
    \mathcal{T}= \frac{ \mathcal{W}_{\rm diss}}{\Omega-n_{\rm orb}}.
\end{equation}
The total dissipative work is the sum of two contributions: the dissipative work of oceanic tidal currents $\mathcal{W}_{\rm diss}^{\rm oc}$, and dissipation in the deforming viscoelastic mantle. In the formalism established thus far, \PAR{we calculated the self-consistently coupled tidal responses of the ocean and solid part for the Earth's half hosting the hemispherical ocean, which corresponds to the effective tidal response of the planet for this hemisphere. The tidal response of the continental hemisphere is simply described by the solid Love numbers introduced in \eq{love_def} since there is no oceanic tide in that case. The coupled solid-oceanic tidal response accounts  for both the direct gravitational tidal forcing generated by the perturber on the solid part and ocean, and for the mutual forcings of the two layers through the variations of the loading exerted by the latter on the former, and the variations of the Earth's self-gravitational potential due to mass redistribution.}{we focused on the former component  accounting for the direct effect of the tidal forcing on the ocean, and we included the effects of loading and self-attraction between the oceanic shell and the deforming solid surface.} \PAR{For simplicity, we ignore the energy dissipated in the solid part in the calculation of the tidal torque, and we only consider that occurring within the oceanic shell, namely $\mathcal{W}_{\rm diss}^{\rm oc}$.}{However, we ignored the dissipation in the mantle itself.} This is justified by \PAR{}{the effective nature of our model, and by }the predominance of the oceanic response over the solid part over the time interval covered by the hemispherical ocean configuration in our model. This hierarchy of contributions is only jeopardized by  the  emerging  significance  of  the  solid  dissipation when moving backwards in time and increasing the Earth's rotational velocity $\Omega$. Solid Earth dissipation would  also be amplified with an early less viscous mantle due to higher Hadean-Archean temperatures \citep{ross1989evolution}. Eventually, a regime transition may lead to the predominance of the mantle's elastic response \citep{lau2015normal,lau2016anelasticity}. In our nominal model of the main text, the switch from the hemispherical ocean configuration to the global ocean configuration occurs mid-Archean, beyond which we self-consistently account for the dissipative contribution of the solid part (Appendix \ref{global_response}). Thus we have only ignored the dissipative contribution of the mantle when it is insignificant.  

The oceanic dissipative work is given by \citep{webb1980tides} 
\begin{align}\nonumber\label{work_friction}
    \mathcal{W}_{\rm diss}^{\rm oc}&=\left\langle \int_\mathcal{O} \Vec{u}(t)\cdot\sigma_{\rm R}\Vec{u}(t) dA\right\rangle\\
    &=\frac{1}{2}\sigma_{\rm R}\sigma^2 \sum_{r=1}^{\infty}\left(\mu_rp_rp_r^*+\nu_rp_{-r}p_{-r}^*\right),
\end{align}
where $\langle \rangle$ denotes time averaging  over the tidal period. This work should be equal to the work done by the tidal force on the ocean \PAN{Similarly, $\mathcal{W}_{\rm tide}$ corresponds actually to the work done by the tidal force on the ocean only, and should be rather denoted by $\mathcal{W}_{\rm oc ;  tide}$ or $\mathcal{W}_{\rm tide}^{\rm oc}$ or $\mathcal{W}_{\rm oc}^{\rm T}$ to be consistent with the notations introduced previously.}{ok}
\begin{align}\nonumber\label{work_moon}
    \mathcal{W}_{\rm tide}^{\rm oc}&=\left\langle\rho_{\rm oc} gH\int_\mathcal{O} \grad\Bar{\zeta}(t)\cdot\Vec{u}(t)dA\right\rangle\\
    &=\frac{1}{2}\rho_{\rm oc} gH\sigma \Im\left\{\sum_{r=1}^\infty p_r\mu_r \Bar{\zeta}_r^* \right\}.
\end{align}
Hence the tidal torque associated with  the lunar  semi-diurnal frequency $\sigma=2(\Omega-n_{\rm M})$, $n_{\rm M}$ being the lunar mean motion, is 
\begin{equation} \label{hemi_lunar_torque}
    \mathcal{T}_{\rm M}= \rho_{\rm oc} gH \Im\left\{\sum_{r=1}^\infty p_r\mu_r \Bar{\zeta}_r^* \right\},
\end{equation}
and we obtain a similar expression for the solar tides $\mathcal{T}_{\rm S}$ when solving the system with the solar tidal frequency component $\sigma=2(\Omega-n_{\rm S})$, $n_{\rm S}$ being the solar mean motion, generating the solar tidal work.

\begin{figure}
    \centering
    \includegraphics[width=\columnwidth]{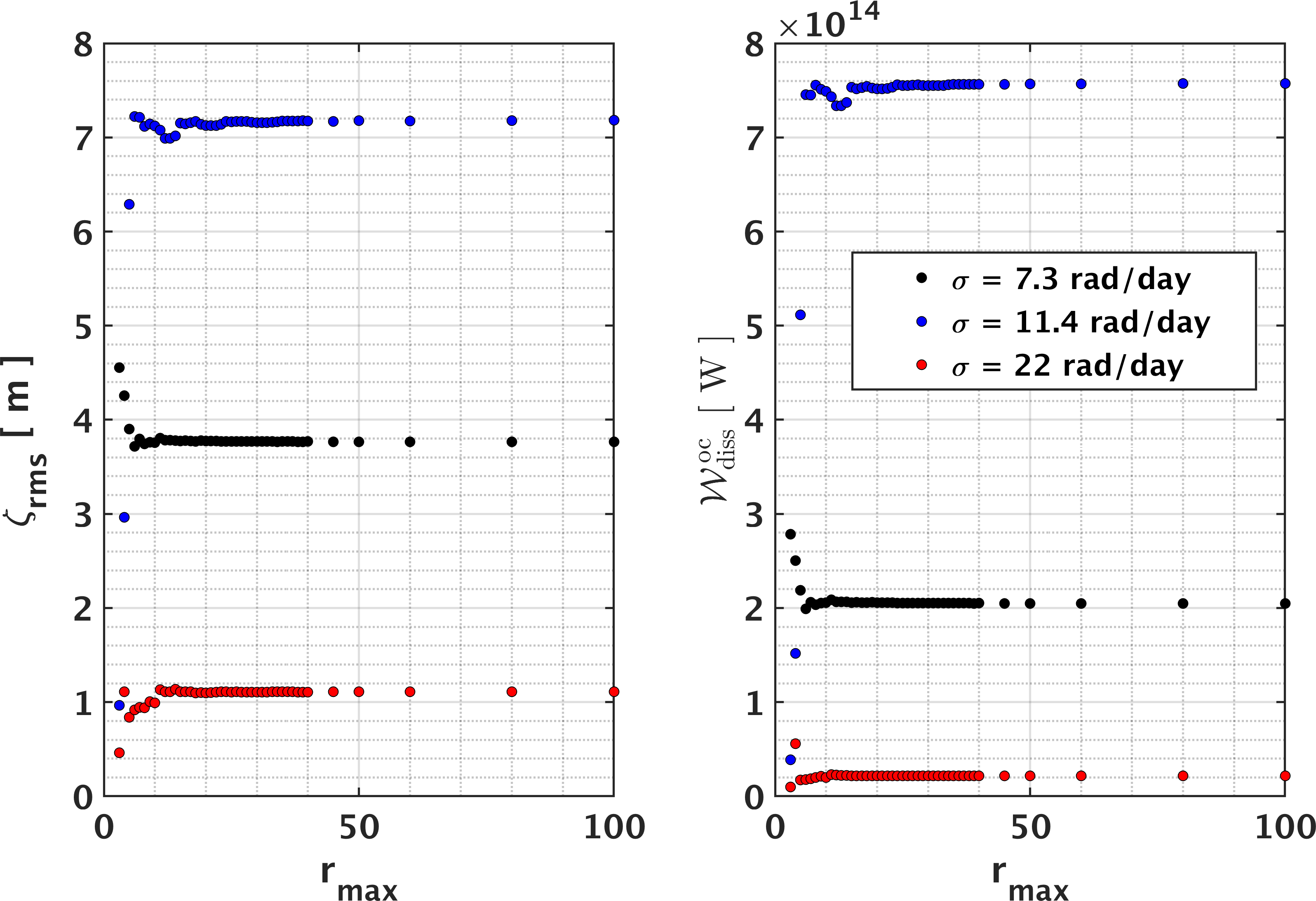}
    \caption{A numerical analysis on the dependence of the tidal response computation on the truncation order $r_{\rm max}$. The response is quantified by the root mean square tidal amplitude $\zeta_{\rm rms}$ (Eq. \ref{rms_amp}) and the dissipative work $\mathcal{W}_{\rm diss}^{\rm oc}$ (Eq. \ref{work_friction}), and plotted for three tidal frequencies $7.3, 11.4, $ and $22$ rad/day that correspond to the vicinity of a tidal resonance, the peak of a resonance, and the background spectrum respectively. }
    \label{convergence_test}
\end{figure}

Besides the tidal torque, the tidal response can  also  be quantified by the root mean square tidal height variation $\zeta_{\rm rms}$ given as
\begin{equation} \label{rms_amp}
    \zeta_{\rm rms} = \sqrt{\frac{H}{\pi R^2} \sum_{r=1}^\infty\mu_r^2p_r^*p_r}.
\end{equation}
As these quantities are computed numerically, a truncation order $r_{\rm max}$ is required. In Fig. \ref{convergence_test} we show the numerical dependence of the tidal response on $r_{\rm max}$. Since the response is dominated by gravity modes, the tidal solution converges fast enough with $r_{\rm max}$. To avoid any truncation effect in our computation, and to properly account for the resonances, we adopted $r_{\rm \max}=50$.

\section{Modelling the tidal response of a global ocean} \label{global_response}

When the global oceanic geometry is encountered in our model, the tidal response is computed based on the analytical formalism described in  \cite{auclair2018oceanic,auclair2019final}. We refer the reader to these references for a complete development of the theory, and we only brief here on the essential steps that lead towards computing the tidal response used in the main text\footnote{We remind the reader that we can proceed with this theory as such only because we are studying dynamics in the coplanar setting. The theory requires further development if one were to account for the Earth's obliquity and lunar inclination.}. In this approach, solving the governing system in \eq{momentum_continuity} is done by expanding the velocity field, the tidal elevation, and the forcing \PAR{gravitational tidal potential in Fourier series of time and longitude}{ longitudinally and spectrally in Fourier series}, with the tidal frequency serving as the expansion frequency. Thus we have
\begin{align}  \label{fourier_decompo}  \nonumber
    &\vec u= \sum_{m,\sigma}\vec u^{\,m,\sigma}(\theta) \exp{i(\sigma t+m\lambda},
    \\\nonumber
    & \zeta= \sum_{m,\sigma}\zeta^{\,m,\sigma}(\theta) \exp{i(\sigma t+m\lambda},\\
    &\zeta_{\rm eq}= \sum_{m,\sigma}\zeta_{\rm eq}^{\,m,\sigma}(\theta) \exp{i(\sigma t+m\lambda}.
\end{align}
Defining the complex tidal frequency $\Tilde{\sigma}$ and the complex spin parameter $\Tilde{\nu}$ as in  \cite{auclair2018oceanic} by
\begin{equation} 
\label{tilde_sigma_tidle_nu}
    \Tilde{\sigma}=\sigma-i\sigma_{\rm R} \hspace{.4cm} \text{and} \hspace{.4cm} \Tilde{\nu}=\frac{2\Omega}{\Tilde{\sigma}},
\end{equation}
and replacing the tidal quantities by their expansions, the governing system reduces to an eigenvalue-eigenfunction problem, known classically (when ignoring friction) as the Laplace tidal equation \citep{lee1997low}. We assume that the Fourier components can be expanded spatially using a set of the latitudinal complex Hough functions \citep{hough} \{$\Theta_n^{m,\Tilde{\nu}}(\theta)$\}, associated with a set of eigenvalues \{$\Lambda_n^{m,\Tilde{\nu}}$\}. To compute these functions and their associated eigenvalues, we adopt the method developed in  \cite{wang2016computation}, where Hough functions are expanded in terms of Associated Legendre Functions
\begin{align} \label{define_hough}
    \nonumber \Theta_n^{m,\Tilde{\nu}}(\theta) &=  \sum_{m\leq l} A_{n,l}^{m,\Tilde{\nu}}P_l^m(\cos\theta), \\
    P_l^m(\cos\theta) &=  \sum_{n}B_{l,n}^{m,\Tilde{\nu}}\Theta_n^{m,\Tilde{\nu}}(\theta),
\end{align}
with $A_{n,l}^{m,\Tilde{\nu}}$ and $B_{l,n}^{m,\Tilde{\nu}}$ being complex \PAR{change of basis}{weighting} coefficients. Using \PAR{the change of basis coefficients $A_{n,l}^{m,\Tilde{\nu}}$}{these weighting coefficients}, the tidal displacement solution is written as\PAR{}{:}
\begin{figure*}
\centering
\includegraphics[width=\textwidth]{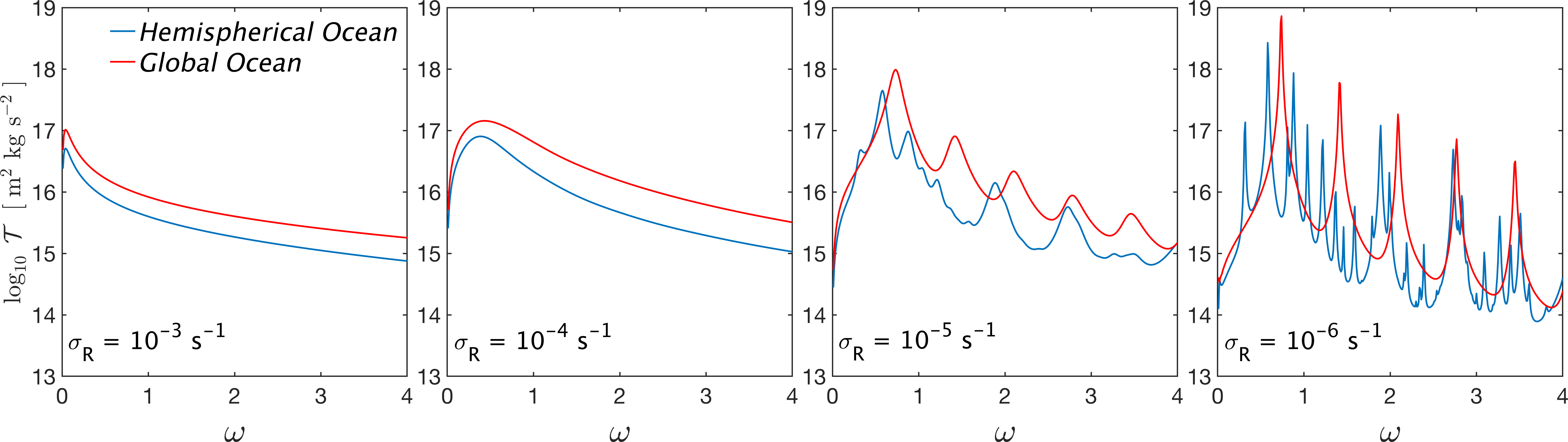}
\caption{Tidal torque between the Earth and the Moon corresponding to the coupled oceanic-solid response in two configurations: a global oceanic shell of thickness $H= 4000$ m (shown in red), and a hemispherical ocean with the same thickness, symmetric around the equator and bounded by  longitudes $\lambda=0$ and $\lambda=\pi$ (in blue). Energy dissipation is quantified by the linear  Rayleigh drag frequency $\sigma_{\rm R}.$ The logarithm of the torque is plotted as a function of the normalized  frequency $\omega= (\Omega - n_{\rm orb})/\Omega_0$, where the Earth's spin rate varies with the tidal forcing frequency $\Omega= n_{\rm orb} + \sigma/2$ at fixed $n_{\rm orb}$, and $\Omega_0$ being the present spin rate of the Earth. } \label{compare_global_hemi_torque}
\end{figure*}
\begin{figure*}
\centering
\includegraphics[width=\textwidth]{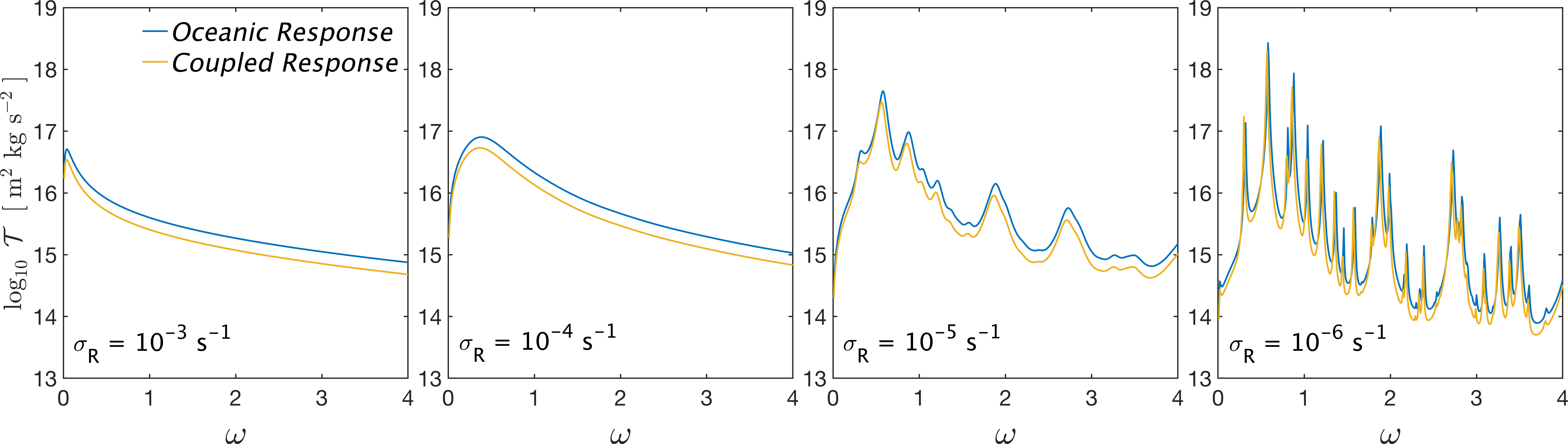}
\caption{Similar to   Fig.\ref{compare_global_hemi_torque}, but comparing the torque of the hemispherical ocean model between a pure oceanic response, and the response of the ocean when accounting for loading and self-attraction effects arising from solid Earth deformation assuming an Andrade rheology. The procedure of this coupling for the hemispheric configuration is detailed in Appendix \ref{section_coupling}. We recall that the energy tidally dissipated in the solid part is ignored in the hemispherical configuration. }
\label{hemispheric_pure_vs_solid}
\end{figure*}
\begin{equation}
    \zeta_l^{m,\sigma} = \sum_n     A_{n,l}^{m,\Tilde{\nu}} \zeta_n^{m,\sigma},
\end{equation}
where the components $\zeta_n^{m,\sigma}$ are solutions to the linear algebraic system\PAR{}{:}
% \begin{equation}
%             \begin{bmatrix}
%                     \sigma\Tilde{\sigma}-\sigma^2_{1,1} & -\sigma^2_{1,n} & -\sigma^2_{1,N} \\
%                     -\sigma^2_{n,1} &\sigma\Tilde{\sigma}-\sigma^2_{n,n}&-\sigma^2_{n,N} \\
%                     -\sigma^2_{N,1} & -\sigma^2_{N,n}&  \sigma\Tilde{\sigma}-\sigma^2_{N,N}
%             \end{bmatrix} \begin{bmatrix}
%                     \zeta_1^{m,\sigma}\\\zeta_n^{m,\sigma}\\\zeta_N^{m,\sigma}
%             \end{bmatrix}
%             =
%         \begin{bmatrix}
%                      \mathcal{F}_1^{m,\sigma}\\ \mathcal{F}_n^{m,\sigma}\\ \mathcal{F}_N^{m,\sigma}
      
%         \end{bmatrix}
% \end{equation}
\begin{equation}  \label{linear_system_hough}
\left(  \sigma\Tilde{\sigma} \boldsymbol{I}_N  - \begin{bmatrix}
                  \sigma^2_{1,1} & \dots& \sigma^2_{1,n} & \dots& \sigma^2_{1,N} \\
                  \vdots &\ddots& \vdots& & \vdots\\
                  \sigma^2_{n,1} & \dots& \sigma^2_{n,n} &\dots & \sigma^2_{n,N} \\
                  \vdots && \vdots&\ddots & \vdots\\
                    \sigma^2_{N,1}&\dots & \sigma^2_{N,n}& \dots &\sigma^2_{N,N}
            \end{bmatrix}  
\right)           
\begin{bmatrix}
                    \zeta_1^{m,\sigma}\\\vdots\\\zeta_n^{m,\sigma}\\\vdots\\\zeta_N^{m,\sigma}
            \end{bmatrix}
            =
        \begin{bmatrix}
                     \mathcal{F}_1^{m,\sigma}\\\vdots\\ \mathcal{F}_n^{m,\sigma}\\\vdots\\ \mathcal{F}_N^{m,\sigma}
      
        \end{bmatrix}\PAR{.}{,}
\end{equation}

In this linear system, $\boldsymbol{I}_N $ denotes the identity matrix of size $N\times N$,  the forcing terms of the studied tidal potential $U_l^{m,\sigma}$ (Eq. \ref{tidal_pot}) \PAR{are expressed as}{being}
\begin{equation} \label{force_vec_component}
     \mathcal{F}_n^{m,\sigma} = -\frac{H \Lambda_n^{m,\Tilde{v}}}{R^2} \sum_{m\leq l}B_{l,n}^{m,\Tilde{\nu}}\gamma_l^T U_l^{m,\sigma},
\end{equation}
and the complex characteristic frequencies \PAR{$\sigma_{n,k}$}{defined} as
 \begin{equation}\label{characteristic_frequencies}
            \sigma_{n,k}= \sqrt{gH \hat{k}_n^2\sum_{l\geq m}\gamma_{l}^L A_{k;l}^{m,\Tilde{\nu}} B_{l,n}^{m,\Tilde{\nu}}},
        \end{equation}
where the horizontal wave-number of the degree-$n$ mode $\hat{k}_n= \sqrt{\Lambda_n^{m,\Tilde{\nu}}}/R$, and the coupling coefficients $\gamma_l^{\rm T}$ and $\gamma_l^{\rm L}$ are defined in \eq{tilt}. Once the solution of this algebraic system is obtained, the self-consistent tidal response of the Earth \PAR{is}{can be} quantified by the total frequency dependent complex \PA{Love number defined, for each order $m$ and degree $l$}, as
\begin{equation} \label{final_love}
    \mathfrak{k}_l^{m,\sigma} = k_l^{\rm T} + \left(1+k_l^{\rm L}\right)\frac{3g}{2l+1}\frac{\rho_{\rm oc}}{\rho_{\rm se}}\frac{\zeta_l^{m,\sigma}}{U_l^{m,\sigma}}.
\end{equation}

The first term of the above expression accounts for the direct tidal gravitational forcing of the solid part by the perturber. The second term is related to the oceanic tidal response, which is coupled to that of the solid part through gravitational and surface loading interactions. We remark that the effective Love number characterizing the full tidal response of the planet (Eq. \ref{final_love}) depends on both the latitudinal and longitudinal harmonic degrees $l$ and $m$, in contrast with  the solid Love number $k_l^{\rm T}$. This results from the fact that Coriolis forces alter the oceanic tidal response, which is not the case for the solid tidal response. The contribution of the component $U_l^{m,\sigma}$ of the tidal potential to the total tidal torque exerted on the Earth scales as the imaginary part of the associated Love number and is expressed as  \citep{efroimsky2009tidal,correia2014deformation}
\begin{equation} \label{global_torque}
    \mathcal{T}_l^m = \frac{3}{2}GM^2 \frac{R^5}{a^6}\Im{\mathfrak{k}_l^{m,\sigma}}.
\end{equation}
Since we restricted our analysis to the study of the dominant semi-diurnal tide, we only consider the quadrupolar potential with $l=m=2$. 

In   Fig. \ref{compare_global_hemi_torque}, we compute the tidal torque for both the hemispheric and global oceanic geometries for a fixed value of $H$ and different orders of magnitude of $\sigma_{\rm R}$. We consider the semi-diurnal lunar gravitational forcing exerted on the Earth. The spectrum of the torque is plotted against the normalized frequency $\omega= (\Omega- n_{\rm orb})/\Omega_0$, where $\Omega_0$ is the present spin rate of the Earth. The distribution of resonances associated with surface-gravity modes distorted by rotation is clearly visible for both geometries. In the global ocean case, \PAR{these resonances are each characterised by the pair of complex frequencies given by}{the spectral positions of these frequencies are computed and they correspond} \citep{auclair2018oceanic}
\begin{equation}\label{resonance_freqs}
    \sigma_n^\pm = i\frac{\sigma_{\rm R}}{2} \pm \sqrt{gH \hat{k}_n^2 - \left(\frac{\sigma_{\rm R}}{2}\right)^2} \PA{,}
\end{equation}
which depicts explicitly the predominance of friction over the rotational distortion of tidal waves in a strong friction regime. This can be verified by visual inspection of   Fig. \ref{compare_global_hemi_torque}. The spectral coverage of the non-resonant background of the torque increases with increasing $\sigma_{\rm R}$. In the opposite limit, resonant peaks are  spelled out intensifying in amplitude as friction is weakened. Besides, when $\sigma_{\rm R}\rightarrow0$, the frequencies $\sigma_n^\pm$ become real and positive, and we recover the eigen-frequencies of \PA{large-wavelength }surface gravity modes travelling around the sphere. It can be also clearly seen from the high friction regime that the  torque of the global ocean is twice that of the hemispherical one, consistent with the simple argument of dissipation increasing proportionally with oceanic area. The same can be deduced if we consider the non-resonant background of the weak friction regime. Comparing the two spectra in this limit reveals the highly irregular nature of the waveforms in the hemispheric response against the fairly regular resonance periodicity in the global configuration. Several resonances can be encountered in the hemispherical configuration spectrum in between two resonant peaks of the global configuration. 
Applied to the Earth-Moon system evolution studied in the main text, we start at the present with a hemispheric ocean, then we switch to a global one. The fitted parameters of $H$ and $\sigma_{\rm R}$ would place the present torque around a resonant peak, then multiple resonances are crossed before settling into the non-resonant background of the hemispherical ocean response. The switch between the configurations occurs right before surfing the next major resonance. 

In   Fig. \ref{hemispheric_pure_vs_solid}, we plot the torque of the hemispheric configuration for two scenarios: accounting for the oceanic response only, and accounting for both the oceanic and solid responses self-consistently. As explained in Appendix \ref{section_coupling}, the effects of self-attraction and loading interactions between the solid mantle and the oceanic shell are evident in attenuating the amplitude of the response and slightly shifting the position of resonances. This delay effect is due to the influence of this coupling on the phase of resonance depths of near-resonant free oscillations \citep{muller2008synthesis}.

\end{appendix}


\begin{thebibliography}{121}
\expandafter\ifx\csname natexlab\endcsname\relax\def\natexlab#1{#1}\fi

\bibitem[{Abramowitz {et~al.}(1988)Abramowitz, Stegun, \&
  Romer}]{abramowitz1988handbook}
Abramowitz, M., Stegun, I.~A., \& Romer, R.~H. 1988, Handbook of mathematical
  functions with formulas, graphs, and mathematical tables

\bibitem[{Alcott {et~al.}(2019)Alcott, Mills, \& Poulton}]{alcott2019stepwise}
Alcott, L.~J., Mills, B.~J., \& Poulton, S.~W. 2019, Science, 366, 1333

\bibitem[{Amante \& Eakins(2009)}]{amante2009etopo1}
Amante, C. \& Eakins, B.~W. 2009, NOAA Technical Memorandum NESDIS NGDC-24

\bibitem[{Andrade(1910)}]{andrade1910viscous}
Andrade, E. N. D.~C. 1910, Proceedings of the Royal Society of London. Series
  A, Containing Papers of a Mathematical and Physical Character, 84, 1

\bibitem[{Arbic \& Garrett(2010)}]{arbic2010coupled}
Arbic, B.~K. \& Garrett, C. 2010, Continental Shelf Research, 30, 564

\bibitem[{Arbic {et~al.}(2009)Arbic, Karsten, \& Garrett}]{arbic2009tidal}
Arbic, B.~K., Karsten, R.~H., \& Garrett, C. 2009, Atmosphere-Ocean, 47, 239

\bibitem[{Arfken \& Weber(1999)}]{arfken1999mathematical}
Arfken, G.~B. \& Weber, H.~J. 1999, Mathematical methods for physicists

\bibitem[{Auclair-Desrotour {et~al.}(2014)Auclair-Desrotour, Le~Poncin-Lafitte,
  \& Mathis}]{auclair2014impact}
Auclair-Desrotour, P., Le~Poncin-Lafitte, C., \& Mathis, S. 2014, Astronomy \&
  Astrophysics, 561, L7

\bibitem[{Auclair-Desrotour {et~al.}(2019)Auclair-Desrotour, Leconte, Bolmont,
  \& Mathis}]{auclair2019final}
Auclair-Desrotour, P., Leconte, J., Bolmont, E., \& Mathis, S. 2019, Astronomy
  \& Astrophysics, 629, A132

\bibitem[{Auclair-Desrotour {et~al.}(2018)Auclair-Desrotour, Mathis, Laskar, \&
  Leconte}]{auclair2018oceanic}
Auclair-Desrotour, P., Mathis, S., Laskar, J., \& Leconte, J. 2018, Astronomy
  \& Astrophysics, 615, A23

\bibitem[{Bell~Jr(1975)}]{bell1975topographically}
Bell~Jr, T. 1975, Journal of Geophysical Research, 80, 320

\bibitem[{Bolmont {et~al.}(2020)Bolmont, Breton, Tobie, Dumoulin, Mathis, \&
  Grasset}]{Bolmont2020}
Bolmont, E., Breton, S.~N., Tobie, G., {et~al.} 2020, Astronomy \&
  Astrophysics, 644, A165

\bibitem[{Bou{\'e} \& Laskar(2006)}]{boue2006precession}
Bou{\'e}, G. \& Laskar, J. 2006, Icarus, 185, 312

\bibitem[{Boulila {et~al.}(2018)Boulila, Laskar, Haq, Galbrun, \&
  Hara}]{boulila2018long}
Boulila, S., Laskar, J., Haq, B.~U., Galbrun, B., \& Hara, N. 2018, Global and
  Planetary Change, 165, 128

\bibitem[{Boyden {et~al.}(2011)Boyden, M{\"u}ller, Gurnis, Torsvik, Clark,
  Turner, Ivey-Law, Watson, \& Cannon}]{boyden2011next}
Boyden, J.~A., M{\"u}ller, R.~D., Gurnis, M., {et~al.} 2011, Next-generation
  plate-tectonic reconstructions using GPlates (Cambridge University Press)

\bibitem[{Carter {et~al.}(2008)Carter, Merrifield, Becker, Katsumata, Gregg,
  Luther, Levine, Boyd, \& Firing}]{carter2008energetics}
Carter, G.~S., Merrifield, M., Becker, J.~M., {et~al.} 2008, Journal of
  Physical Oceanography, 38, 2205

\bibitem[{{Castelnau} {et~al.}(2008){Castelnau}, {Duval}, {Montagnat}, \&
  {Brenner}}]{Castelnau}
{Castelnau}, O., {Duval}, P., {Montagnat}, M., \& {Brenner}, R. 2008, Journal
  of Geophysical Research (Solid Earth), 113, B11203

\bibitem[{{Castillo-Rogez} {et~al.}(2011){Castillo-Rogez}, {Efroimsky}, \&
  {Lainey}}]{Castillo-Rogez}
{Castillo-Rogez}, J.~C., {Efroimsky}, M., \& {Lainey}, V. 2011, Journal of
  Geophysical Research (Planets), 116, E09008

\bibitem[{Correia {et~al.}(2014)Correia, Bou{\'e}, Laskar, \&
  Rodr{\'\i}guez}]{correia2014deformation}
Correia, A.~C., Bou{\'e}, G., Laskar, J., \& Rodr{\'\i}guez, A. 2014, Astronomy
  \& Astrophysics, 571, A50

\bibitem[{Correia \& Laskar(2010)}]{correia_tidal_2010}
Correia, A. C.~M. \& Laskar, J. 2010, in Exoplanets (Tucson, AZ: University of
  Arizona Press), 239--266

\bibitem[{Crease(1966)}]{crease1966tables}
Crease, J. 1966, Tables of the integral. $\int P_n^m(z)P_r^s(z) dz$

\bibitem[{{\'C}uk {et~al.}(2016){\'C}uk, Hamilton, Lock, \&
  Stewart}]{cuk2016tidal}
{\'C}uk, M., Hamilton, D.~P., Lock, S.~J., \& Stewart, S.~T. 2016, Nature, 539,
  402

\bibitem[{{\'C}uk {et~al.}(2019){\'C}uk, Hamilton, \& Stewart}]{cuk2019early}
{\'C}uk, M., Hamilton, D.~P., \& Stewart, S.~T. 2019, Journal of Geophysical
  Research: Planets, 124, 2917

\bibitem[{Daher {et~al.}(2021)Daher, Arbic, Williams, Ansong, Boggs,
  M{\"u}ller, Schindelegger, Austermann, Cornuelle, Crawford,
  {et~al.}}]{daher2021long}
Daher, H., Arbic, B.~K., Williams, J.~G., {et~al.} 2021, Journal of Geophysical
  Research: Planets, 126, e2021JE006875

\bibitem[{{Darwin}(1879)}]{Darwin1879}
{Darwin}, G.~H. 1879, Philosophical Transactions of the Royal Society of London
  Series I, 170, 447

\bibitem[{de~Azarevich \& Azarevich(2017)}]{de2017lunar}
de~Azarevich, V. L.~L. \& Azarevich, M.~B. 2017, Geo-Marine Letters, 37, 333

\bibitem[{Dhuime {et~al.}(2012)Dhuime, Hawkesworth, Cawood, \&
  Storey}]{dhuime2012change}
Dhuime, B., Hawkesworth, C.~J., Cawood, P.~A., \& Storey, C.~D. 2012, Science,
  335, 1334

\bibitem[{Dong \& Lemus(2002)}]{dong2002overlap}
Dong, S.-H. \& Lemus, R. 2002, Applied mathematics letters, 15, 541

\bibitem[{Dziewonski \& Anderson(1981)}]{dziewonski1981preliminary}
Dziewonski, A.~M. \& Anderson, D.~L. 1981, Physics of the earth and planetary
  interiors, 25, 297

\bibitem[{Efroimsky(2012)}]{efroimsky2012tidal}
Efroimsky, M. 2012, The Astrophysical Journal, 746, 150

\bibitem[{Efroimsky \& Williams(2009)}]{efroimsky2009tidal}
Efroimsky, M. \& Williams, J.~G. 2009, Celestial Mechanics and Dynamical
  Astronomy, 104, 257

\bibitem[{Eriksson \& Simpson(2000)}]{eriksson2000quantifying}
Eriksson, K.~A. \& Simpson, E.~L. 2000, Geology, 28, 831

\bibitem[{Fang {et~al.}(2020)Fang, Wu, Fang, Shi, Zhang, Yang, Li, \&
  Cao}]{fang_cyclostratigraphy_2020}
Fang, J., Wu, H., Fang, Q., {et~al.} 2020, Palaeogeography, Palaeoclimatology,
  Palaeoecology, 540, 109530

\bibitem[{Farhat \& Touma(2021)}]{farhat2021laplace}
Farhat, M.~A. \& Touma, J.~R. 2021, Monthly Notices of the Royal Astronomical
  Society, 507, 6078

\bibitem[{Farrell(1972)}]{farrell1972deformation}
Farrell, W. 1972, Reviews of Geophysics, 10, 761

\bibitem[{{Fienga} {et~al.}(2021){Fienga}, {Deram}, {Di Ruscio}, {Viswanathan},
  {Camargo}, {Bernus}, {Gastineau}, \& {Laskar}}]{INPOP21a}
{Fienga}, A., {Deram}, P., {Di Ruscio}, A., {et~al.} 2021, Notes Scientifiques
  et Techniques de l'Institut de M{\'e}canique C{\'e}leste, 110

\bibitem[{{Findley} {et~al.}(1977){Findley}, {Lai}, {Onaran}, \&
  {Christensen}}]{findley}
{Findley}, W.~N., {Lai}, J.~S., {Onaran}, K., \& {Christensen}, R.~M. 1977,
  Journal of Applied Mechanics, 44, 364

\bibitem[{{Fox-Kemper} {et~al.}(2003){Fox-Kemper}, {Ferrari}, \&
  {Pedlosky}}]{FoxKemper2003}
{Fox-Kemper}, B., {Ferrari}, R., \& {Pedlosky}, J. 2003, Journal of Physical
  Oceanography, 33, 478

\bibitem[{Garrett \& Munk(1971)}]{GARRETT1971493}
Garrett, C. \& Munk, W. 1971, Deep Sea Research and Oceanographic Abstracts,
  18, 493

\bibitem[{{Gent} \& {McWilliams}(1983)}]{GMW1983}
{Gent}, P.~R. \& {McWilliams}, J.~C. 1983, Dynamics of Atmospheres and Oceans,
  7, 67

\bibitem[{Gerkema \& Zimmerman(2008)}]{gerkema2008introduction}
Gerkema, T. \& Zimmerman, J. 2008, Lecture Notes, Royal NIOZ, Texel, 207

\bibitem[{Gerstenkorn(1967)}]{gerstenkorn1967controversy}
Gerstenkorn, H. 1967, Icarus, 7, 160

\bibitem[{Goldreich(1966)}]{goldreich1966history}
Goldreich, P. 1966, Reviews of Geophysics, 4, 411

\bibitem[{Green {et~al.}(2017)Green, Huber, Waltham, Buzan, \&
  Wells}]{green2017explicitly}
Green, J., Huber, M., Waltham, D., Buzan, J., \& Wells, M. 2017, Earth and
  Planetary Science Letters, 461, 46

\bibitem[{Griffiths \& Peltier(2009)}]{griffiths2009modeling}
Griffiths, S.~D. \& Peltier, W.~R. 2009, Journal of Climate, 22, 2905

\bibitem[{Gurnis {et~al.}(2012)Gurnis, Turner, Zahirovic, DiCaprio, Spasojevic,
  M{\"u}ller, Boyden, Seton, Manea, \& Bower}]{gurnis2012plate}
Gurnis, M., Turner, M., Zahirovic, S., {et~al.} 2012, Computers \& Geosciences,
  38, 35

\bibitem[{{Han} \& {Huang}(2020)}]{HH2020helmholtz}
{Han}, L. \& {Huang}, R.~X. 2020, Journal of Physical Oceanography, 50, 679

\bibitem[{Hansen(1982)}]{hansen1982secular}
Hansen, K.~S. 1982, Reviews of Geophysics, 20, 457

\bibitem[{Hawkesworth {et~al.}(2020)Hawkesworth, Cawood, \&
  Dhuime}]{hawkesworth2020evolution}
Hawkesworth, C., Cawood, P.~A., \& Dhuime, B. 2020, Frontiers in earth science,
  8

\bibitem[{{Hough}(1898)}]{hough}
{Hough}, S.~S. 1898, Philosophical Transactions of the Royal Society of London
  Series A, 191, 139

\bibitem[{Huang {et~al.}(2020)Huang, Gao, Jones, Tao, Carroll, Ibarra, Wu, \&
  Wang}]{huang2020astronomical}
Huang, H., Gao, Y., Jones, M.~M., {et~al.} 2020, Palaeogeography,
  Palaeoclimatology, Palaeoecology, 550, 109735

\bibitem[{Johnson \& Wing(2020)}]{johnson2020limited}
Johnson, B.~W. \& Wing, B.~A. 2020, Nature Geoscience, 13, 243

\bibitem[{Kaula(2013)}]{kaula2013theory}
Kaula, W.~M. 2013, Theory of satellite geodesy: applications of satellites to
  geodesy (Courier Corporation)

\bibitem[{Klatt {et~al.}(2021)Klatt, Chennu, Arbic, Biddanda, \&
  Dick}]{klatt2021possible}
Klatt, J.~M., Chennu, A., Arbic, B.~K., Biddanda, B., \& Dick, G.~J. 2021,
  Nature Geoscience, 14, 564

\bibitem[{Lantink {et~al.}(2021)Lantink, Davies, \& Hilgen}]{Lantink2021}
Lantink, M., Davies, J., \& Hilgen, F. 2021, in review

\bibitem[{Laskar(2005)}]{laskar2005note}
Laskar, J. 2005, Celestial Mechanics and Dynamical Astronomy, 91, 351

\bibitem[{Laskar {et~al.}(2004)Laskar, Robutel, Joutel, Gastineau, Correia, \&
  Levrard}]{laskar2004long}
Laskar, J., Robutel, P., Joutel, F., {et~al.} 2004, Astronomy \& Astrophysics,
  428, 261

\bibitem[{Lau {et~al.}(2016{\natexlab{a}})Lau, Faul, Mitrovica, Al-Attar,
  Tromp, \& Garapi{\'c}}]{lau2016anelasticity}
Lau, H.~C., Faul, U., Mitrovica, J.~X., {et~al.} 2016{\natexlab{a}},
  Geophysical journal international, ggw401

\bibitem[{Lau {et~al.}(2016{\natexlab{b}})Lau, Mitrovica, Austermann, Crawford,
  Al-Attar, \& Latychev}]{lau2016inferences}
Lau, H.~C., Mitrovica, J.~X., Austermann, J., {et~al.} 2016{\natexlab{b}},
  Journal of Geophysical Research: Solid Earth, 121, 6991

\bibitem[{Lau {et~al.}(2015)Lau, Yang, Tromp, Mitrovica, Latychev, \&
  Al-Attar}]{lau2015normal}
Lau, H.~C., Yang, H.-Y., Tromp, J., {et~al.} 2015, Geophysical Journal
  International, 202, 1392

\bibitem[{Lee \& Saio(1997)}]{lee1997low}
Lee, U. \& Saio, H. 1997, The Astrophysical Journal, 491, 839

\bibitem[{Levrard \& Laskar(2003)}]{levrard2003climate}
Levrard, B. \& Laskar, J. 2003, Geophysical Journal International, 154, 970

\bibitem[{Longuet-Higgins(1968)}]{longuet1968eigenfunctions}
Longuet-Higgins, M.~S. 1968, Philosophical Transactions of the Royal Society of
  London. Series A, Mathematical and Physical Sciences, 262, 511

\bibitem[{Longuet-Higgins \& Pond(1970)}]{longuet1970free}
Longuet-Higgins, M.~S. \& Pond, G.~S. 1970, Philosophical Transactions of the
  Royal Society of London. Series A, Mathematical and Physical Sciences, 266,
  193

\bibitem[{MacDonald(1967)}]{macdonald1967evidence}
MacDonald, G. 1967, Proceedings of the Royal Society of London. Series A.
  Mathematical and Physical Sciences, 296, 298

\bibitem[{Matsuyama(2014)}]{matsuyama2014tidal}
Matsuyama, I. 2014, Icarus, 242, 11

\bibitem[{Matthews {et~al.}(2016)Matthews, Maloney, Zahirovic, Williams, Seton,
  \& Mueller}]{matthews2016global}
Matthews, K.~J., Maloney, K.~T., Zahirovic, S., {et~al.} 2016, Global and
  Planetary Change, 146, 226

\bibitem[{Maurice {et~al.}(2020)Maurice, Tosi, Schwinger, Breuer, \&
  Kleine}]{maurice2020long}
Maurice, M., Tosi, N., Schwinger, S., Breuer, D., \& Kleine, T. 2020, Science
  advances, 6, eaba8949

\bibitem[{Mavromatis \& Alassar(1999)}]{mavromatis1999generalized}
Mavromatis, H. \& Alassar, R. 1999, Applied mathematics letters, 12, 101

\bibitem[{Merdith {et~al.}(2021)Merdith, Williams, Collins, Tetley, Mulder,
  Blades, Young, Armistead, Cannon, Zahirovic, {et~al.}}]{merdith2021extending}
Merdith, A.~S., Williams, S.~E., Collins, A.~S., {et~al.} 2021, Earth-Science
  Reviews, 214, 103477

\bibitem[{Meyers \& Malinverno(2018)}]{meyers2018proterozoic}
Meyers, S.~R. \& Malinverno, A. 2018, Proceedings of the National Academy of
  Sciences, 115, 6363

\bibitem[{Mignard(1979)}]{mignard1979evolution}
Mignard, F. 1979, The Moon and the planets, 20, 301

\bibitem[{Mojzsis {et~al.}(1996)Mojzsis, Arrhenius, McKeegan, Harrison, Nutman,
  \& Friend}]{mojzsis1996evidence}
Mojzsis, S.~J., Arrhenius, G., McKeegan, K., {et~al.} 1996, Nature, 384, 55

\bibitem[{Motoyama {et~al.}(2020)Motoyama, Tsunakawa, \&
  Takahashi}]{motoyama2020tidal}
Motoyama, M., Tsunakawa, H., \& Takahashi, F. 2020, Icarus, 335, 113382

\bibitem[{M{\"u}ller(2008{\natexlab{a}})}]{muller2008large}
M{\"u}ller, M. 2008{\natexlab{a}}, A large spectrum of free oscillations of the
  World Ocean including the full ocean loading and self-attraction effects,
  Vol.~14 (Springer Science \& Business Media)

\bibitem[{M{\"u}ller(2008{\natexlab{b}})}]{muller2008synthesis}
M{\"u}ller, M. 2008{\natexlab{b}}, Ocean Modelling, 20, 207

\bibitem[{Munk \& MacDonald(1960)}]{munk1960rotation}
Munk, W.~H. \& MacDonald, G.~J. 1960, Cambridge [Eng.] University Press

\bibitem[{Neron~de Surgy \& Laskar(1997)}]{neron1997long}
Neron~de Surgy, O. \& Laskar, J. 1997, Astronomy and Astrophysics, 318, 975

\bibitem[{Ogilvie(2014)}]{ogilvie2014tidal}
Ogilvie, G.~I. 2014, Annual Review of Astronomy and Astrophysics, 52, 171

\bibitem[{Ooe(1989)}]{ooe1989effects}
Ooe, M. 1989, Journal of Physics of the Earth, 37, 345

\bibitem[{Palmer {et~al.}(1986)Palmer, Shutts, \&
  Swinbank}]{palmer1986alleviation}
Palmer, T., Shutts, G., \& Swinbank, R. 1986, Quarterly Journal of the Royal
  Meteorological Society, 112, 1001

\bibitem[{Peck {et~al.}(2001)Peck, Valley, Wilde, \& Graham}]{peck2001oxygen}
Peck, W.~H., Valley, J.~W., Wilde, S.~A., \& Graham, C.~M. 2001, Geochimica et
  Cosmochimica Acta, 65, 4215

\bibitem[{Petit \& Luzum(2010)}]{petit2010iers}
Petit, G. \& Luzum, B. 2010, IERS conventions (2010), Tech. rep., Bureau
  International des Poids et mesures sevres (france)

\bibitem[{Platzman(1983)}]{platzman1983world}
Platzman, G.~W. 1983, Science, 220, 602

\bibitem[{Platzman(1984)}]{platzman1984normal}
Platzman, G.~W. 1984, Journal of physical oceanography, 14, 1532

\bibitem[{Proudman(1920{\natexlab{a}})}]{proudman1920dynamical1}
Proudman, J. 1920{\natexlab{a}}, Proceedings of the London Mathematical
  Society, 2, 1

\bibitem[{Proudman(1920{\natexlab{b}})}]{proudman1920dynamical}
Proudman, J. 1920{\natexlab{b}}, Proceedings of the London Mathematical
  Society, 2, 51

\bibitem[{Regge(1958)}]{regge1958symmetry}
Regge, T. 1958, Il Nuovo Cimento (1955-1965), 10, 544

\bibitem[{Renaud \& Henning(2018)}]{renaud2018increased}
Renaud, J.~P. \& Henning, W.~G. 2018, The Astrophysical Journal, 857, 98

\bibitem[{Riley {et~al.}(1999)Riley, Hobson, \& Bence}]{riley1999mathematical}
Riley, K.~F., Hobson, M.~P., \& Bence, S.~J. 1999, Mathematical methods for
  physics and engineering

\bibitem[{Ross \& Schubert(1989)}]{ross1989evolution}
Ross, M. \& Schubert, G. 1989, Journal of Geophysical Research: Solid Earth,
  94, 9533

\bibitem[{Rubincam(2016)}]{rubincam2016tidal}
Rubincam, D.~P. 2016, Icarus, 266, 24

\bibitem[{Sonett \& Chan(1998)}]{sonett1998neoproterozoic}
Sonett, C. \& Chan, M.~A. 1998, Geophysical Research Letters, 25, 539

\bibitem[{S{\o}rensen {et~al.}(2020)S{\o}rensen, Nielsen, Thibault, Zhao,
  Schovsbo, \& Dahl}]{sorensen2020astronomically}
S{\o}rensen, A.~L., Nielsen, A.~T., Thibault, N., {et~al.} 2020, Earth and
  Planetary Science Letters, 548, 116475

\bibitem[{Strauss(2007)}]{strauss2007partial}
Strauss, W.~A. 2007, Partial differential equations: An introduction (John
  Wiley \& Sons)

\bibitem[{Sun {et~al.}(2019)Sun, Xu, Cawood, Tang, Zhao, Li, \&
  Zhang}]{sun2019crustal}
Sun, C., Xu, W., Cawood, P.~A., {et~al.} 2019, Scientific reports, 9, 1

\bibitem[{Tobie {et~al.}(2019)Tobie, Grasset, Dumoulin, \&
  Mocquet}]{Tobie2019aap}
Tobie, G., Grasset, O., Dumoulin, C., \& Mocquet, A. 2019, Astronomy \&
  Astrophysics, 630, A70

\bibitem[{Tobie {et~al.}(2005)Tobie, Mocquet, \& Sotin}]{Tobie2005icarus}
Tobie, G., Mocquet, A., \& Sotin, C. 2005, Icarus, 177, 534

\bibitem[{Touma \& Wisdom(1994)}]{touma1994evolution}
Touma, J. \& Wisdom, J. 1994, The Astronomical Journal, 108, 1943

\bibitem[{Touma \& Wisdom(1998)}]{touma1998resonances}
Touma, J. \& Wisdom, J. 1998, The Astronomical Journal, 115, 1653

\bibitem[{Touma \& Wisdom(2001)}]{touma2001nonlinear}
Touma, J. \& Wisdom, J. 2001, The Astronomical Journal, 122, 1030

\bibitem[{Tremaine {et~al.}(2009)Tremaine, Touma, \&
  Namouni}]{tremaine2009satellite}
Tremaine, S., Touma, J., \& Namouni, F. 2009, The astronomical journal, 137,
  3706

\bibitem[{Tyler(2011)}]{tyler2011tidal}
Tyler, R. 2011, Icarus, 211, 770

\bibitem[{Tyler(2021)}]{tyler2021tidal}
Tyler, R.~H. 2021, The Planetary Science Journal, 2, 70

\bibitem[{Vallis(2017)}]{vallis2017atmospheric}
Vallis, G.~K. 2017, Atmospheric and oceanic fluid dynamics (Cambridge
  University Press)

\bibitem[{Varshalovich {et~al.}(1988)Varshalovich, Moskalev, \&
  Khersonskii}]{Varshalovich}
Varshalovich, D.~A., Moskalev, A.~N., \& Khersonskii, V.~K. 1988, Quantum
  Theory of Angular Momentum (WORLD SCIENTIFIC)

\bibitem[{Walker \& Zahnle(1986)}]{walker1986}
Walker, J. C.~G. \& Zahnle, K.~J. 1986, Nature, 320, 600, aDS Bibcode:
  1986Natur.320..600W

\bibitem[{Waltham(2015)}]{waltham_milankovitch_2015}
Waltham, D. 2015, Journal of Sedimentary Research, 85, 990, aDS Bibcode:
  2015JSedR..85..990W

\bibitem[{Wang {et~al.}(2016)Wang, Boyd, \& Akmaev}]{wang2016computation}
Wang, H., Boyd, J.~P., \& Akmaev, R.~A. 2016, Geoscientific Model Development,
  9, 1477

\bibitem[{{Watterson}(2001)}]{Watterson2001}
{Watterson}, I.~G. 2001, Journal of Atmospheric and Oceanic Technology, 18, 691

\bibitem[{Webb(1973)}]{webb1973age}
Webb, D. 1973, in Deep Sea Research and Oceanographic Abstracts, Vol.~20,
  Elsevier, 847--852

\bibitem[{Webb(1980)}]{webb1980tides}
Webb, D. 1980, Geophysical Journal International, 61, 573

\bibitem[{Webb(1982)}]{webb1982tides}
Webb, D. 1982, Geophysical Journal International, 70, 261

\bibitem[{Wilde {et~al.}(2001)Wilde, Valley, Peck, \&
  Graham}]{wilde2001evidence}
Wilde, S.~A., Valley, J.~W., Peck, W.~H., \& Graham, C.~M. 2001, Nature, 409,
  175

\bibitem[{Williams(1990)}]{williams1990tidal}
Williams, G.~E. 1990, Journal of Physics of the Earth, 38, 475

\bibitem[{Williams(1997)}]{williams1997precambrian}
Williams, G.~E. 1997, Geophysical Research Letters, 24, 421

\bibitem[{Williams(2000)}]{williams2000geological}
Williams, G.~E. 2000, Reviews of Geophysics, 38, 37

\bibitem[{Williams \& Boggs(2016)}]{williams2016secular}
Williams, J.~G. \& Boggs, D.~H. 2016, Celestial Mechanics and Dynamical
  Astronomy, 126, 89

\bibitem[{Wood {et~al.}(2019)Wood, Liu, Bowyer, Wilby, Dunn, Kenchington,
  Cuthill, Mitchell, \& Penny}]{wood2019integrated}
Wood, R., Liu, A.~G., Bowyer, F., {et~al.} 2019, Nature ecology \& evolution,
  3, 528

\bibitem[{Zahel(1980)}]{zahel1980mathematical}
Zahel, W. 1980, Physics of the Earth and Planetary Interiors, 21, 202

\bibitem[{Zhong {et~al.}(2020)Zhong, Wu, Fan, Fang, Shi, Zhang, Yang, Li, \&
  Cao}]{zhong_late_2020}
Zhong, Y., Wu, H., Fan, J., {et~al.} 2020, Palaeogeography, Palaeoclimatology,
  Palaeoecology, 540, 109520

\end{thebibliography}
\end{document}